\documentclass[onecolumn,noshowpacs,nofootinbib,11pt]{revtex4-1}

\usepackage{float,xcolor,upgreek,yfonts,tikz}
\usepackage{color}
\newcommand{\bes}{\begin{subequations}}
\newcommand{\ees}{\end{subequations}}
\usepackage{graphicx,bigints}
\usepackage{graphicx,float}\usepackage[all]{xy}
\usepackage{amsmath,upgreek}

   \usepackage{caption}
   
 \usepackage{booktabs,makecell}
\usepackage{subcaption}
\usepackage{amssymb}

\usepackage{alphabeta}
 \usepackage{cancel}
\usepackage{dcolumn}
\usepackage{textgreek}
\usepackage{multirow} 
\usepackage{units}
\usepackage{overpic}
\usepackage{physics}
\usepackage{bm}

\newcommand{\beq}{\begin{eqnarray}}
\newcommand{\eeq}{\end{eqnarray}}
\newcommand{\be}{\begin{equation}}
\newcommand{\ee}{\end{equation}}

\newcommand{\bea}{\begin{eqnarray}}
\newcommand{\eea}{\end{eqnarray}}

\newcommand{\ba}{\begin{eqnarray}}
\newcommand{\ea}{\end{eqnarray}}
\usepackage[colorlinks,hyperindex,unicode]{hyperref}
\definecolor{green1}{RGB}{0,128,0} 
\hypersetup{hidelinks,backref=true,pagebackref=true,hyperindex=true,colorlinks=true,breaklinks=true,urlcolor= blue}
\hypersetup{%
  colorlinks = true,
  linkcolor  = blue,
  citecolor = cyan,
}
\usepackage{bookmark,textgreek}
\usepackage{hyperref,color,xcolor}
\hypersetup{hidelinks,hyperindex=true,colorlinks=true,breaklinks=true,urlcolor= blue}
\hypersetup{%
  colorlinks = true,
  linkcolor  = blue
}
\newcommand\orcidcasadio{{\href{https://orcid.org/0000-0002-1330-7787}{\orcidicon}}}

\newcommand\orcidchristian{{\href{https://orcid.org/0009-0009-5593-0479}{\orcidicon}}}

\newcommand\orcidroldao{{\href{https://orcid.org/0000-0003-3978-532X}{\orcidicon}}}
\newcommand{\orcidicon}{%
	\begin{tikzpicture}
	\draw[lime, fill=lime] (0,0)
		circle [radius=0.16]
		node[white] {{\fontfamily{qag}\selectfont \tiny ID}};
	\draw[white, fill=white] (-0.0625,0.095)
		circle [radius=0.007];
	\end{tikzpicture}	\hspace{-2mm}
}

\begin{document}
\title{Gravitational decoupling and aerodynamics: black holes and analog gravity in a jet propulsion lab}

\author{R.~Casadio\orcidcasadio}
\affiliation{Dipartimento di Fisica e Astronomia, Universit\`a di Bologna, via Irnerio~46, 40126 Bologna, Italy}
\affiliation{I.N.F.N., Sezione di Bologna, I.S.~FLAG, viale B.~Pichat~6/2, 40127 Bologna, Italy}
\email{casadio@bo.infn.it}

\author{C. Noberto Souza\orcidchristian}
\affiliation{Center for Natural and Human Sciences, Federal University of ABC, 09210-580, Santo Andr\'e, Brazil}
\email{christian.noberto@ufabc.edu.br}

\author{R. da Rocha\orcidroldao}
\affiliation{Center of Mathematics, Federal University of ABC, 09210-580, Santo Andr\'e, Brazil}
\email{roldao.rocha@ufabc.edu.br}

\medbreak
\begin{abstract} 
{\color{black}
A connection is established between transonic sound waves propagating along a de Laval nozzle and 
quasinormal modes emitted from hairy black holes obtained with the gravitational decoupling method
applied to the Reissner--Nordstr\"om geometry.
Aerodynamical features provide an analogue setup to test experimentally perturbations of fluid flows
in a de~Laval nozzle producing quasinormal modes.
In particular, nozzle shape, pressure, Mach number, temperature, density, and thrust coefficient profiles
are determined as functions of the black hole parameters for several multipole numbers.
The black hole quasinormal mode frequencies are also investigated for different overtones, 
evaluating the quality factor of the nozzle.}  
\end{abstract}


\keywords{Gravitational decoupling; black holes; de Laval nozzle; self-gravitating compact objects; analog gravity; aerodynamics.}

\maketitle
\section{Introduction}

{\color{black}
The detection of gravitational waves (GW) emitted when black hole and neutron star binaries merge
is a most relevant success of fundamental physics.  
The temporal evolution of coalescing binary systems of compact stellar distributions essentially
embraces three stages.
An initial inspiral process, during which the frequency of GWs essentially equals the revolution frequency,
is followed by the formation of a remnant as the byproduct of the coalescence of the two components.
Finally, the remnant emits GWs during the ringdown phase until it reaches a configuration with some degree
of stability. 
Coalescing binary systems in the strong gravity regime, involving black holes with stellar mass,
and also degenerate stellar distributions, represent the most favourable candidates for generating
GWs that can be detected by radar interferometry.
The evolution of detectors like LIGO, KAGRA, and Virgo~\cite{LIGOScientific:2017vwq}, will eventually
be able to probe a large set of theories of gravity beyond general relativity (GR). 

The gravitational decoupling (GD)~\cite{Ovalle:2017fgl,Ovalle:2019qyi,Ovalle:2018ans,Ovalle:2013xla}
of Einstein's field equations has proven an effective theoretical method for building possible stellar distributions
in many circumstances~\cite{Casadio:2017sze,Casadio:2023mgl,Estrada:2019aeh,Gabbanelli:2019txr,Leon:2023nbj,
daRocha:2020jdj,Contreras:2022vec,Maurya:2021zvb,Singh:2021iwv,Maurya:2021tca,
Cavalcanti:2022adb,Pant:2023irv,Maurya:2023szc,Jasim:2021kga,Cavalcanti:2016mbe,
Ramos:2021drk,Rincon:2019jal,daRocha:2021aww,daRocha:2021sqd,Tello-Ortiz:2019gcl,Panyasiripan:2024kyu,Rehman:2023eor,Cavalcanti:2022xet,Morales:2018urp,
Heras:2021xxz,Panotopoulos:2018law,Singh:2019ktp,Albalahi:2024vpy,Sharif:2023ecm}.
The GD allows for describing the inner region of self-gravitational compact structures containing
realistic as well as exotic fluids with anisotropic distributions.
Einstein's field equations are highly non-linear, implying that available analytical solutions
that are also physically sound is scarce, except for the case of very specific scenarios. 
The GD extension is based on the splitting of the source energy-momentum into two components.
The first one is chosen to generate a known solution of GR, whereas the second 
component corresponds to an additional source, which can carry any type of charge,
including tidal and gauge ones, or hairy fields associated with gravity beyond GR.
The GD method is particularly suited for describing anisotropic compact stars~
\cite{Ovalle:2023ref,Gabbanelli:2018bhs,
PerezGraterol:2018eut,Andrade:2023wux,Heras:2018cpz,Hensh:2019rtb,
Contreras:2019iwm,Tello-Ortiz:2021kxg,Sharif:2020vvk}. 
It has also allowed to shed new light on some aspects of the AdS/CFT correspondence,
with the tools of holographic quantum entanglement entropy~\cite{Meert:2020sqv,Meert:2021khi,daRocha:2019pla,daRocha:2020gee}. 
Anisotropic degenerate quark and neutron stars were reported in
Refs.~\cite{Sharif:2022vvi,Contreras:2021xkf,Sharif:2020lbt,Maurya:2020ebd},
whereas black holes with hairy charges were recently detailed in Refs.~\cite{Ovalle:2020kpd,
Ovalle:2021jzf,Avalos:2023ywb,Contreras:2021yxe,Sultana:2020pcc,Ditta:2023arv,Zhang:2022niv}. 

One of the most fundamental features of black holes is given by their quasinormal (QN) resonant frequencies
in response to perturbations.
Crucial aspects of GD-extended black holes have already been disclosed by investigating their QN
modes~\cite{Cavalcanti:2022cga,Yang:2022ifo,Li:2022hkq,Avalos:2023jeh}. 
QN modes do not depend on the cause of the perturbation, but solely on the
externally observable parameters of the system.
In this work, we show how GD-extended black hole solutions can be mapped into analog models of gravity
in order to test some of their possible features in a laboratory.
Such an approach was introduced in Ref.~\cite{Unruh:1980cg}, where acoustic waves propagating
through inviscid inhomogeneous fluid flows were shown to emulate wave equations for scalar fields
in black hole curved backgrounds.
In transonic fluid flows, although sound waves are allowed to propagate from the subsonic to the
supersonic region, they cannot propagate against the direction of flow.
Hence, the critical sonic point where the velocity of sound equals the fluid velocity acts as an acoustic horizon,
very much like an event horizon for sound waves.
For a fluid in a pipe, this horizon can form at the nozzle throat, where the tube is narrowest~\cite{Visser:1997ux}. 
Many analog gravity models have been proposed and a variety of experiments have been 
performed or designed to observe the analog of QN ringing, superradiance, and the Hawking
radiation~\cite{Torres:2016iee,Torres:2019sbr,Euve:2015vml,MunozdeNova:2018fxv,Mosna:2016nrt}.
Besides the importance of experimental evidence, analog models of gravity are relevant in the
theoretical realm to advance our understanding and new intuition about real black hole
physics~\cite{daRocha:2017lqj,daRocha:2017tiz}.
Ref.~\cite{Okuzumi:2007hf} showed that the QN ringing in sound waves is emitted from acoustic
black holes, analogously to black holes radiating the QN ringing in GWs. 
Ref.~\cite{Cuyubamba:2013iwa} proposed a de~Laval nozzle as the acoustic analog of a massive
scalar field propagating in a Schwarzschild black hole background, presenting the possibility
of observing the QN ringing of a scalar field with effective mass in the laboratory. 
Ref.~\cite{deOliveira:2023qxe} addressed QN modes of a conformally coupled scalar field on
a BTZ black hole background analog to a de~Laval nozzle, whose ending correlates to the
spatial infinity of the BTZ black hole solution. 

Since some of the gravitational excitations in black hole backgrounds can be described in
analogy to QN modes of sound waves in a nozzle, one can indirectly explore GD-hairy black hole
solutions in the context of aerodynamics.
In this work, we will precisely study QN modes of GD-extended hairy Reissner--Nordstr\"om
solutions using transonic waves in a de~Laval nozzle.
Given the lack of observational evidence for black hole hair, such experiments in aerodynamics 
could help to understand the physical features of GD-black hole solutions. 
Sec.~\ref{sec1} is devoted to reviewing the GD method to obtain GD-extended hairy
Reissner--Nordstr\"om solutions.
In Sec.~\ref{sec3}, the connection between perturbations on black hole geometries
and sound waves inside a de Laval nozzle is explored, describing the conditions and
constraint equations in which the analogy is valid.
Sec.~\ref{sec4} is dedicated to showing how GD-extended black hole parameters map
into the nozzle shape, the profiles of the pressure, Mach number, temperature,
density, and the thrust coefficient.
The QN mode frequencies are computed with the Mashhoon method and,
subsequently, the quality factor of the analogue de~Laval nozzle is calculated.
Finally, Sec.~\ref{sec5} discusses the main results and contains the concluding remarks.
}

{\color{black}
\section{GD-extended black hole solutions}
\label{sec1}
The GD method allows to extend known solutions of GR in order to accommodate for
more complex sources, splitting it into terms that can be more easily worked
out~\cite{Ovalle:2017fgl,Ovalle:2019qyi},
including the case of GD-hairy black holes~\cite{Ovalle:2020kpd}.
The Einstein field equations are expressed as usual as
\begin{equation}
\label{corr2}
R_{\mu\nu}-\frac{1}{2}R g_{\mu\nu}
=
\upkappa^2\,\breve{T}_{\mu\nu}.
\end{equation}
The energy-momentum source in Eq. (\ref{corr2}) must satisfy the conservation
law $\nabla_\mu\,\breve{T}^{\mu\nu}=0$, because the Bianchi identity holds for the
Einstein tensor, and can be split into 
\begin{equation}
\label{emt}
\breve{T}_{\mu\nu}
=
{T}^{\rm}_{\mu\nu}
+
\alpha\,\vartheta_{\mu\nu},
\end{equation}
where ${T}^{\mu}_{\ \nu}={\rm diag}[\rho,-p,-p,-p]$ represents a perfect fluid with energy density
$\rho$ and isotropic pressure $p$.
The tensor $\vartheta_{\mu\nu}$ can describe any exotic source,
like dark matter, dark energy, towers of Kaluza--Klein bosons or fermions,
as well as gravitational contributions beyond GR, like the embedding in extra-dimensions,
stringy or effective quantum gravity
corrections~\cite{Ovalle:2010zc,Antoniadis:1998ig,daRocha:2017cxu,daRocha:2012pt,Casadio:2019usg,Abdalla:2009pg,Casadio:2023iqt}.

Static and spherically symmetric stellar distributions can be described by a metric in
Schwarzschild-like coordinates
\begin{equation}
ds^{2}
=
e^{\upsigma (r)}dt^{2}-e^{\upbeta (r)}dr^{2}
-r^{2}d\theta^2-r^2\sin^2\theta d\phi^2. 
\label{metric}
\end{equation}
The Einstein field equations~(\ref{corr2}) then explicitly read
\bes
\begin{eqnarray}
\label{ec1}
\!\!\!\!\!\!\!\!\!\!\!\!\upkappa^2\!
\left(
{T}_0^{\ 0}+\vartheta_0^{\ 0}
\right)
&\!=\!&
\frac 1{r^2}
-
e^{-\upbeta(r) }\left( \frac1{r^2}-\frac{\upbeta'(r)}r\right),
\\
\label{ec2}
\!\!\!\!\!\!\!\!\upkappa^2\!
\left({T}_1^{\ 1}+\vartheta_1^{\ 1}\right)
&\!=\!&
\frac 1{r^2}
-
e^{-\upbeta(r) }\left( \frac 1{r^2}+\frac{\upsigma'(r)}r\right),
\\
\label{ec3}
\!\!\!\!\!\!\!\!\!\!\!\!\!\!\!\upkappa^2\!
\left({T}_2^{\ 2}\!+\!\vartheta_2^{\ 2}\right)
&\!=\!&
\!-\frac {1}{4}e^{-\upbeta(r) }
\left[2\upsigma''(r)\!+\!\upsigma'^2(r)\!-\!\upbeta'(r)\upsigma'(r)
\!+\!\frac{2}{r}\left(\upsigma'(r)\!-\!\upbeta'(r)\right)\right],
\end{eqnarray}
\ees
with primes denoting derivatives with respect to $r$. 
Eqs.~(\ref{ec1}) -- (\ref{ec3}) define the effective energy density, 
radial and tangential pressures, respectively given by~\cite{Ovalle:2019qyi}
{\bes
\beq
\breve{\rho}
&=&
\rho+
\alpha\vartheta_0^{\ 0},\label{efecden}\\
\breve{p}_{\scalebox{.63}{\textsc{rad}}}
&=&
p
-\alpha\vartheta_1^{\ 1},
\label{efecprera}\\
\breve{p}_{\scalebox{.63}{\textsc{tan}}}
&=&
p
-\alpha\vartheta_2^{\ 2}, 
\label{efecpretan}
\eeq\ees}
with anisotropy factor
\beq
\Updelta = 
\breve{p}_{\scalebox{.63}{\textsc{tan}}}-\breve{p}_{\scalebox{.63}{\textsc{rad}}}
=
\alpha
\left(
\vartheta_1^{\ 1}
-
\vartheta_2^{\ 2}
\right)
 .
\eeq 

We can now consider a solution to the field equations~\eqref{corr2} generated 
by a given fluid with energy-momentum tensor ${T}_{\mu\nu}$~\cite{Ovalle:2020kpd}
\begin{equation}
ds^{2}
=
e^{\upxi(r)}dt^{2}
-e^{\upzeta(r)}dr^{2}
-r^{2}d\theta^2-r^2\sin^2\theta d\phi^2
,
\label{pfmetric}
\end{equation}
where the radial metric component 
\begin{equation}
\label{standardGR}
e^{-\upzeta(r)}
= 
1-\frac{\upkappa^2}{r}\int_0^r \mathtt{r}^2\,\rho(\mathtt{r})\, \dd\mathtt{r}
=
1-\frac{2m(r)}{r}
\end{equation}
defines the Misner--Sharp mass function $m(r)$.
We can next deform the metric as  
\bes
\begin{eqnarray}
\label{gd1}
\upxi(r)
&\mapsto &
\upsigma(r)=\upxi(r)+\alpha h(r),
\\
\label{gd2}
e^{-\upzeta(r)} 
&\mapsto &
e^{-\upbeta(r)}=e^{-\upzeta(r)}+\alpha g(r), 
\end{eqnarray}
\ees
where $h(r)$  and $g(r)$ are the geometric deformations for the time and radial  metric
components, respectively.
Eqs.~(\ref{gd1}, \ref{gd2}) split the system (\ref{ec1}) -- (\ref{ec3}) into two sets:
the Einstein field equations for ${T}_{\mu\nu}$, which are solved by the kernel metric~(\ref{pfmetric}),
and the coupled system 
\bes
\begin{eqnarray}
\label{ec1d}
\!\!\!\!\!\upkappa^2\,\vartheta_0^{\ 0}
&=&
-\alpha\left(\frac{g(r)}{r^2}+\frac{g'(r)}{r}\right),
\\
\label{ec2d}
\!\!\!\!\!\upkappa^2\,\vartheta_1^{\ 1}
+\frac{\alpha}{r}{e^{-\upzeta(r)}h'(r)}
&\!=\!&
-\alpha g(r)\left(\frac{1}{r^2}+\frac{\upsigma'(r)}{r}\right),
\\
\label{ec3d}
\!\!\!\!\!\!\!\!\!\!\!\!\!\upkappa^2\vartheta_2^{\ 2}\!+\!\alpha{g}\left(2\upsigma''(r)\!+\!\upsigma'^2(r)\!+\!\frac{2\upsigma'(r)}{r}\right)\!
&\!=\!&\!
-\frac{\alpha}{4}g'(r)\!\left(\upsigma'(r)\!+\!\frac{2}{r}\right)\!+\!W(r),
\end{eqnarray}
\ees
where \cite{Ovalle:2017fgl}
\beq
W(r) = \alpha e^{-\upzeta(r)}\left(2h(r)''+h'^2(r)+\frac{2h'(r)}{r}+2\upxi'(r)h'(r)-\upzeta'(r)h'(r)\right)
\ .
\eeq 
\par
The so-called tensor-vacuum is a region in which ${T}_{\mu\nu}=0$ but $\vartheta_{\mu\nu}\neq 0$,
corresponding to the vacuum concerning ordinary matter in GR, and provides the fundamental
setup for describing (the exterior of) hairy black holes. 
Eqs. (\ref{ec1}) -- (\ref{ec2}) correspond to a negative value for the radial pressure, as
\begin{equation}
\breve{p}_{\scalebox{.63}{\textsc{rad}}}
=
-\breve{\rho}.
\label{schwcon}
\end{equation}
It yields the expression 
\begin{equation}
\label{fg}
\alpha\,g(r)
=
\left(1-\frac{2M}{r}\right)\left(e^{\alpha\,h(r)}-1\right)
,
\end{equation}
and the metric~\eqref{metric} describes a deformation of the Schwarzschild metric:
\begin{eqnarray}
\label{hairyBH}
ds^{2}
&\!=\!&
e^{\alpha h(r)}\left(1-\frac{2M}{r}\right)
dt^{2}
\!-\!e^{-\alpha h(r)}\left(1-\frac{2M}{r}\right)^{-1}
dr^2-r^{2}d\theta^2-r^2\sin^2\theta d\phi^2. 
\end{eqnarray}
\par
Outside the Schwarzschild event horizon $r\gtrsim 2M$, the tensor-vacuum 
can be realized, for example, by assuming $\vartheta_0^{\ 0}
=c\,\vartheta_1^{\ 1}+d\,\vartheta_2^{\ 2}$, 
with $c$, $d\in\mathbb{R}$. 
Eqs.~(\ref{ec1d}) -- (\ref{ec3d}) then yield  
\begin{eqnarray}
\label{master}
\!\!\!\!\!\!\!\!\!\!\!\!\!\!\!\!\!\!&&b\,r\,(r-2M)\,k''(r)+2\,\left[(c+d-1)\,r-2\,(c-1)\,M\right]
k'(r)
+2\,(c-1)\,(k(r)-1)=0,
\end{eqnarray}
for 
$ 
k(r)
=
e^{\alpha\,h(r)}$.  Eq. (\ref{master}) has solutions of the following type \cite{Ovalle:2020kpd}:
\begin{equation}
\label{master2}
e^{\alpha\,h(r)}
=
1+\frac{1}{r-2M}
\left[\ell_0+r\left(\frac{\ell}{r}\right)^{N}
\right]
,
\end{equation}
where $\ell_0=\alpha\ell$ is a primary hair charge, whereas $
N
=
2\left(c-1\right)/d> 1$ ensures asymptotic flatness, corresponding to  
\bes
\beq
\breve{\rho}
&=&
\vartheta_0^{\ 0}
=
\frac{\alpha}{\upkappa^2(N-1)}\frac{\ell^N}{r^{N+2}},
\label{efecdenx}\\
\breve{p}_{\scalebox{.63}{\textsc{rad}}}
&=&
-\vartheta_1^{\ 1}
=
-\breve{\rho},\\
\breve{p}_{\scalebox{.63}{\textsc{tan}}}
&=&
-\vartheta_2^{\ 2}
=
\frac{N}{2}\,\breve{\rho}
.
\label{efecptanx}
\eeq
\ees
\par
The dominant energy conditions, 
\begin{eqnarray}
\breve{\rho}
\geq
|\breve{p}_{\scalebox{.63}{\textsc{rad}}}|,\qquad\qquad\qquad
\breve{\rho}
\geq
|\breve{p}_{\scalebox{.63}{\textsc{tan}}}|,
\label{dom2} 
\end{eqnarray}
using Eqs. (\ref{efecden}, \ref{efecpretan}) are respectively equivalent to 
\bes
\begin{eqnarray}
\label{dom6}
-r(r-2M)k''(r)
-4(r-M)k'(r)
-2k(r)+2
&\geq&
0,
\qquad
\\
\label{dom61}
r\,(r-2M)\,k''(r)
+4\,M\,k'(r)
-2\,k(r)+2
&\geq&
0.
\end{eqnarray} 
\ees
One can therefore solve the equality part of Eq. \eqref{dom6}, obtaining 
\begin{equation}
\label{dominantg}
k(r)
=
1
-
\frac{1}{r-2M}
\left(
\alpha\,\ell
+\alpha\,M\,e^{-r/M}
-\frac{Q^2}{r}
\right)
,
\end{equation}
where $Q$ is an integration constant that could represent tidal or gauge charges.
The saturated case in condition \eqref{dom61} with Eq.~\eqref{dominantg} reads
\begin{equation}
\frac{4\,Q^2}{r^2}
\ge
\frac{\alpha}{M}\,(r+2M)\,e^{-r/M}
.\label{sat}
\end{equation}
The GD-extended hairy Reissner--Nordstr\"om metric of Ref.~\cite{Ovalle:2020kpd} is then obtained
for the value of $Q$ that saturates the inequality (\ref{sat}) by replacing Eq.~\eqref{dominantg} into~\eqref{hairyBH}
and is given by
\begin{eqnarray}
\label{eq23}
e^{\upsigma(r)}
=
e^{-\upbeta(r)}&=&
1-\frac{2\mathcal{M}}{r}
+\frac{Q^2}{r^2}
-\frac{\alpha}{r}
\left(\mathcal{M}-\frac{\alpha\ell}{2}\right)
e^{-2r/(2\mathcal{M}-\alpha\ell)}
\ ,
\end{eqnarray}
where $\mathcal{M} = M+\alpha\ell/2$.
The parameter $\ell$ measures the increase of the entropy with respect to the Schwarzschild
area law, $S=4\pi {M^2}$, due to hairy effects~\cite{Ovalle:2020kpd,Casadio:2012rf}.
The inequality $\ell \leq 2{\cal M}/\alpha$ is also consistent with asymptotic flatness. 
Finally, we recall that the largest zero of Eq.~\eqref{eq23} defines the event horizon~\cite{Contreras:2021yxe}.

\section{Black holes and the de~Laval nozzle}
\label{sec3}

This section addresses the conditions for which sound waves in a fluid flowing within a de~Laval nozzle
can mimic scalar perturbations of the GD-extended Reissner--Nordstr\"om black hole.
We are particularly interested in the QN ringing modes.
Perturbing a real black hole can induce the emission of GWs, with a possible initial burst of 
strong-field radiation, followed by a typically longer period of damping oscillations dominated by the QN
modes.

For simplicity, we shall just consider a massless scalar field $\Psi\qty(x^\mu)$ obeying the Klein--Gordon
equation in a background described by the metric $g_{\mu\nu}$, that is
\begin{equation}
\label{Klein-Gordon}
    \frac{1}{\sqrt{-g}} \partial_\mu \qty(\sqrt{-g} g^{\mu \nu} \partial_\nu) \Psi \qty(x^\mu) = 0.
\end{equation}
With the solution~\eqref{eq23}, the Klein--Gordon equation reduces to 
\begin{equation}
\begin{split}
\label{Klein-Gordon in metric with values}
- \frac{1}{B} \frac{\partial^2 \Psi}{\partial t^2} + \frac{1}{A} \qty(\frac{2}{r} + \frac{1}{2B} \frac{\dd B}{\dd r} - \frac{1}{2A} \frac{\dd A}{\dd r}) \pdv{\Psi}{r} + \frac{1}{A} \pdv[2]{\Psi}{r} +\frac{1}{r^2} \qty[\frac{1}{\sin \theta} \frac{\partial}{\partial \theta}\qty(\sin \theta \pdv{\Psi}{\theta}) + \frac{1}{\sin ^2 \theta} \pdv[2]{\Psi}{\phi}] = 0,
 \end{split}
\end{equation}
where we defined $B\qty(r) = e^{\upsigma (r)}$ and $A\qty(r) = e^{\upbeta (r)}$. 
The field $\Psi\qty(x^\mu)$ in Eq. \eqref{Klein-Gordon in metric with values} can be separated into
the modes
\begin{equation}\label{ztr}
    \Psi_{lm}\qty(t,r,\theta,\phi)=Z_l\qty(t,r) Y_{l}^{m}\qty(\theta,\phi),
\end{equation}
where 
\be
Y_{l}^m
= 
(-1)^m
\sqrt{\frac{(2\,{l}+1)}{4\pi}\frac{({l}-m)!}{({l}+m)!}}
\,P_{l}^m(\cos\theta)\,e^{im \phi}
\ee
are spherical harmonics of degree ${l}$ and order $m$, with 
$P_{l}^m$ the associated Legendre polynomials.}
The $Z_l\qty(t,r)$ part  in Eq.~\eqref{ztr} can be expressed as $Z_l\qty(t,r)=\frac{1}{r}N\qty(t,r)$, yielding the classical wave equation 
\begin{equation}
    \label{wave equation}
    -\frac{\partial^2}{\partial t^2}{N(t,r_\star)} + \frac{\partial^2}{\partial r_\star^2}{N(t,r_\star)} = V_{\textsc{eff}}(r_\star) N(t,r_\star),
\end{equation}
with the effective potential
\begin{equation}
\label{final effective potential}
    V_{\textsc{eff}}(r) = B(r) \frac{l\qty(l+1)}{r^2} + \frac{1}{2r} \dv{}{r}\left(\frac{B\qty(r)}{A\qty(r)}\right),
\end{equation}
where the variable $r_\star(r)$ represents the tortoise coordinate in a general background
\begin{equation}
    \label{tortoise coordinate}
    \dv{r_\star}{r} = \qty(\frac{B}{A})^{-1/2},
\end{equation}
which will be particularly regarded as the GD-extended hairy Reissner--Nordstr\"om metric. 
The effective potential \eqref{final effective potential} has the dimension of the inverse of the square length, designating the curvature scattering of sound wave perturbations along the de Laval nozzle analog to the acoustic black hole \cite{okuzumi2007quasinormal}. 

For periodic oscillations of frequency $\omega$, a transformation in the temporal variable can be performed as $N(r,t) = \exp\left(-i \omega t\right) R(r)$, yielding 
\begin{equation}
    \label{quasinormal modes equation}
    \qty[\dv[2]{}{r_\star} + \omega^2 -  V_{\textsc{eff}}(r_\star)] R(r_\star) = 0,
\end{equation}
describing a periodic mode with frequency $\omega$. 


On the other hand, the Navier--Stokes equations \begin{equation}\label{eq:nse}
     \pdv{\vb{v}}{t} + \qty(\vb{v}\cdot \vb{\nabla})\vb{v} = - \frac{\nabla p}{\uprho} - \nabla \phi + \frac{1}{\uprho}\qty[\eta \nabla^2 \vb{v} + \qty(\xi + \frac{1}{3}\eta) \nabla\qty(\nabla\cdot\vb{v})]
\end{equation}
can describe the dynamics of a generic fluid flow, 
where $\eta$ is the dynamic shear viscosity, $\xi$ represents the second viscosity, $\uprho$ is the fluid density, and $\phi$ is a scalar potential. 
Assuming an inviscid flow, Eq. \eqref{eq:nse} becomes the Euler equation. If the fluid flow is isentropic, the velocity can be written in terms of the velocity potential as $\vb{v} = - \nabla\psi$. With these conditions, the set of equations that describe the fluid flow read 
\begin{align}
\label{conj eq}
        \pdv{(\uprho {\scalebox{.85}{\textsc{A}}})}{t} + \nabla\cdot{(\uprho {\scalebox{.85}{\textsc{A}}}\vb{v})} = 0, &&
        -\pdv{\psi}{t} + h + \frac{1}{2} (\nabla \psi)^2 + \phi = 0, \end{align}
where ${\scalebox{.85}{\textsc{A}}}$ is the cross-sectional area of the de Laval nozzle and $h$ denotes the specific enthalpy. To find the wave dynamics, Eq. \eqref{conj eq} must be perturbed, to wit:
\begin{equation}
\label{eq:soundpropagation}
    -\pdv{}{t}\qty{\frac{\uprho {\scalebox{.85}{\textsc{A}}}}{c_s^2}\qty[\pdv{(\delta \psi)}{t} +\vb{v}\cdot \qty(\nabla \delta \psi)]}
    +  \nabla\cdot{
    \qty{ 
     \uprho {\scalebox{.85}{\textsc{A}}} \nabla \delta \psi
     - \qty{\frac{\uprho {\scalebox{.85}{\textsc{A}}}}{c_s^2}\qty[\pdv{(\delta \psi)}{t} +\vb{v}\cdot \qty(\nabla \delta \psi)]} \vb{v}
    }} = 0.
\end{equation}
The Venturi effect governs the behaviour of de Laval nozzles. For fluid flows passing through a strangled part of some tube, which has variable cross-section ${\scalebox{.85}{\textsc{A}}}(x)$, the fluid pressure is then reduced, and the flow velocity increases. For the case of nonviscous, adiabatic, and isentropic fluids, consisting of an ideal gas, with an equation of state $p=\uprho R T$, for $T$ denoting 
 the temperature and $R$ the molar gas constant. Other fundamental quantities that characterise the ideal gas are the heat capacities, to constant pressure ($C_p$) and constant volume ($C_V$), with the property $R = C_p - C_V$. One denotes the adiabatic index  $\gamma= C_p/C_V$.   An isentropic fluid flow has the additional property 
  \begin{eqnarray}\label{isen}
     p = \uprho^\gamma=T^{\frac{\gamma}{\gamma-1}}\,,
     \end{eqnarray}      
     encoding the shock-free property and flow continuity.
The Mach number, 
$
\mathtt{M}(x)= \|{{\bf v}(x)}\|/{c_s(x)},
$ is a relevant dimensionless quantity in aerodynamics,
measuring the ratio of flow velocity to the local speed of sound, for  $c_s^2={\dd p}/{\dd \uprho}=\gamma RT$ denoting the (local) speed of sound and $x$ denotes the longitudinal nozzle coordinate.
The mass flux $\dd m/\dd t$ represents the mass of fluid flowing through the cross-section of the nozzle per unit of time. An important approximation to be taken is to assume that the nozzle radius $r = r(x)$ varies sufficiently slowly along the longitudinal direction of the nozzle axis, $x$. In such a way, the perturbations of the fluid flow in the de Laval nozzle can be considered quasi-one-dimensional. Then, rewriting Eq. \eqref{eq:soundpropagation} in the one-dimensional direction of propagation, denoting by $v_x \equiv v$ the only non-vanishing component of {\bf v}, the following equation is obtained, governing the perturbation of the scalar field:
\begin{align}
\label{eq:one-dimensional wave equation}
    \qty[\qty(\pdv{}{t} + \pdv{v}{x}) \frac{\uprho {\scalebox{.85}{\textsc{A}}}}{c^2_s} \qty(\pdv{}{t} + v\pdv{}{x}) - \pdv{}{x} \qty(\uprho {\scalebox{.85}{\textsc{A}}} \pdv{}{x})] \delta\psi = 0.
\end{align}
Now, analogously to the modes in Eq. \eqref{quasinormal modes equation}, if the solutions are set to be stationary, one can apply a Fourier transformation to express it in terms of the frequency $\omega$, as
\begin{equation*}
    \delta\psi(x,t) = \frac{1}{2\pi} \int \dd \omega e^{-i\omega t} \delta\psi_\omega(x). 
\end{equation*}
Replacing it into Eq. \eqref{eq:one-dimensional wave equation}, a time-independent differential equation for the new function $\delta\psi_\omega$ can be written as
\begin{multline}
\label{eq: time-independent differential equation}
\frac{1}{2\pi} \int \dd \omega  e^{-i \omega t}\left\{\uprho {\scalebox{.85}{\textsc{A}}}\qty(1-\frac{v^{2}}{c_{s}^{2}}) \dv[2]{}{x} +\qty[\dv{(\uprho {\scalebox{.85}{\textsc{A}}})}{x}+ 2 i \omega \frac{\uprho {\scalebox{.85}{\textsc{A}}} v}{c_{s}^{2}}-\dv{}{x} \qty(\frac{\uprho {\scalebox{.85}{\textsc{A}}} v^{2}}{c_{s}^{2}})] \dv{}{x}\right. \\
\left.+\qty[\omega^{2} \frac{\uprho {\scalebox{.85}{\textsc{A}}}}{c_{s}^{2}}+i \omega \dv{}{x}\qty(\frac{\uprho {\scalebox{.85}{\textsc{A}}} v}{c_{s}^{2}})] \right\}\delta\psi_{\omega} =0.
\end{multline}
Some auxiliary quantities can be employed, to simplify the analysis  \cite{okuzumi2007quasinormal, Abdalla2007}.   The first one encodes a transfer function, defined as
\begin{equation}
\label{eq:transfer function}
    H_\omega (x) = \sqrt{g_{\scalebox{.63}{\textsc{c}}}} \int \dd t e^{i\omega \qty[t - F(x)]} \delta \psi(t,x),
\end{equation}
where $g_{\scalebox{.63}{\textsc{c}}} ={\uprho {\scalebox{.85}{\textsc{A}}}}/{c_s}$ and $
    F(x) = \int \dd x {\abs{v}}/{(c^2_s - v^2)}.$ 
The second one consists of a change of coordinate $x = x(x_\star)$ based on the definition of Eq. \eqref{tortoise coordinate} for the metric of an acoustic black hole with Mach number $\mathtt{M}$. The tortoise coordinate for the canonical acoustic black hole can be written as
\begin{equation}
\label{eq: acoustic BN tortoise coordinate}
    \dv{x_\star}{x} = \frac{c_{s0}}{c_s(1 - \mathtt{M}^2)},
\end{equation}
where $c_{s0}$ denotes the stagnation sound speed, 
and $x_\star$ is the acoustic analog, on the aerodynamics side, of the tortoise coordinate, with $\lim_{x\to\infty}x_\star(x) = +\infty = -\lim_{x\to 0}x_\star(x)$. 
Therefore Eq. \eqref{eq: time-independent differential equation} assumes the form of the Schrödinger equation \eqref{quasinormal modes equation}:
\begin{equation}
    \label{eq:schrodinger acoustic BN}
    \qty[\dv[2]{}{x_\star} + \frac{\omega^2}{c_{s0}^2} -  V_{\textsc{eff}}(x_\star)] H_\omega(x_\star) = 0,
\end{equation}
where the effective potential now reads 
\begin{equation}
    \label{eq: effective potential acoustic BN}
    V_{\textsc{eff}}(x_\star) = \frac{1}{2g_{\scalebox{.63}{\textsc{c}}}} \dv[2]{g_{\scalebox{.63}{\textsc{c}}}}{x_\star} - \frac{1}{4g_{\scalebox{.63}{\textsc{c}}}^2}\qty(\dv{g_{\scalebox{.63}{\textsc{c}}}}{x_\star})^2.
\end{equation}
Now that the equations and potentials for both sides -- gravity and aerodynamics -- have been defined, the procedure to apply them in an experimental context will be addressed. First, a clear relation between $\uprho$ and $g_{\scalebox{.63}{\textsc{c}}}$ has to be established, and, second, the constraint equations that connect both Schrödinger-type equations (\ref{quasinormal modes equation}, \ref{eq:schrodinger acoustic BN}), governing the QN modes, have to be investigated.

As long as the wave dynamics \eqref{eq:schrodinger acoustic BN} and the effective potential \eqref{eq: effective potential acoustic BN} are expressed in terms of $g_{\scalebox{.63}{\textsc{c}}}$, it is convenient that the de Laval nozzle area be expressed in terms of $g_{\scalebox{.63}{\textsc{c}}}$ as well. The longitudinal section area ${\scalebox{.85}{\textsc{A}}}$ along the $x$-axis, relative to the throat area ${\scalebox{.85}{\textsc{A}}}_\star$, for a perfect fluid and isentropic flow, can be described by the following equations:
\begin{eqnarray}
    \frac{{\scalebox{.85}{\textsc{A}}}}{{\scalebox{.85}{\textsc{A}}}_\star} &=& \frac{1}{\mathtt{M}} \qty[\frac{2}{\gamma+1}\qty(1+\frac{\gamma-1}{2}\mathtt{M}^2)]^{\qty(\gamma+1)/2\qty(\gamma-1)} = 
    \label{eq: area em funcao da pressao ou densidade}
    \frac{\qty(\frac{\gamma-1}{2})^{1/2} \qty(\frac{2}{\gamma+1})^{\qty(\gamma+1)/2\qty(\gamma-1)}}{\qty(\frac{\uprho}{\uprho_0})\qty[1 - \qty(\frac{\uprho}{\uprho_0})^{(\gamma-1)}]^{1/2}}.
\end{eqnarray}
Analogously to Ref. \cite{Abdalla2007},  hereon the area ${\scalebox{.85}{\textsc{A}}}$ and the density $\uprho$ will be measured in units of the throat cross-sectional area, ${\scalebox{.85}{\textsc{A}}}_\star$, and total density, $\uprho_0$, respectively, in such a way that ${\scalebox{.85}{\textsc{A}}}/{\scalebox{.85}{\textsc{A}}}_\star \mapsto {\scalebox{.85}{\textsc{A}}}$ and $\uprho/\uprho_0 \mapsto \uprho$. Having fixed the scale to be used, with the definition \eqref{eq: area em funcao da pressao ou densidade}, the expressions for $g_{\scalebox{.63}{\textsc{c}}}$ and ${\scalebox{.85}{\textsc{A}}}$ can be written as:
\begin{align}
\label{eq: g e a igual}
    &&
    g_{\scalebox{.63}{\textsc{c}}} = \frac{1}{2}\uprho^{\frac{3-\gamma}{2}} {\scalebox{.85}{\textsc{A}}},
    &&
    {\scalebox{.85}{\textsc{A}}}^{-1} = \uprho \qty[1-\uprho^{\qty(\gamma-1)}]^{1/2}.
    &&
\end{align}
Therefore, $\uprho$ can be put in terms of $g_{\scalebox{.63}{\textsc{c}}}$, as:
\begin{align}
\label{eq: rho em funcao de g}
    \uprho^{1-\gamma} = 2g_{\scalebox{.63}{\textsc{c}}}^2\qty(1-\sqrt{1-g_{\scalebox{.63}{\textsc{c}}}^{-2}}).
\end{align}
Then, applying Eq. \eqref{eq: rho em funcao de g} in Eq. \eqref{eq: g e a igual}, the cross-sectional area can be expressed as a function of $g_{\scalebox{.63}{\textsc{c}}}$
\begin{equation}
\label{eq: area em funcao de g}
    {\scalebox{.85}{\textsc{A}}} = \frac{\sqrt{2}\qty[2g_{\scalebox{.63}{\textsc{c}}}^2\qty(1-\sqrt{1-g_{\scalebox{.63}{\textsc{c}}}^{-2}})]^{\frac{1}{\gamma-1}}}{\qty(1-\sqrt{1-g_{\scalebox{.63}{\textsc{c}}}^{-2}})^{1/2}}.
\end{equation}
Using the relations for an isentropic flow,
\begin{align}
\label{eq: rel local total}
    \frac{T_0}{T} = 1 + \frac{\gamma -1 }{2} \mathtt{M}^2,
    &&
    \frac{p_0}{p} = \qty(1 + \frac{\gamma -1 }{2} \mathtt{M}^2)^{\frac{\gamma}{\gamma -1}},
    &&
    \frac{\uprho_0}{\uprho} = \qty(1 + \frac{\gamma -1 }{2} \mathtt{M}^2)^{\frac{1}{\gamma -1}},
\end{align}
with Eq. \eqref{eq: rho em funcao de g}, $g_{\scalebox{.63}{\textsc{c}}}$ can be expressed in terms of the Mach number and the heat capacity ratio $\gamma$, as:
\begin{equation}\label{eq: mach em funcao de g}
g_{\scalebox{.63}{\textsc{c}}} = \frac{1}{\mathtt{M}}\qty(1+\frac{\gamma-1}{2}\mathtt{M}^2)\frac{1}{\sqrt{2(\gamma-1)}}.   
\end{equation}
Since $\mathtt{M}=1$ at the event horizon $r_h$, corresponding to the throat of the de Laval nozzle, $g_{\scalebox{.63}{\textsc{c}}}$ needs to be finite and greater than unit, as Eq. \eqref{eq: rho em funcao de g} requires. Thus, for the air ($\gamma\approx1.403$):
\begin{equation}
\label{eq: cond contorno g}
\lim_{r\to r_h}    g_{\scalebox{.63}{\textsc{c}}} = \frac{\gamma+1}{2\sqrt{2}\sqrt{\gamma-1}}=\frac{3}{\sqrt{5}}>1.
\end{equation}
This requirement serves as a useful boundary condition for numerical  integrations hereon.

Both Schrödinger-type equations (\ref{quasinormal modes equation}, \ref{eq:schrodinger acoustic BN}) were formulated based on an effective potential and the tortoise coordinate. To find a function $g_{\scalebox{.63}{\textsc{c}}}$ that produces on the aerodynamical side the same effective potential as the spherical black hole wave equation in Eq. \eqref{eq: effective potential acoustic BN}, the tortoise coordinates of both equations must be equal. Thus, from Eqs. \eqref{eq: acoustic BN tortoise coordinate}, \eqref{eq: rho em funcao de g}, and \eqref{eq: mach em funcao de g} one obtains:
\begin{equation}
\label{eq: coordenadas iguais}
    \dd r_\star = \dd x_\star = 
    \frac{c_{s0}}{c_s} \frac{\dd x}{(1 - \mathtt{M}^2)} =
    \frac{\uprho^{\frac{1-\gamma}{2}}}{1-\mathtt{M}^2}\dd x =
    \frac{\qty[2g_{\scalebox{.63}{\textsc{c}}}^2\qty(1-\sqrt{1-g_{\scalebox{.63}{\textsc{c}}}^{-2}})]^{1/2}}{1- \frac{2}{\gamma-1}\qty[2g_{\scalebox{.63}{\textsc{c}}}^2\qty(1-\sqrt{1-g_{\scalebox{.63}{\textsc{c}}}^{-2}})-
1]} \dd x.
\end{equation}
The de Laval nozzle radius can be expressed by Eq. \eqref{tortoise coordinate}. Defining \beq
\mathcal{F}(r) \equiv \sqrt{B(r)/A(r)}\eeq to simplify the calculation, Eq. \eqref{eq: coordenadas iguais} becomes:
\begin{equation}
\label{eq:xemfuncaoder123}
    \dv{x}{r} = \frac{\gamma+1-4g_{\scalebox{.63}{\textsc{c}}}(r)^2\qty(1-\sqrt{1-g_{\scalebox{.63}{\textsc{c}}}(r)^{-2}})}
    {\mathcal{F}(r)\qty(\gamma-1)\qty[2g_{\scalebox{.63}{\textsc{c}}}(r)^2 \qty(1-\sqrt{1-g_{\scalebox{.63}{\textsc{c}}}(r)^{-2}})]^{1/2}}.
\end{equation}
To determine the function $g_{\scalebox{.63}{\textsc{c}}}(r)$, first, Eq. \eqref{eq: effective potential acoustic BN} can be rewritten, making the dependence on $r$ clear, using Eq. \eqref{tortoise coordinate}:
\begin{equation}
    \label{eq: pot em funcao de r}
    \frac{1}{2g_{\scalebox{.63}{\textsc{c}}}(r)} \left(\mathcal{F}(r)\mathcal{F}'(r)g_{\scalebox{.63}{\textsc{c}}}'(r)+\mathcal{F}(r)^2g_{\scalebox{.63}{\textsc{c}}}''(r)\right)- \frac{\mathcal{F}^2(r)g_{\scalebox{.63}{\textsc{c}}}^{'2}(r)}{4g_{\scalebox{.63}{\textsc{c}}}^2(r)} = V_{\textsc{eff}}(r).
\end{equation}
The potential term in Eq. \eqref{eq: pot em funcao de r}, to complete the constraint, will be replaced by Eq. \eqref{final effective potential}. 

Now, to proceed, it will be necessary to solve Eq. \eqref{eq: pot em funcao de r} with the potential given in Eq. \eqref{final effective potential} and boundary conditions given by Eq. \eqref{eq: cond contorno g}.
Eq. \eqref{eq: pot em funcao de r} cannot be directly solved by Runge--Kutta (RK) algorithms, as there are singularities with $g_{\scalebox{.63}{\textsc{c}}}$. To circumvent the singularities, the substitution 
\beq\label{gc11}
g_{\scalebox{.63}{\textsc{c}}}\qty(r) \equiv \lambda^2\qty(r)
\eeq can be used, and Eq. \eqref{eq: pot em funcao de r} assumes the form
\begin{align}\label{eq: lambda}
\lambda''(r) + \frac{\mathcal{F}'(r)}{\mathcal{F}(r)}\lambda'(r) - \frac{V_{\textsc{eff}}(r)}{\mathcal{F}^2(r)}\lambda(r) = 0.
\end{align}
Before numerically proceeding, a significant test of consistency 
can be implemented, by considering the effective potential \eqref{final effective potential} associated with the GD-extended hairy Reissner--Nordstr\"om metric  \eqref{eq23} in the limit where $\alpha\to0$ and $Q\to0$, corresponding to the Schwarzschild-like solution. 
In this limit, one obtains
 \beq\label{gc12}
g_{\scalebox{.63}{\textsc{c}}}=\frac{\gamma+1}{2\sqrt{2}\sqrt{\gamma-1}}\frac{\Gamma(2+l){}_2F_1(1-l,2+l,3,r)}{6(l-1)!}r^4,
\eeq
complying with the particular result in Ref. \cite{Abdalla2007} for the Schwarzschild solution. 
Eq. \eqref{eq: lambda} is not well defined at the zeros of the equation $\mathcal{F}(r) = 0$, corresponding to $r_h$, so the analysis must take place in the region $r > r_h$.
Taking it into account, Eq. \eqref{eq: lambda}
is solvable, at least near $r\sim r_h$, by the Frobenius method  \cite{neuringer1978frobenius}, with 
\begin{eqnarray}
    \lambda\qty(r) &=& \sum_{k=0}^\infty a_k \qty(r-r_h)^k, \label{eq:frob1} \end{eqnarray}
where $a_k$ are constants yet to be determined.
Using Eq. (\ref{eq:frob1}), Eq. \eqref{eq: lambda} becomes a recurrence equation:
\begin{equation}\label{eq: recurrence eq}
    \qty(k +2)\qty(k+1) a_{k+2} = a_k\frac{V_{\textsc{eff}}(r)}{\mathcal{F}^2(r)}  - \qty(k+1)a_{k+1} \frac{\mathcal{F}'(r)}{\mathcal{F}(r)},
\end{equation}
where, with Eq. \eqref{eq: cond contorno g},
\beq\label{eq:condinilamb}
a_0 &=& \lim_{r\to r_h}\lambda(r) = \sqrt{\frac{\gamma+1}{2\sqrt{2}\sqrt{\gamma-1}}},\nonumber\\
a_1 &=& 0.
\eeq
Eq. \eqref{eq: lambda}, solved with the boundary conditions \eqref{eq:condinilamb}, has solutions:
\begin{eqnarray}
    \lambda\qty(r) &=& \sqrt{\frac{\gamma+1}{2\sqrt{2}\sqrt{\gamma-1}}} \qty[1 +\frac{1}{2} \frac{V_{\textsc{eff}}(r)}{\mathcal{F}^2(r)} \qty(r-r_h)^2 - \frac{1}{6} \frac{V_{\textsc{eff}}(r)}{\mathcal{F}^2(r)} \frac{\mathcal{F}'(r)}{\mathcal{F}(r)} \qty(r-r_h)^3 + \cdots].
\end{eqnarray}
These conditions are used to allow for the numerical integration of $\lambda\qty(r)$ using RK methods, thus finding $g_{\scalebox{.63}{\textsc{c}}}$ as a function of $r$, that reproduces the black hole geometry. Afterward, it will be necessary to explicitly express $g_{\scalebox{.63}{\textsc{c}}}$ as a function of the longitudinal coordinate $x$ of the de Laval nozzle. This is implemented by using Eq. \eqref{eq:xemfuncaoder123}, and subsequently employing the results in Eq. \eqref{eq: area em funcao de g} to plot the profile of the nozzle.
The results heretofore obtained  in this section regard arbitrary static spherically symmetric spacetimes, for completeness. Hereon we will take the GD-extended hairy Reissner--Nordstr\"om metric  \eqref{eq23} and implement the analog gravitational system, analyzing the properties of the analog GD-de Laval nozzle and addressing the QN mode frequencies as well.


{\color{black}
\section{de Laval nozzle analog of GD-extended Reissner--Nordstr\"om black hole}

\label{sec4}

Having established the analogy between a de~Laval nozzle and the GD-extended
Reissner--Nordstr\"om metric~\eqref{eq23}, acoustic experiments are not only
restricted to the qualitative point of view but in addition, up to the experimental accuracy,
numerical exactness can be probed by observing and measuring features of the fluid
propagation in the de Laval nozzle.
Since QN modes are governed by the wave equation~\eqref{quasinormal modes equation}
on the gravity side, and by Eq.~\eqref{eq:schrodinger acoustic BN} for acoustic black hole,
the numerical correspondence means that the effective potential encoding perturbations
of some aerodynamic system match the effective potential associated with the GD-extended
hairy Reissner--Nordstr\"om black hole.
Therefore sound waves propagating in a de~Laval nozzle provide a physical system
in the laboratory with the same effective potential as the GD-extended Reissner--Nordstr\"om black hole.}

\begin{figure}[H]
    \centering
    \begin{subfigure}{0.44\textwidth}
        \centering
        \includegraphics[width=\linewidth]{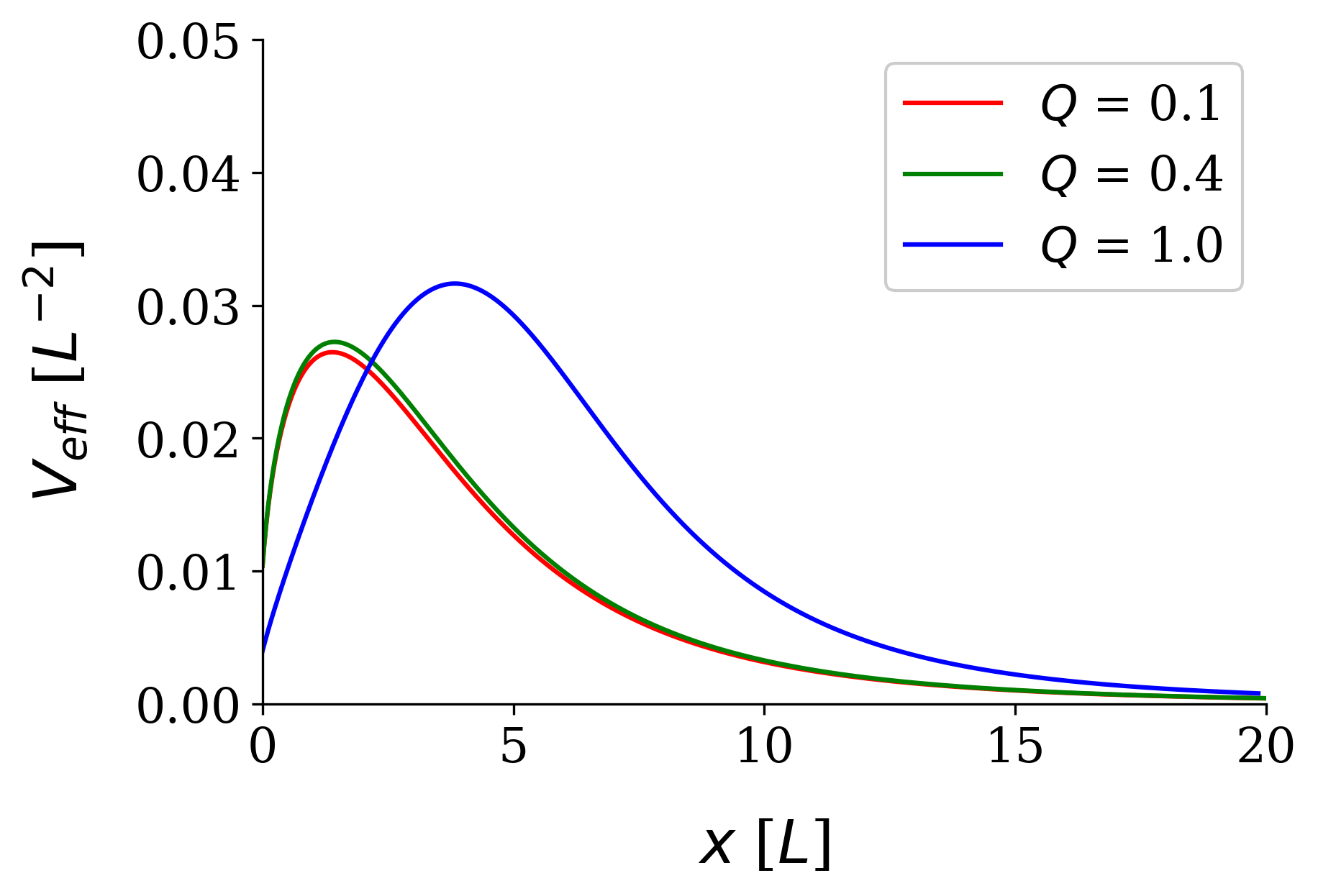}
        \caption{$\alpha = 0.1$ and $l=0$.}
        \label{fig:Va0}
    \end{subfigure}\qquad\qquad
    \begin{subfigure}{0.44\textwidth}
        \centering
        \includegraphics[width=\linewidth]{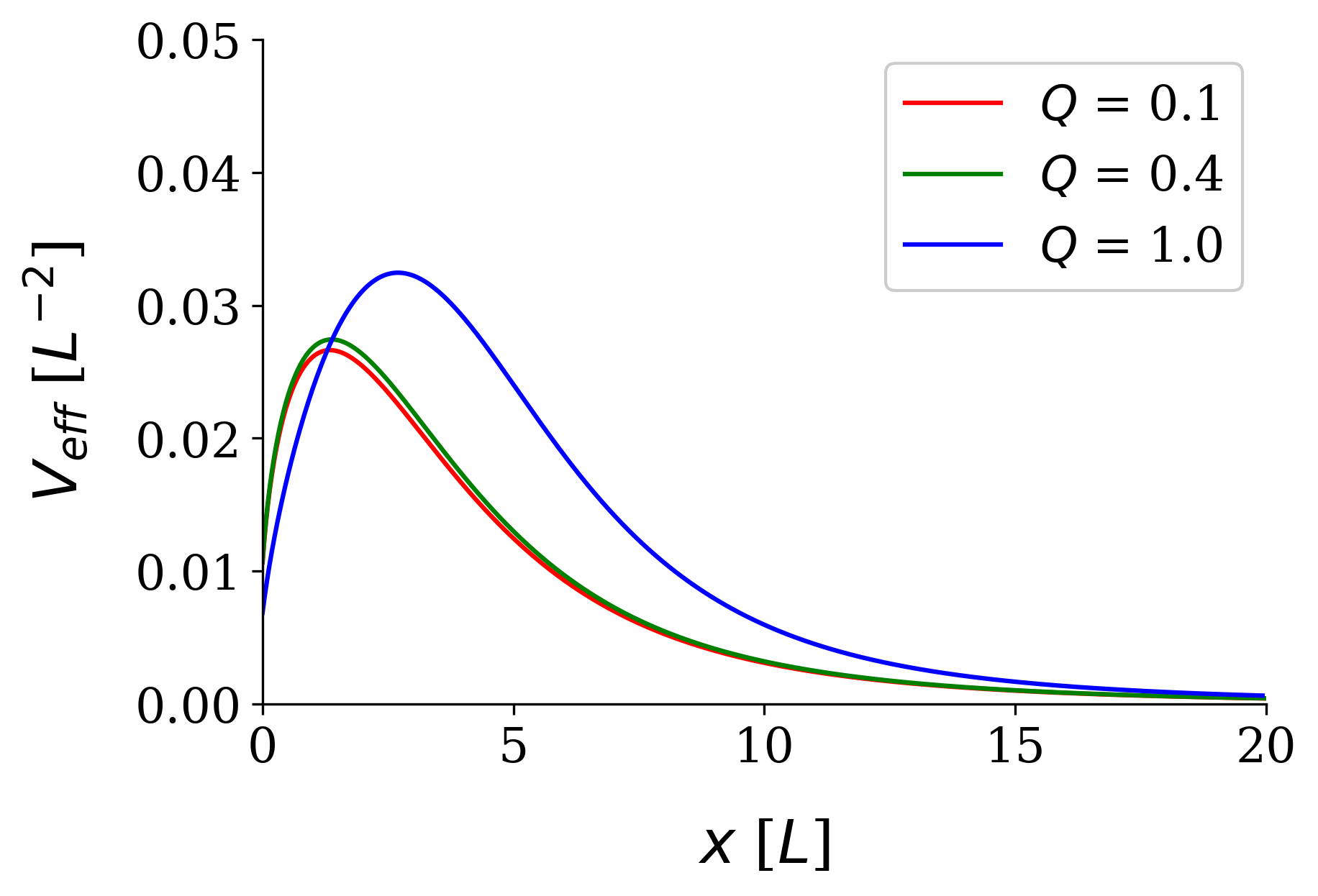}
        \caption{$\alpha = 0.5$ and $l=0$.}
        \label{fig:Vb0}
    \end{subfigure}\medbreak\medbreak
      \begin{subfigure}{0.44\textwidth}
        \centering
        \includegraphics[width=\linewidth]{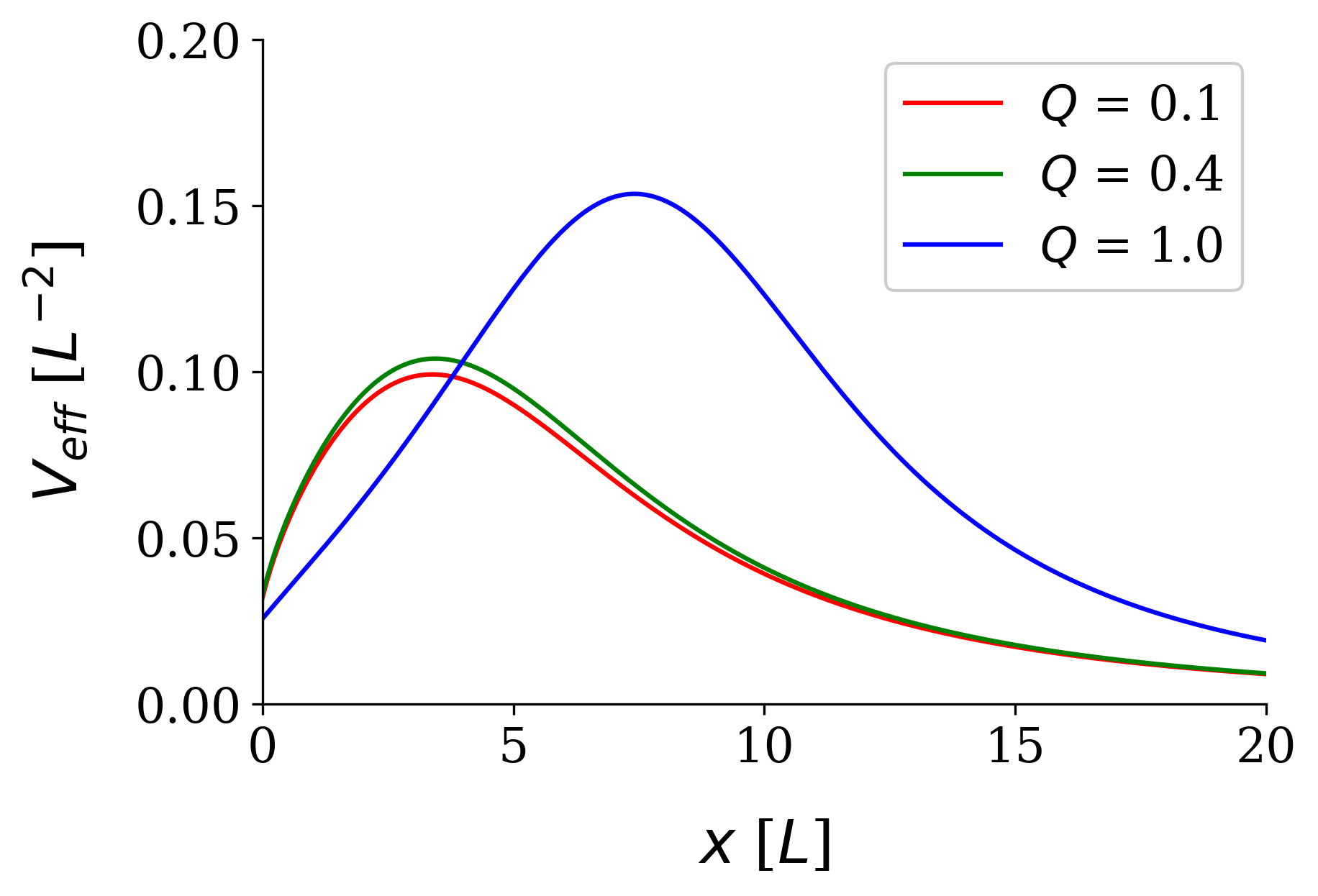}
        \caption{$\alpha = 0.1$ and $l=1$.}
        \label{fig:Va1}
    \end{subfigure}\qquad\qquad
    \begin{subfigure}{0.44\textwidth}
        \centering
        \includegraphics[width=\linewidth]{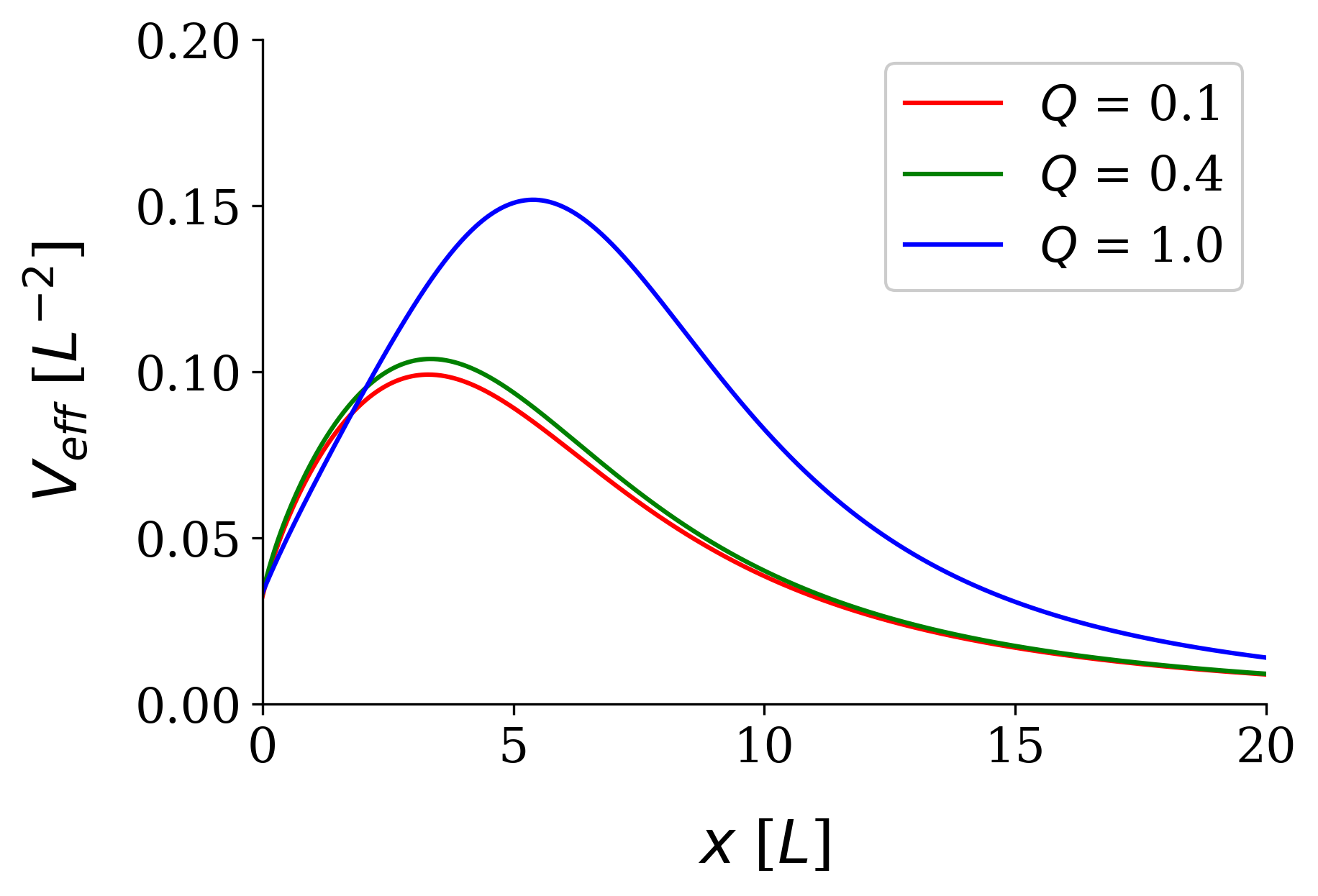}
        \caption{$\alpha = 0.5$ and $l=1$.}
        \label{fig:Vb1}
    \end{subfigure} \medbreak\medbreak
     \begin{subfigure}{0.44\textwidth}
        \centering\includegraphics[width=\linewidth]{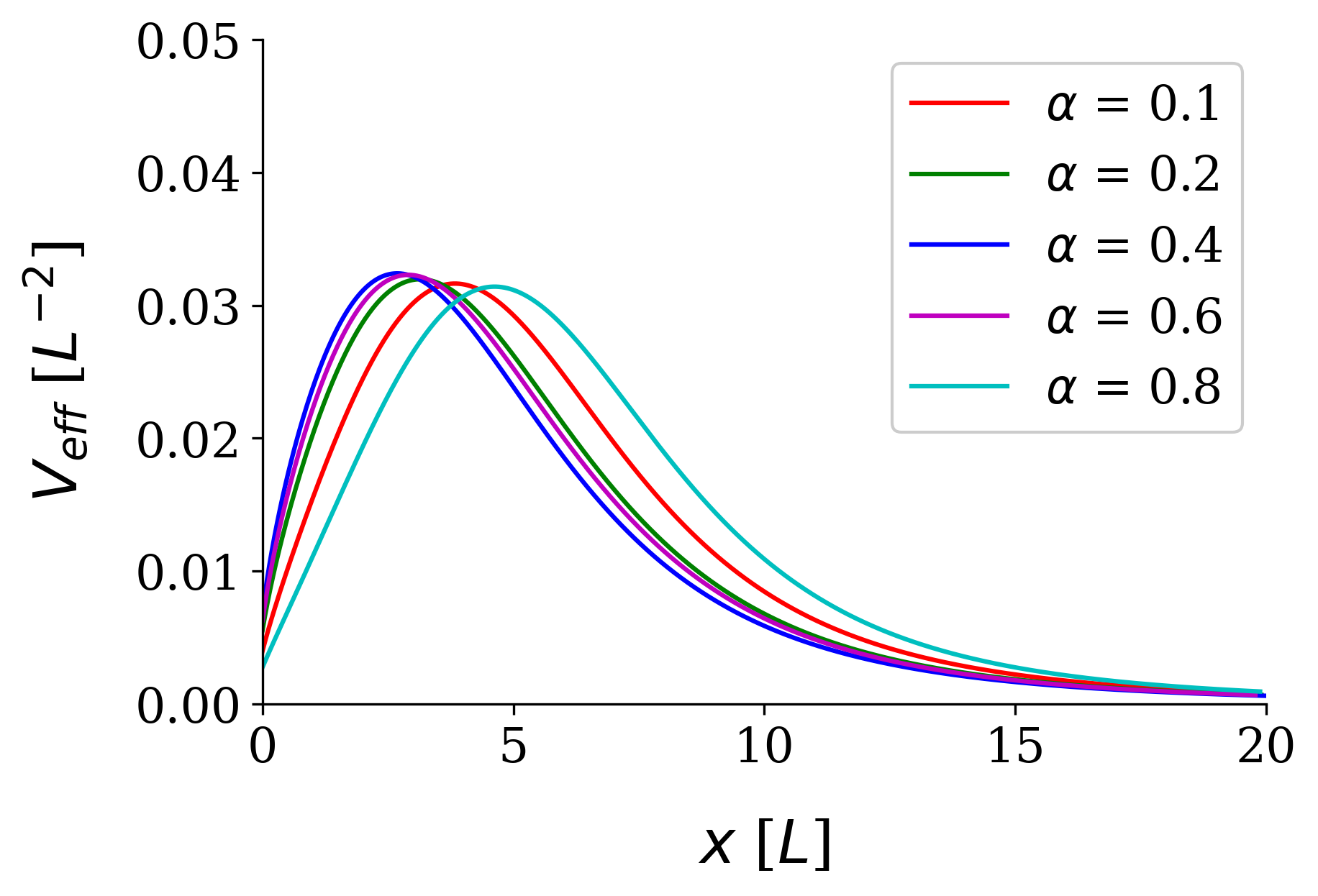}
        \caption{$Q = 1$ and $l=0$.}
        \label{fig:Vc0}
    \end{subfigure}\qquad\qquad
    \begin{subfigure}{0.44\textwidth}
        \centering\includegraphics[width=\linewidth]{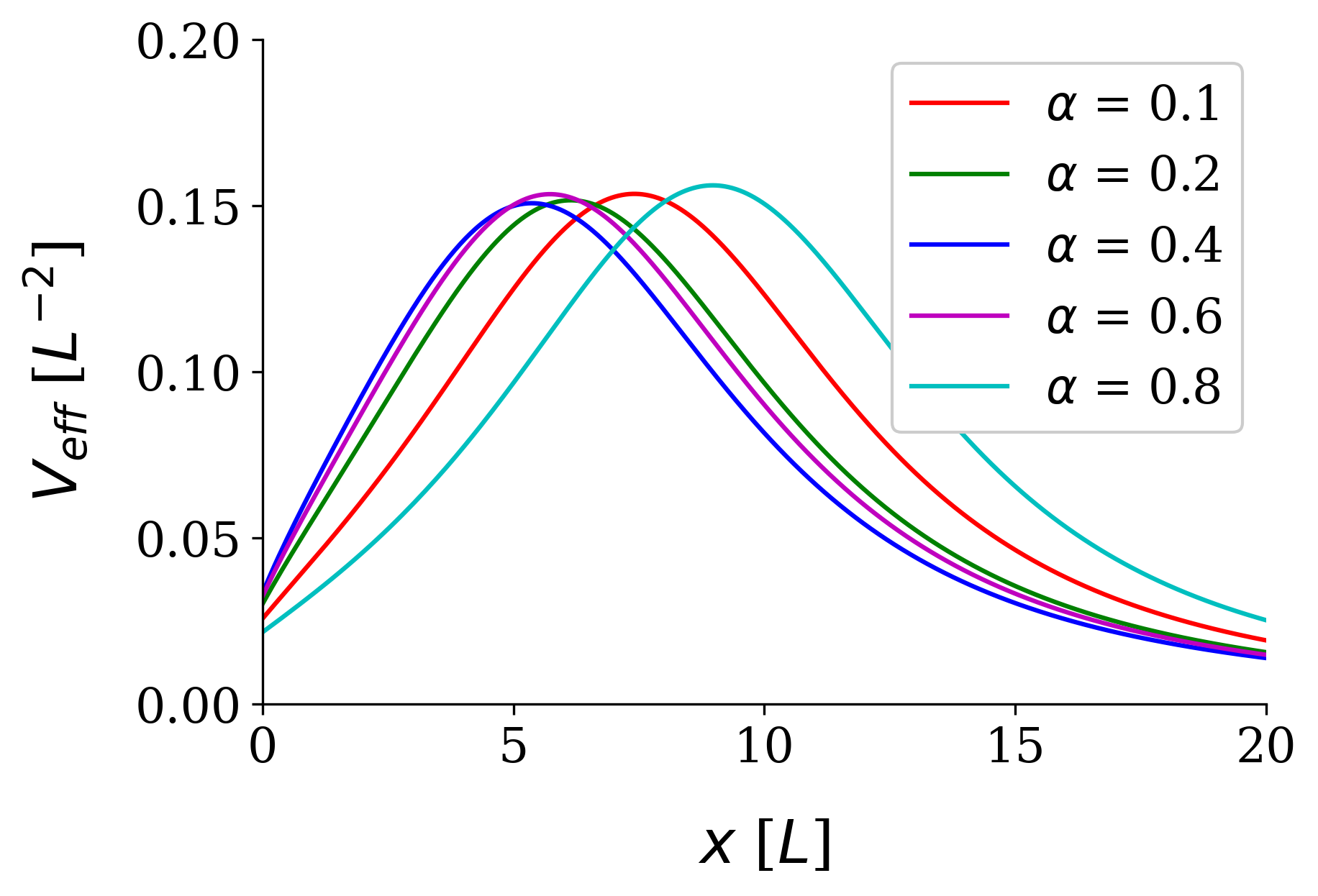}
        \caption{$Q = 1$ and $l=1$.}
        \label{fig:Vc1}
    \end{subfigure}
    \caption{Effective potential as a function of the longitudinal direction of the de Laval nozzle, for the GD-extended hairy Reissner--Nordstr\"om metric  \eqref{eq23}.}
    \label{fig:Veff}
\end{figure}

\begin{figure}[H]
    \centering
    \begin{subfigure}{0.42\textwidth}
        \centering
        \includegraphics[width=\linewidth]{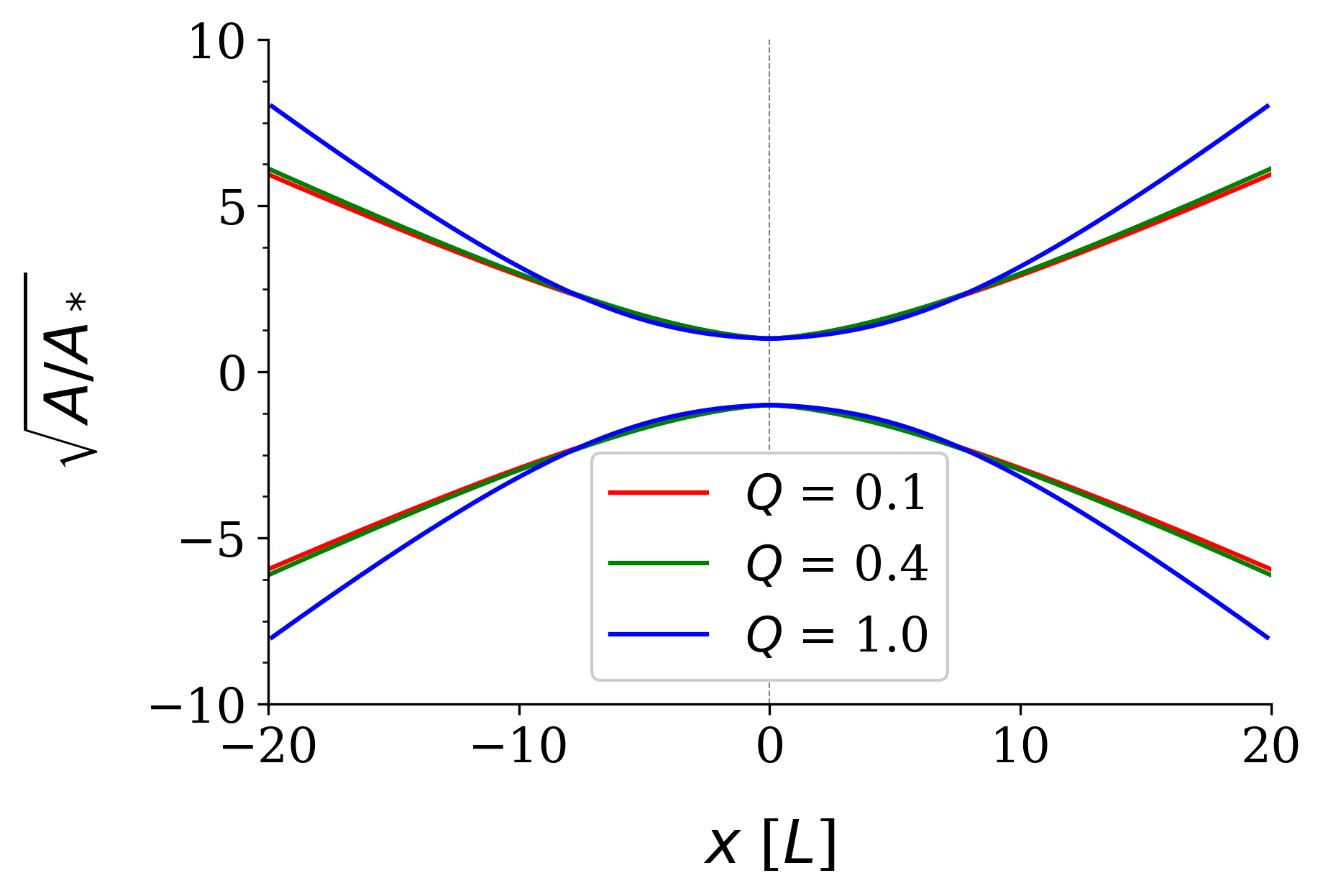}
        \caption{$\alpha = 0.1$ and $l=0$.}
        \label{fig:Aa0}
    \end{subfigure}\qquad\qquad
    \begin{subfigure}{0.42\textwidth}
        \centering
        \includegraphics[width=\linewidth]{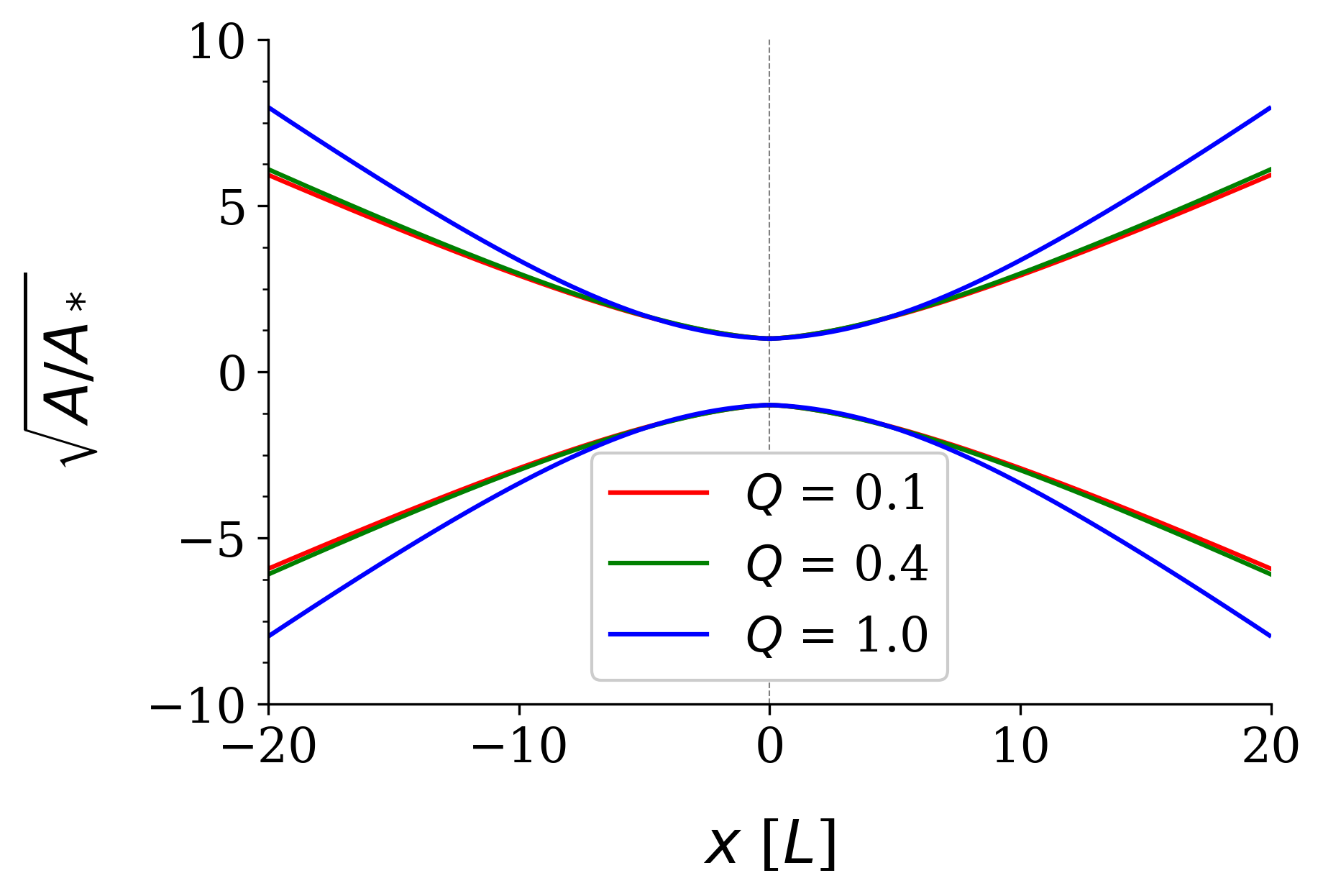}
        \caption{$\alpha = 0.5$ and $l=0$.}
        \label{fig:Ab0}  \end{subfigure}\medbreak\medbreak
    \begin{subfigure}{0.42\textwidth}
        \centering
        \includegraphics[width=\linewidth]{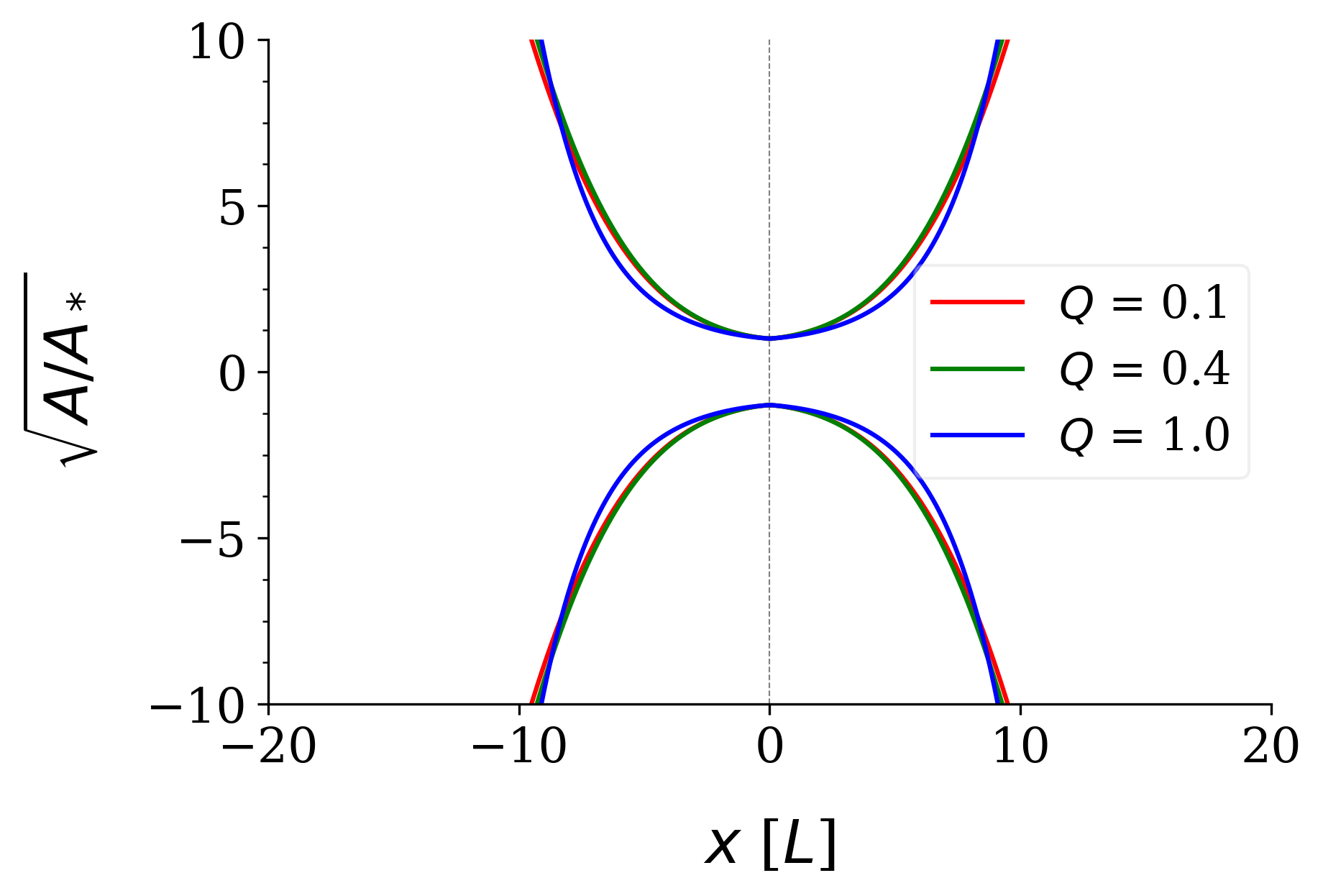}
        \caption{$\alpha = 0.1$ and $l=1$.}
        \label{fig:Aa1}
    \end{subfigure}\qquad\qquad
    \begin{subfigure}{0.42\textwidth}
        \centering
        \includegraphics[width=\linewidth]{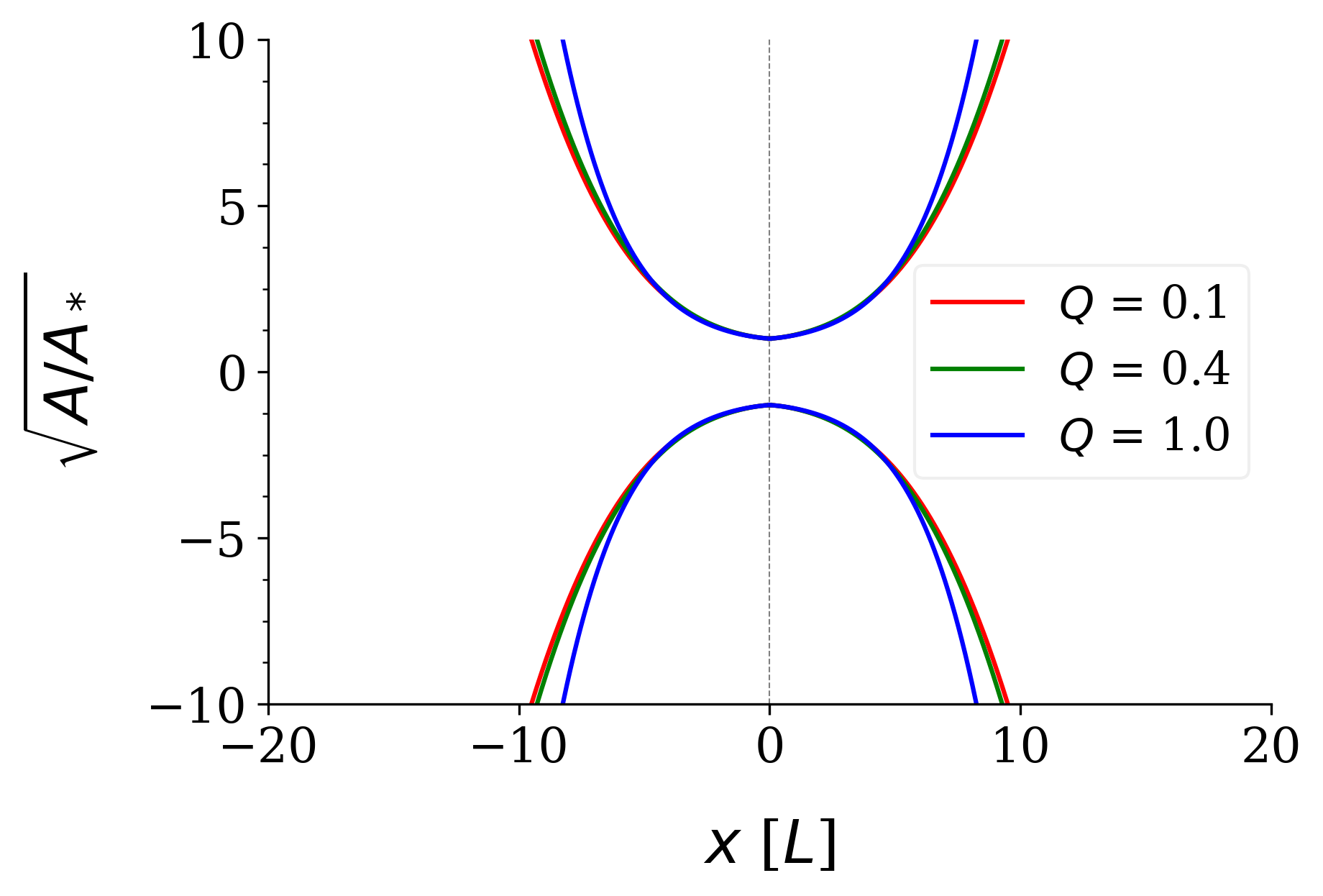}
        \caption{$\alpha = 0.5$ and $l=1$.}
        \label{fig:Ab1}
    \end{subfigure}\medbreak\medbreak
    \begin{subfigure}{0.42\textwidth}
        \centering\includegraphics[width=\linewidth]{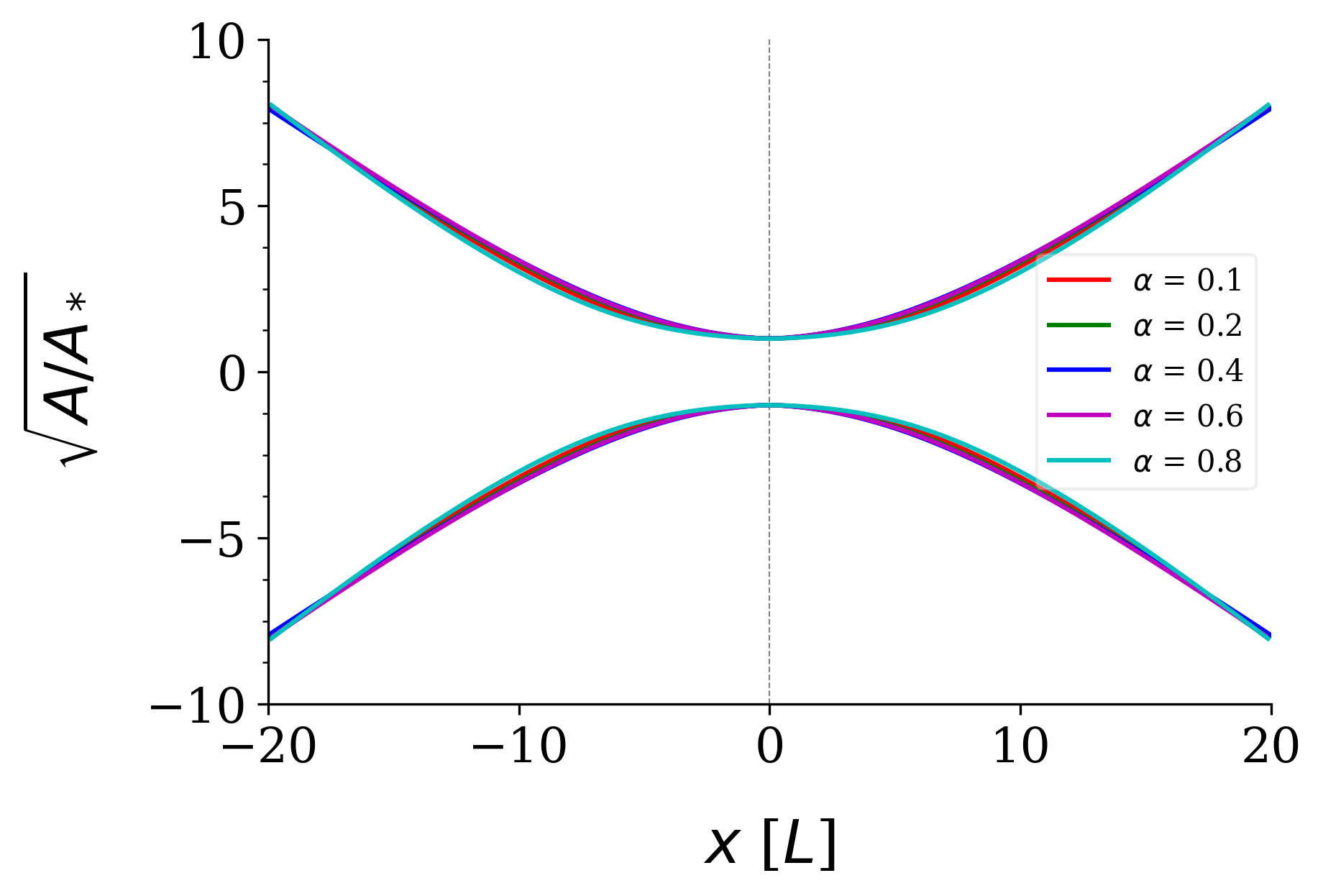}
        \caption{$Q = 1$ and $l=0$.}
        \label{fig:Ac0}
    \end{subfigure}\qquad\qquad
    \begin{subfigure}{0.42\textwidth}
        \centering\includegraphics[width=\linewidth]{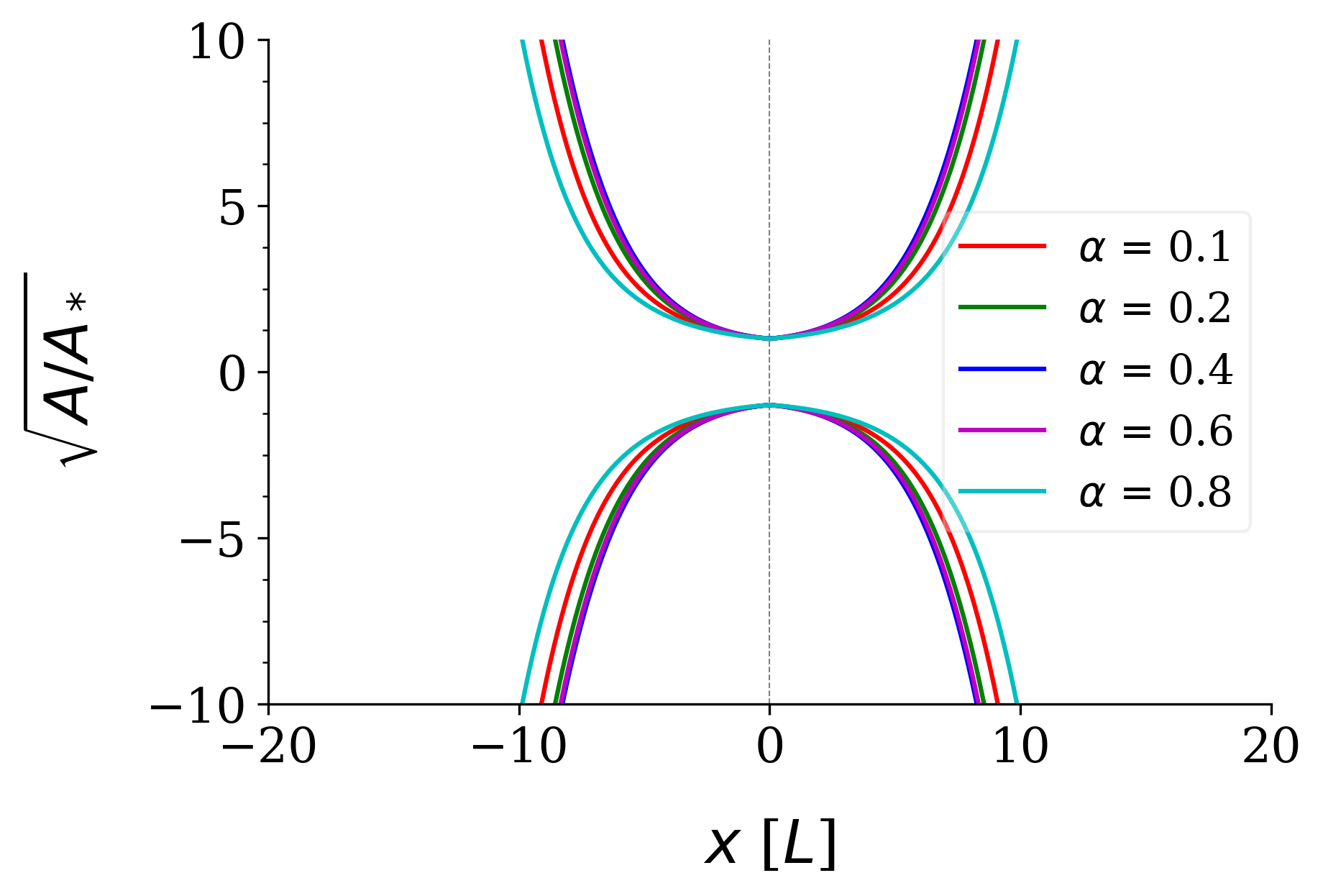}
        \caption{$Q = 1$ and $l=1$.}
        \label{fig:Ac1}
    \end{subfigure}
    \caption{Nozzle shape as a function of the longitudinal direction of the de Laval nozzle, for the GD-extended hairy Reissner--Nordstr\"om metric  \eqref{eq23}.}
    \label{fig:shape}
\end{figure}

\begin{figure}[H]
    \centering
    \begin{subfigure}{0.44\textwidth}
        \centering
        \includegraphics[width=\linewidth]{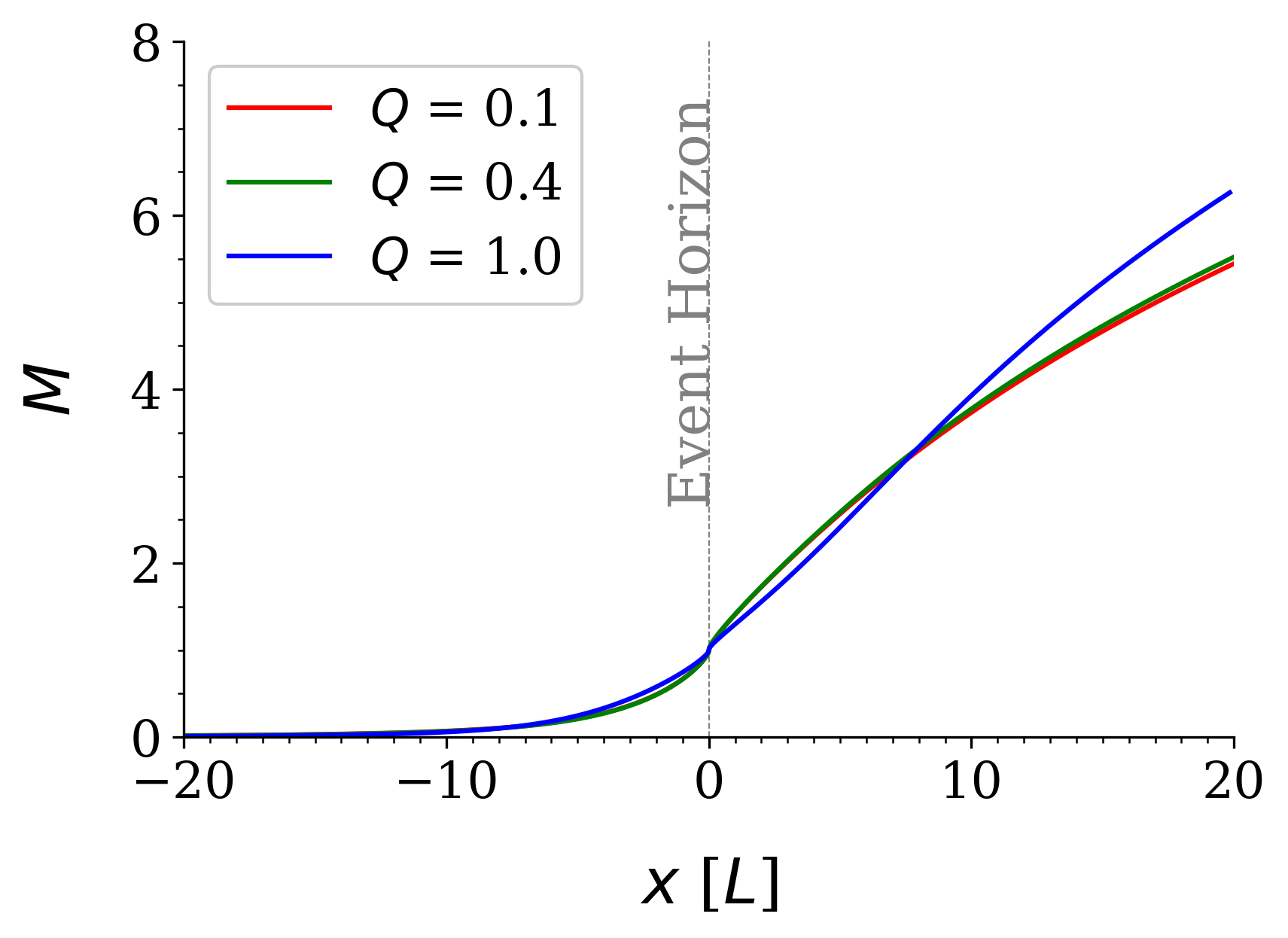}
        \caption{$\alpha = 0.1$ and $l=0$.}
        \label{fig:Ma0}
    \end{subfigure}\qquad\qquad
    \begin{subfigure}{0.44\textwidth}
        \centering
        \includegraphics[width=\linewidth]{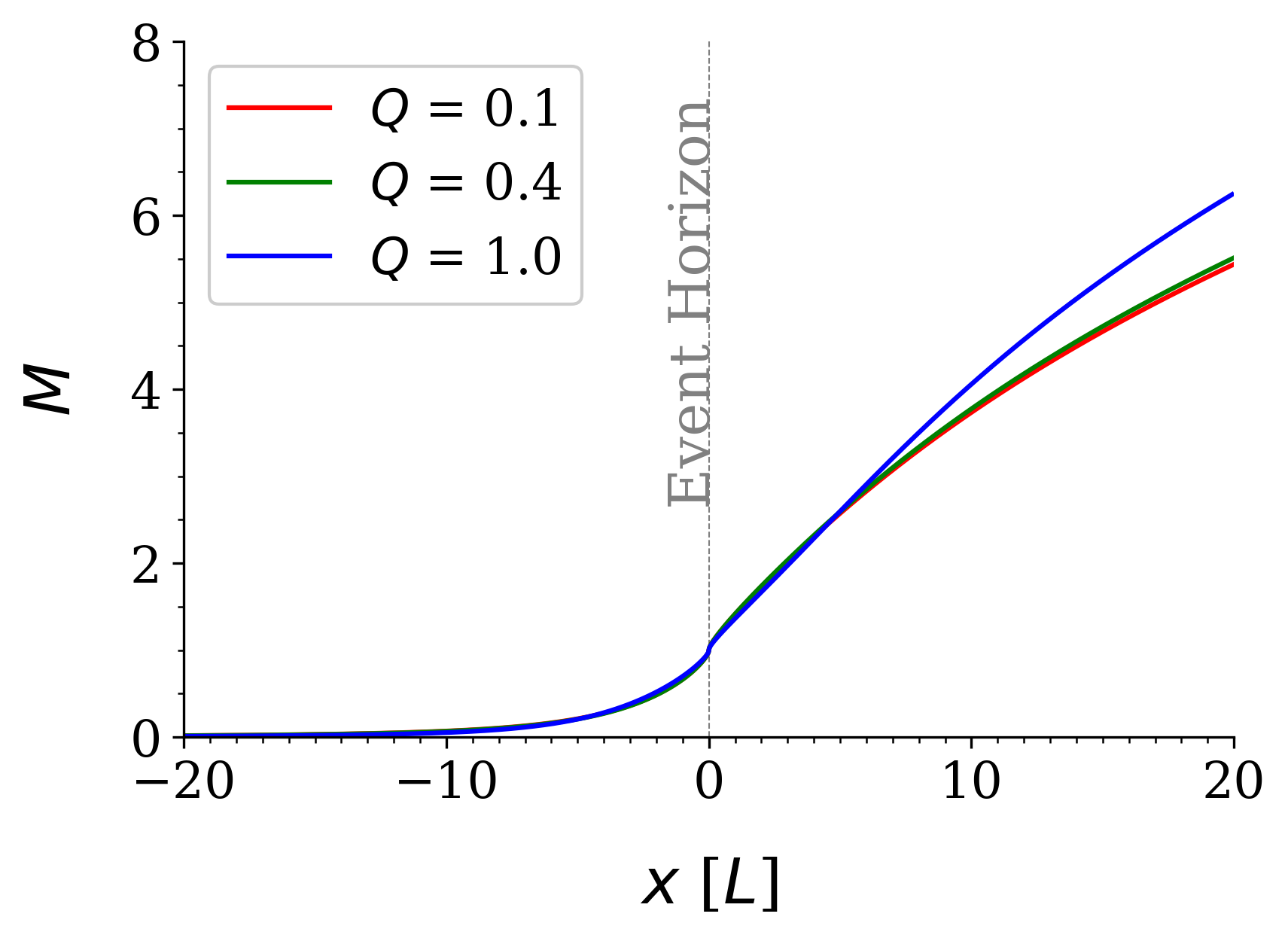}
        \caption{$\alpha = 0.5$ and $l=0$.}
        \label{fig:Mb0}
    \end{subfigure}\medbreak\medbreak
    \begin{subfigure}{0.44\textwidth}
        \centering
        \includegraphics[width=\linewidth]{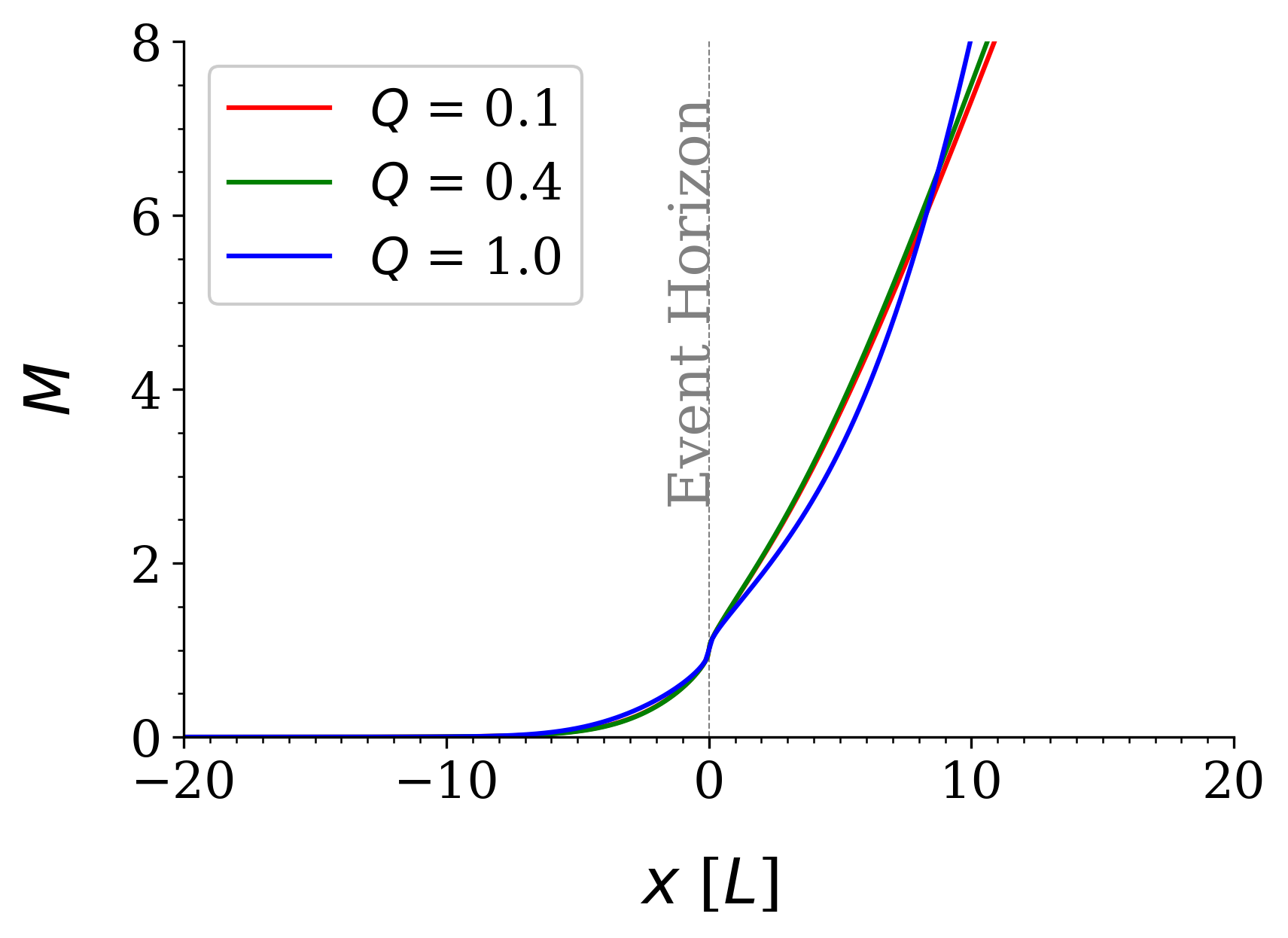}
        \caption{$\alpha = 0.1$ and $l=1$.}
        \label{fig:Ma1}
    \end{subfigure}\qquad\qquad
    \begin{subfigure}{0.44\textwidth}
        \centering
        \includegraphics[width=\linewidth]{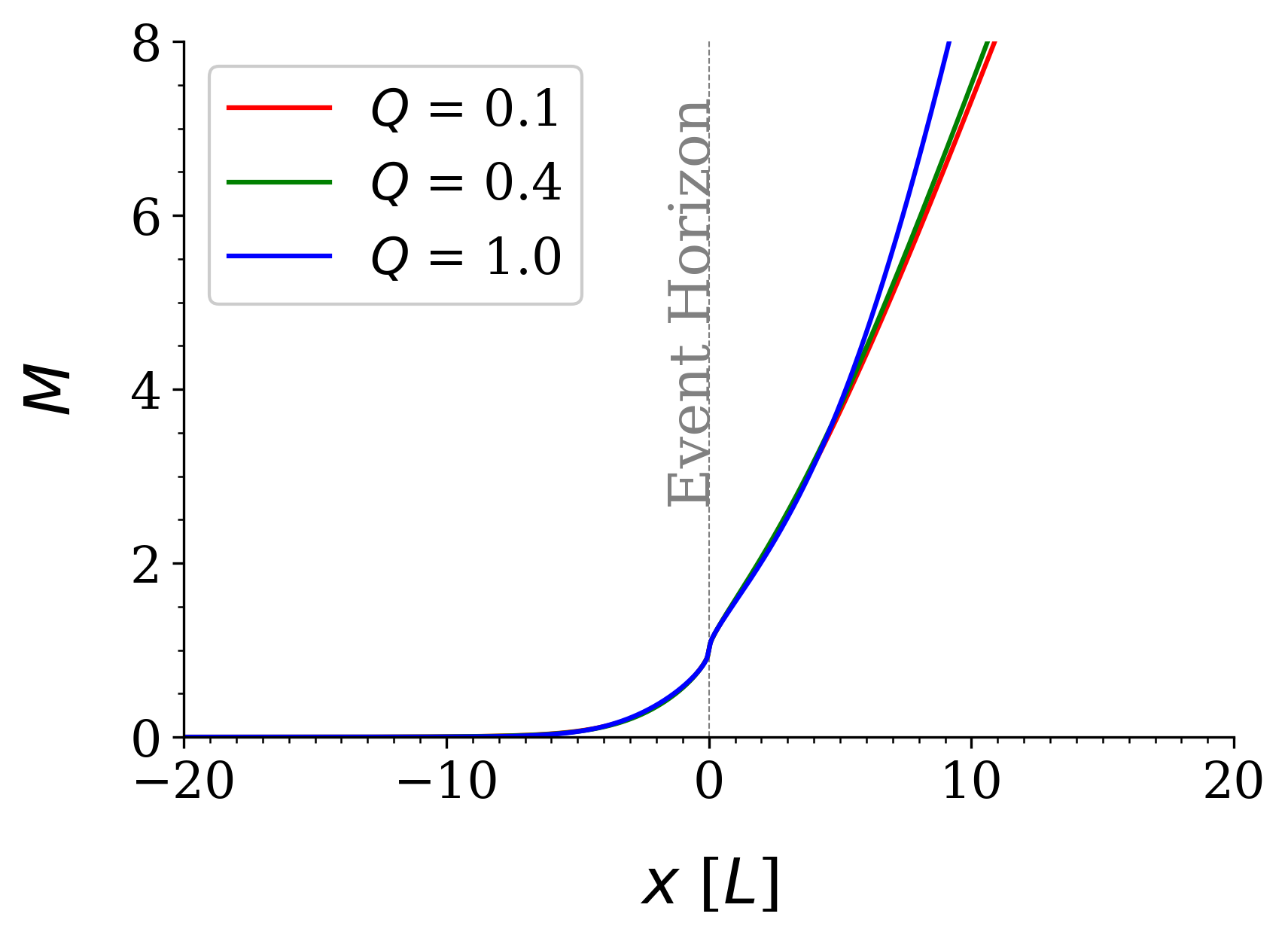}
        \caption{$\alpha = 0.5$ and $l=1$.}
        \label{fig:Mb1}
    \end{subfigure}\medbreak\medbreak
        \begin{subfigure}{0.44\textwidth}
        \centering
        \includegraphics[width=\linewidth]{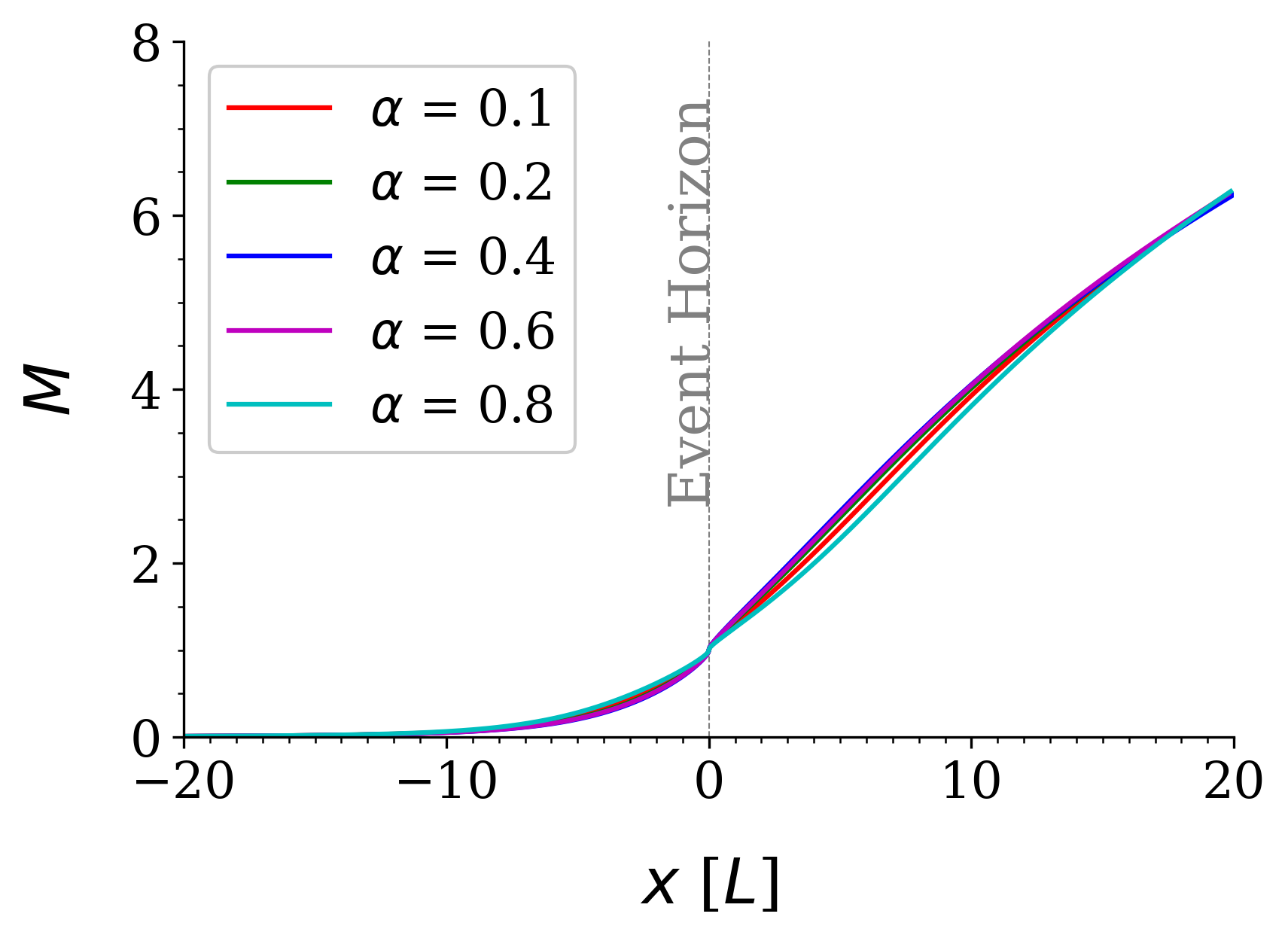}
        \caption{$Q = 1$ and $l=0$.}
        \label{fig:Mc0}
    \end{subfigure}\qquad\qquad
    \begin{subfigure}{0.44\textwidth}
        \centering
        \includegraphics[width=\linewidth]{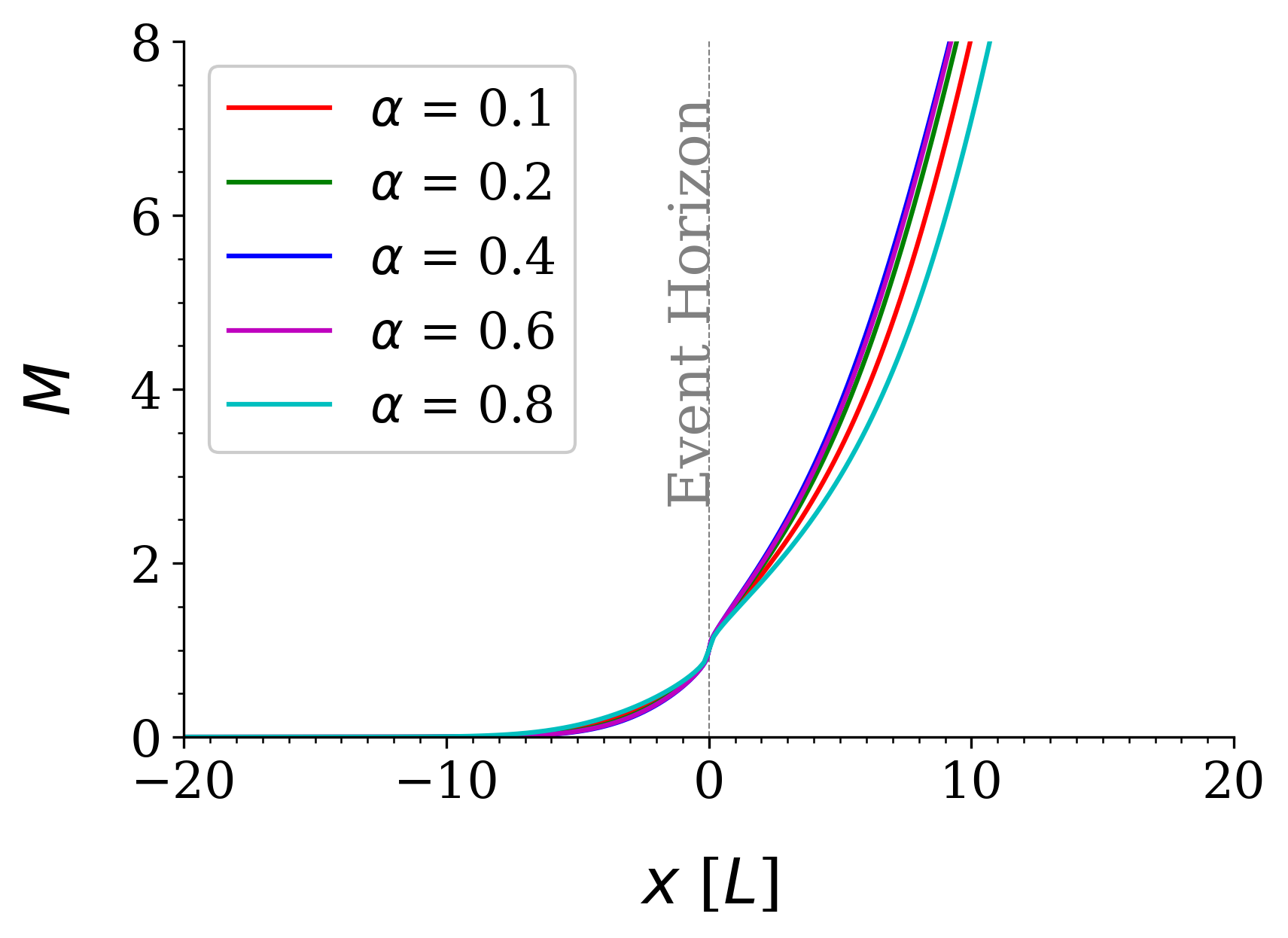}
        \caption{$Q = 1$ and $l=1$.}
        \label{fig:Mc1}
    \end{subfigure}
    \caption{Mach number as a function of the longitudinal direction of the de Laval nozzle, for the GD-extended hairy Reissner--Nordstr\"om metric  \eqref{eq23}.}
    \label{fig:mach}
\end{figure}

\begin{figure}[H]
    \centering
    \begin{subfigure}{0.45\textwidth}
        \centering
        \includegraphics[width=\linewidth]{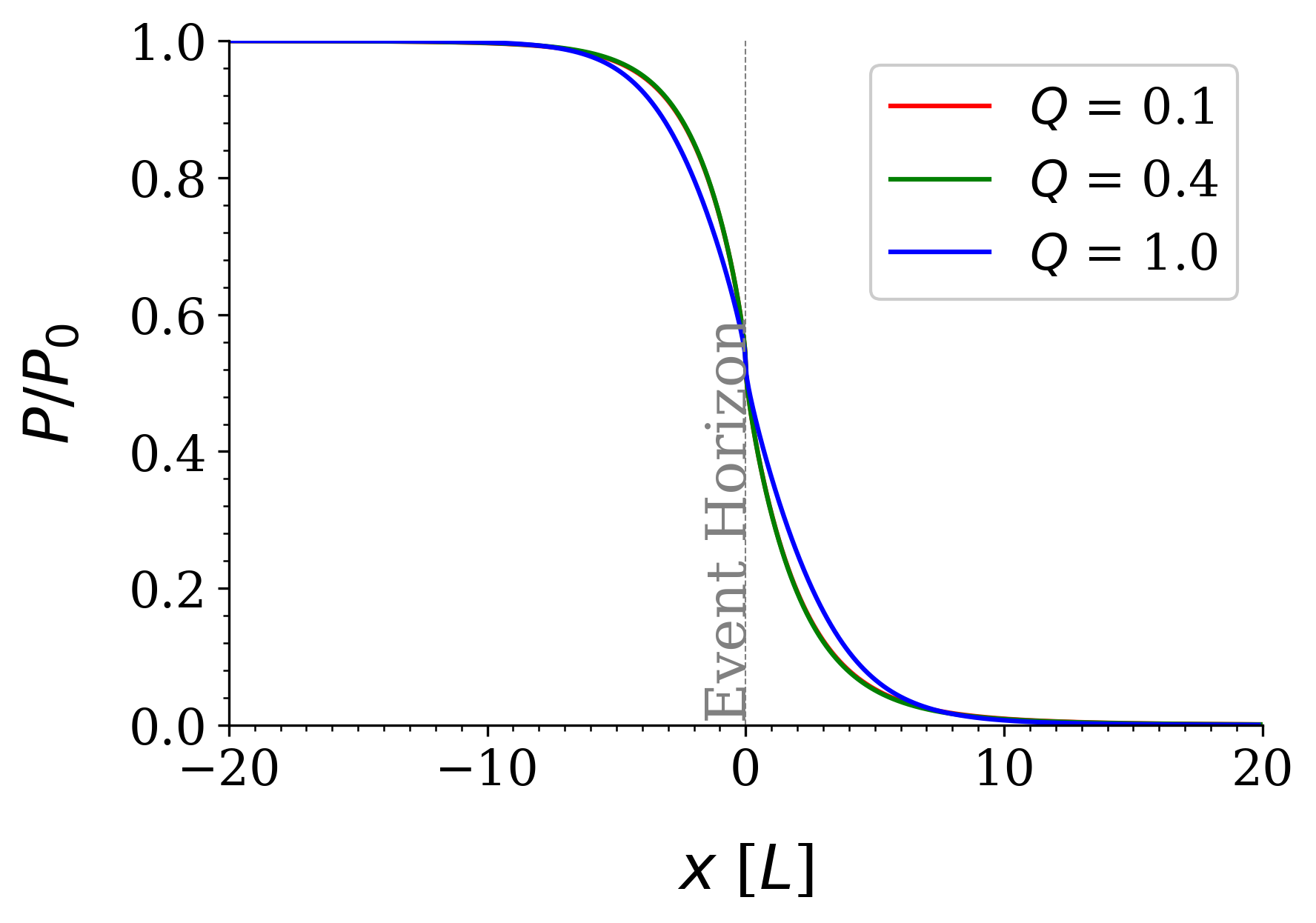}
        \caption{$\alpha = 0.1$ and $l=0$.}
        \label{fig:Pa0}
    \end{subfigure}]\qquad\qquad
    \begin{subfigure}{0.45\textwidth}
        \centering
        \includegraphics[width=\linewidth]{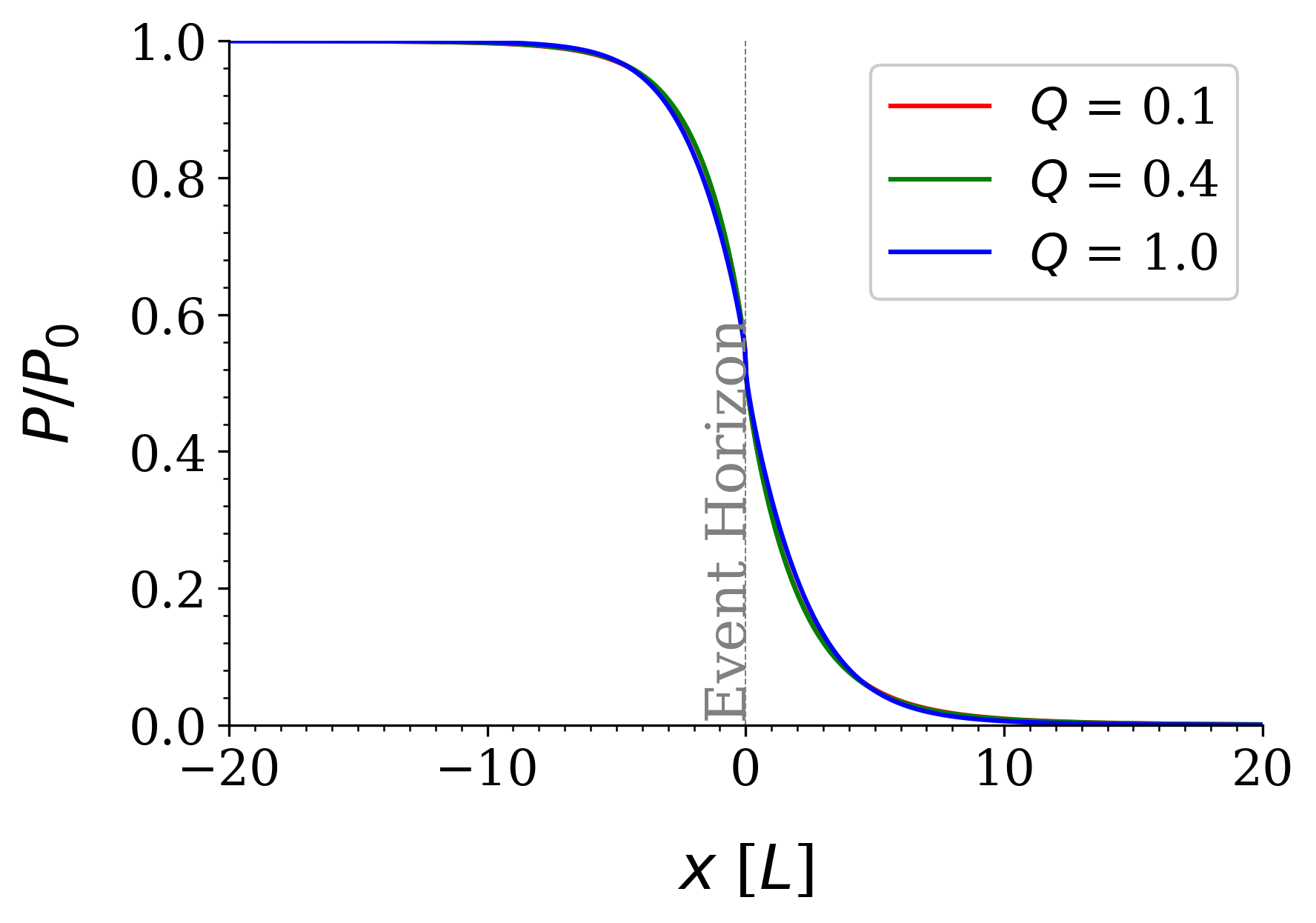}
        \caption{$\alpha = 0.5$ and $l=0$.}
        \label{fig:Pb0}
    \end{subfigure}
\medbreak\medbreak   \begin{subfigure}{0.45\textwidth}
        \centering
        \includegraphics[width=\linewidth]{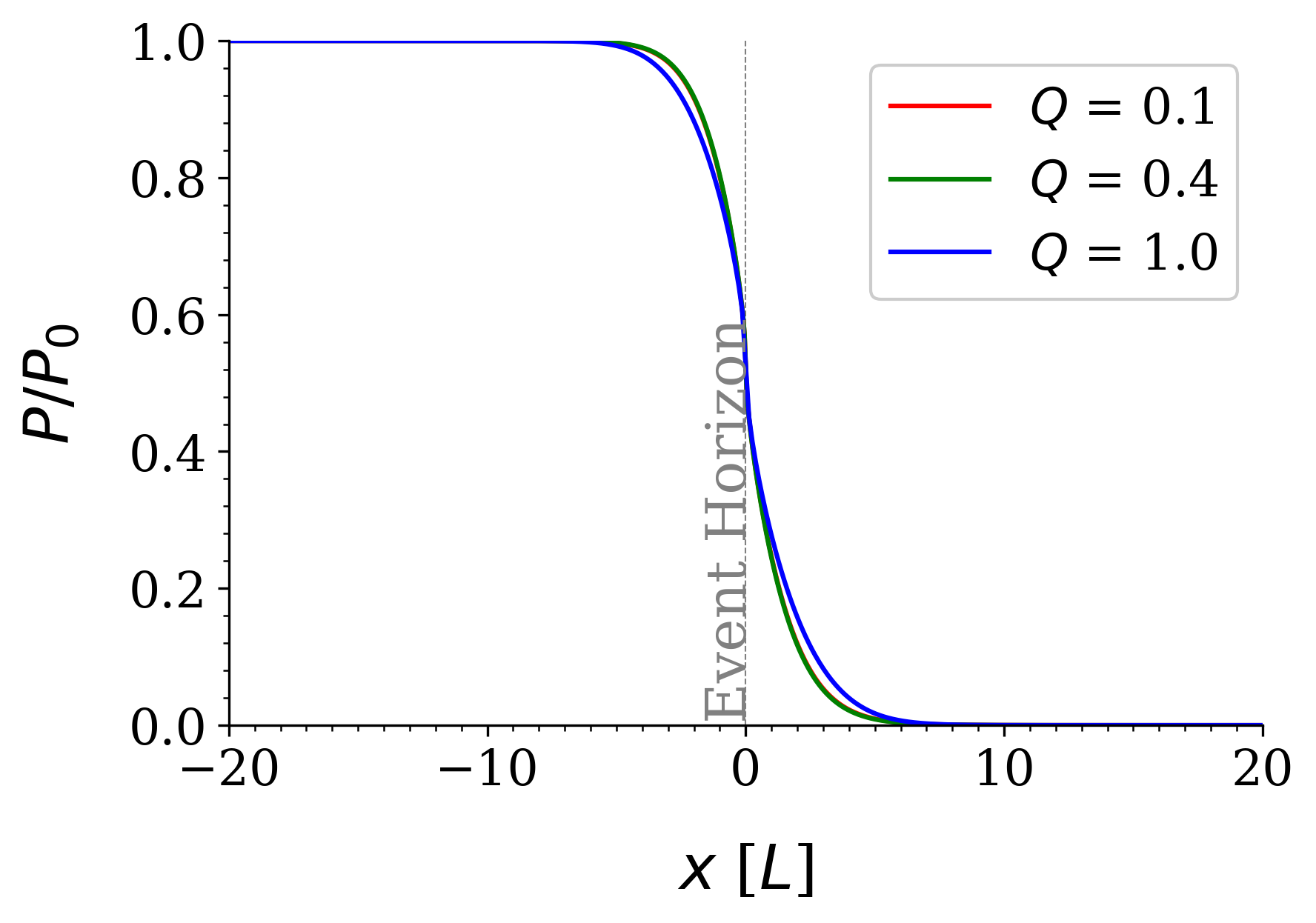}
        \caption{$\alpha = 0.1$ and $l=1$.}
        \label{fig:Pa1}
    \end{subfigure}\qquad\qquad
    \begin{subfigure}{0.45\textwidth}
        \centering
        \includegraphics[width=\linewidth]{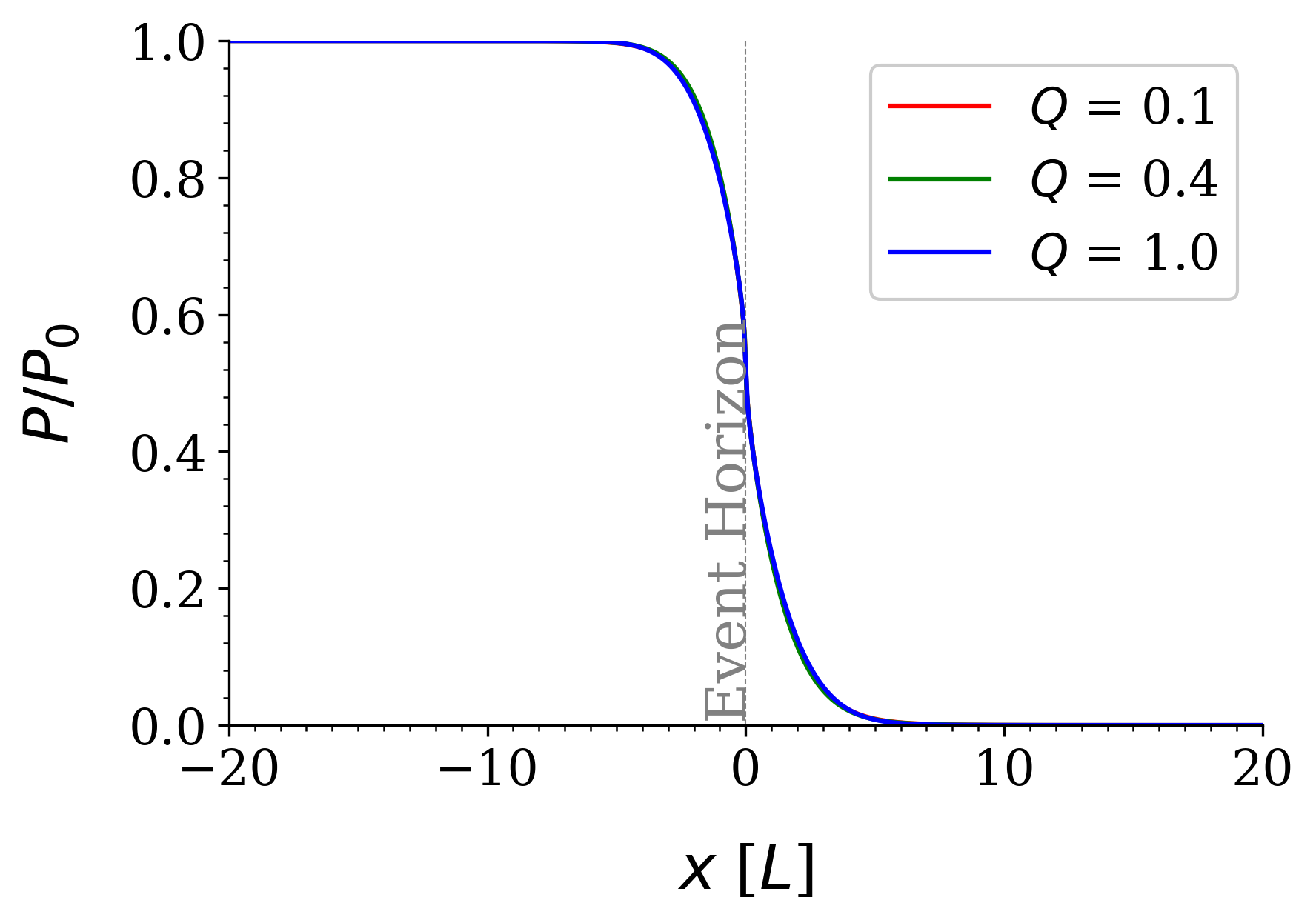}
        \caption{$\alpha = 0.5$ and $l=1$.}
        \label{fig:Pb1}
    \end{subfigure}\medbreak\medbreak
        \begin{subfigure}{0.45\textwidth}
        \centering
        \includegraphics[width=\linewidth]{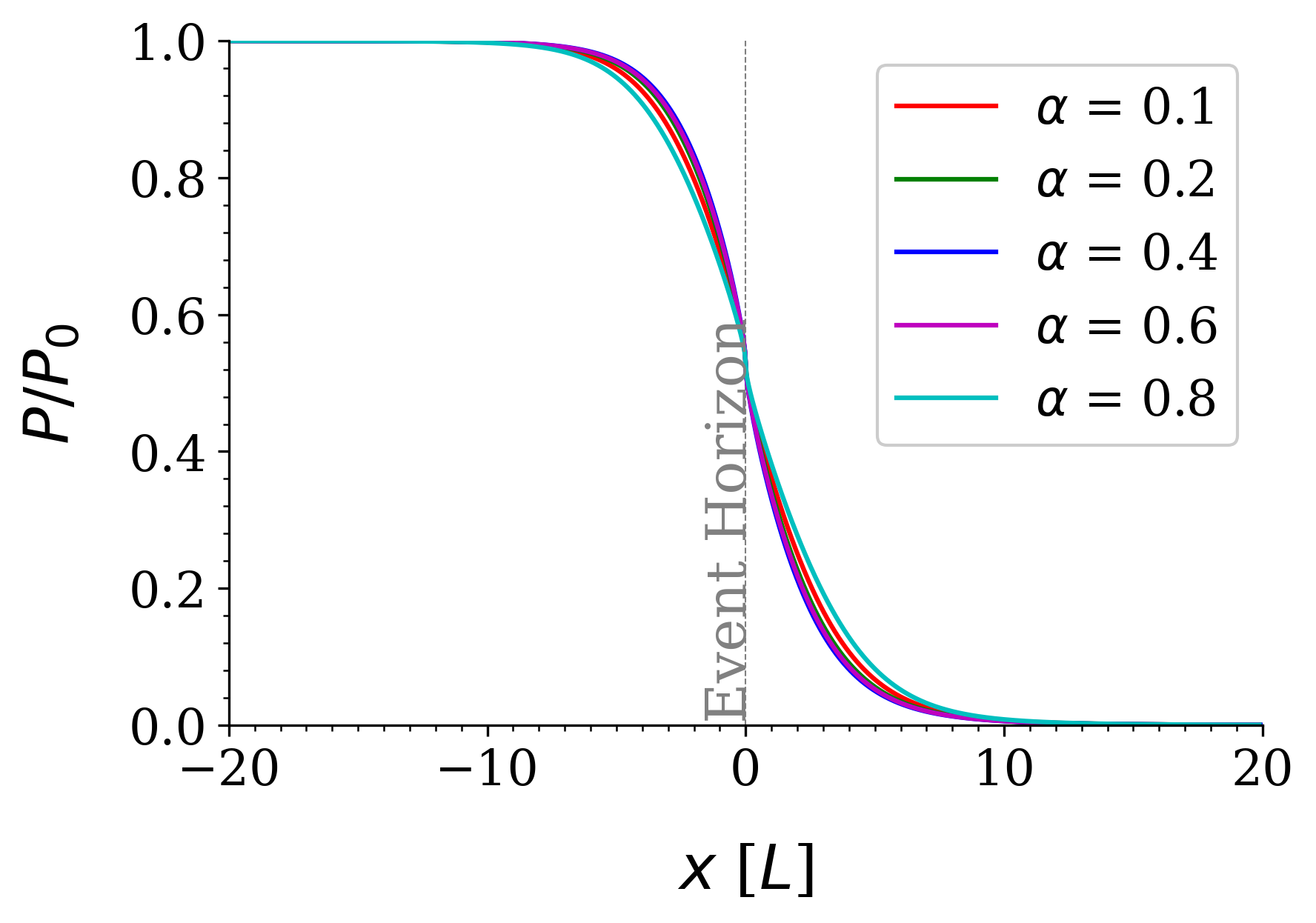}
        \caption{$Q = 1$ and $l=0$.}
        \label{fig:Pc0}
    \end{subfigure}
    \qquad\qquad
    \begin{subfigure}{0.45\textwidth}
        \centering
        \includegraphics[width=\linewidth]{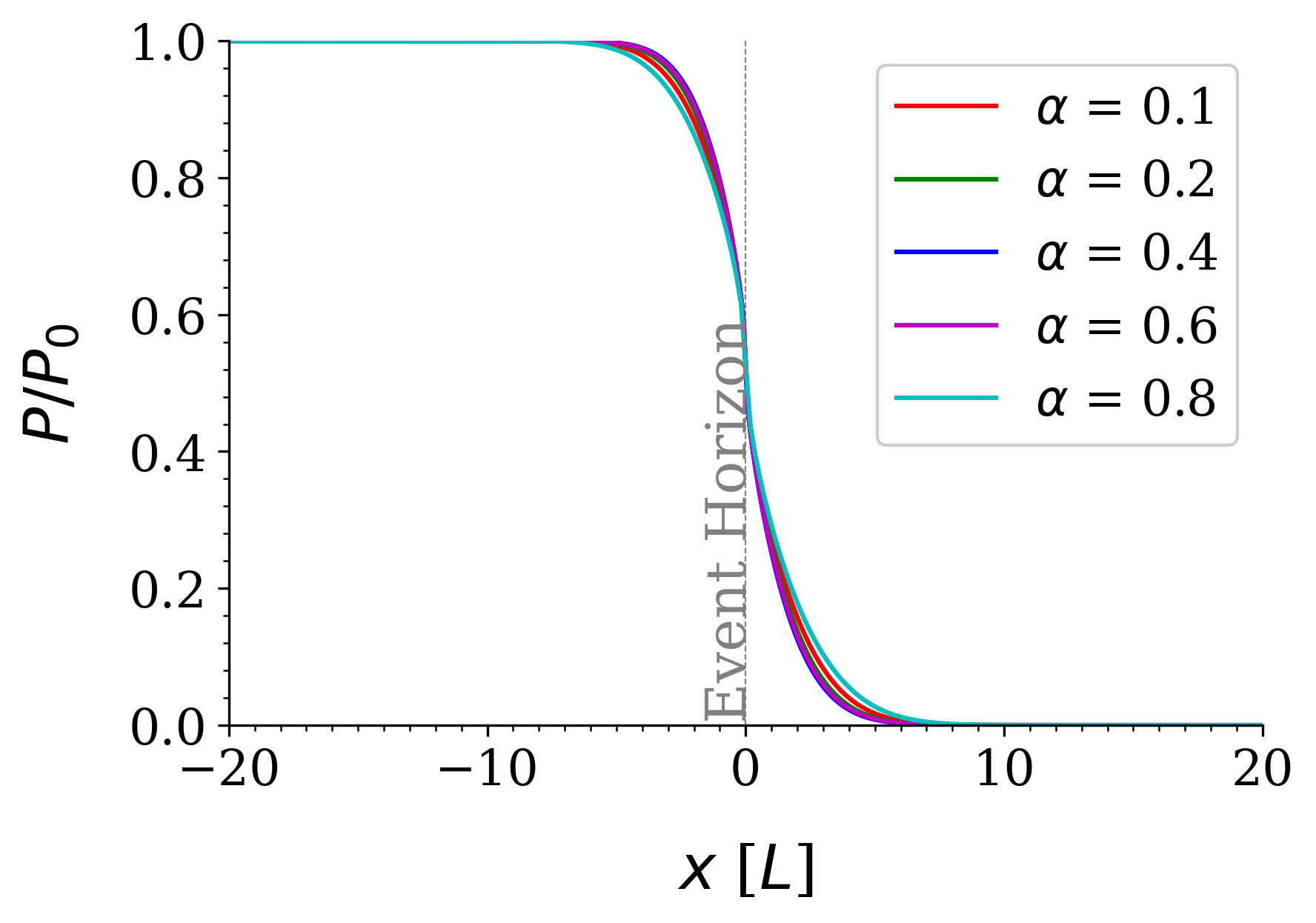}
        \caption{$Q = 1$ and $l=1$.}
        \label{fig:Pc1}
    \end{subfigure}
    \caption{Relative pressure as a function of the longitudinal direction of the de Laval nozzle, for the GD-extended hairy Reissner--Nordstr\"om metric  \eqref{eq23}.}
    \label{fig:pressure}
\end{figure}

\begin{figure}[H]
    \centering
    \begin{subfigure}{0.42\textwidth}
        \centering
        \includegraphics[width=\linewidth]{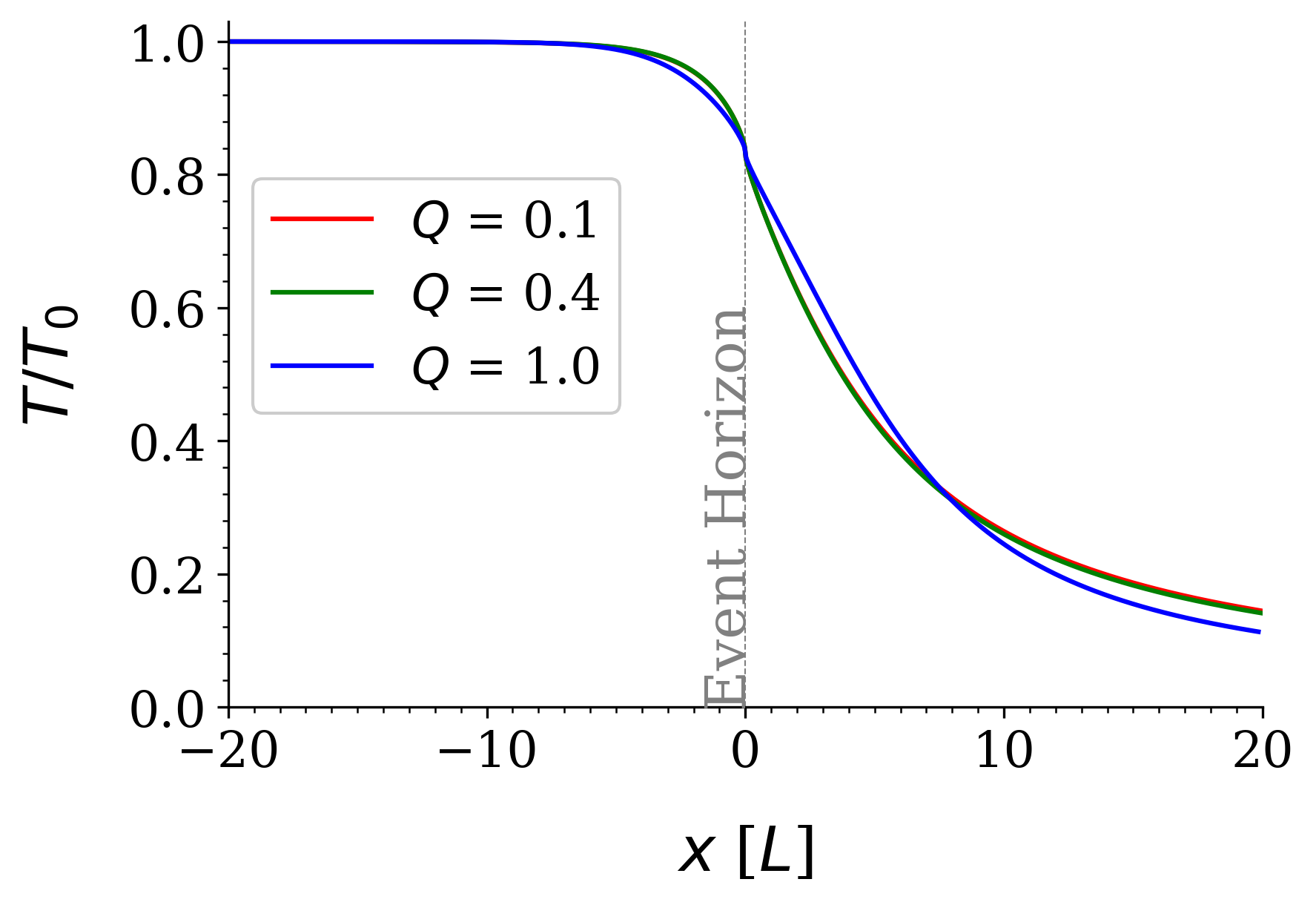}
        \caption{$\alpha = 0.1$ and $l=0$.}
        \label{fig:Ta0}
    \end{subfigure}\qquad\qquad
    \begin{subfigure}{0.42\textwidth}
        \centering
        \includegraphics[width=\linewidth]{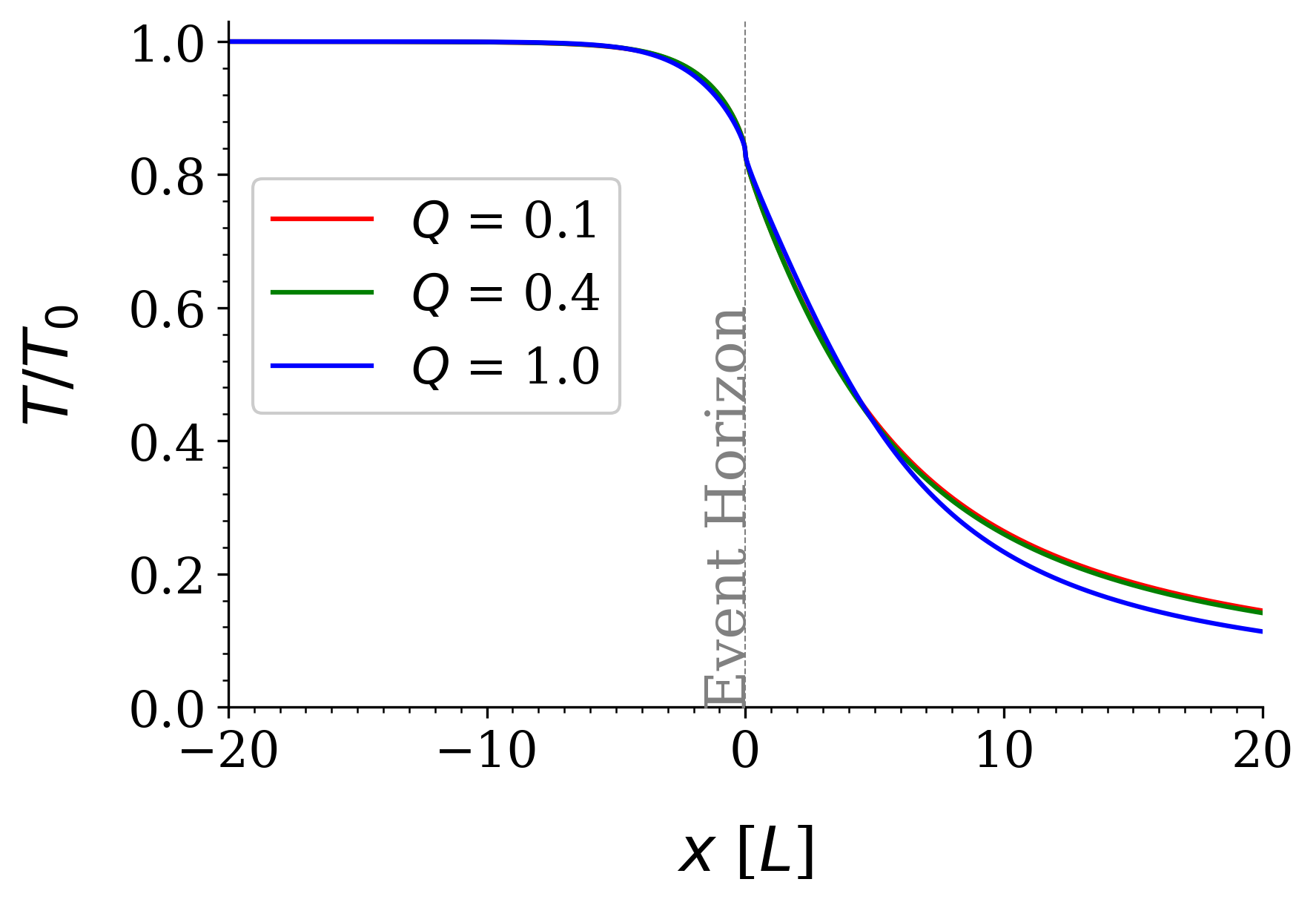}
        \caption{$\alpha = 0.5$ and $l=0$.}
        \label{fig:Tb0}
    \end{subfigure}
\medbreak\medbreak   \begin{subfigure}{0.42\textwidth}
        \centering
        \includegraphics[width=\linewidth]{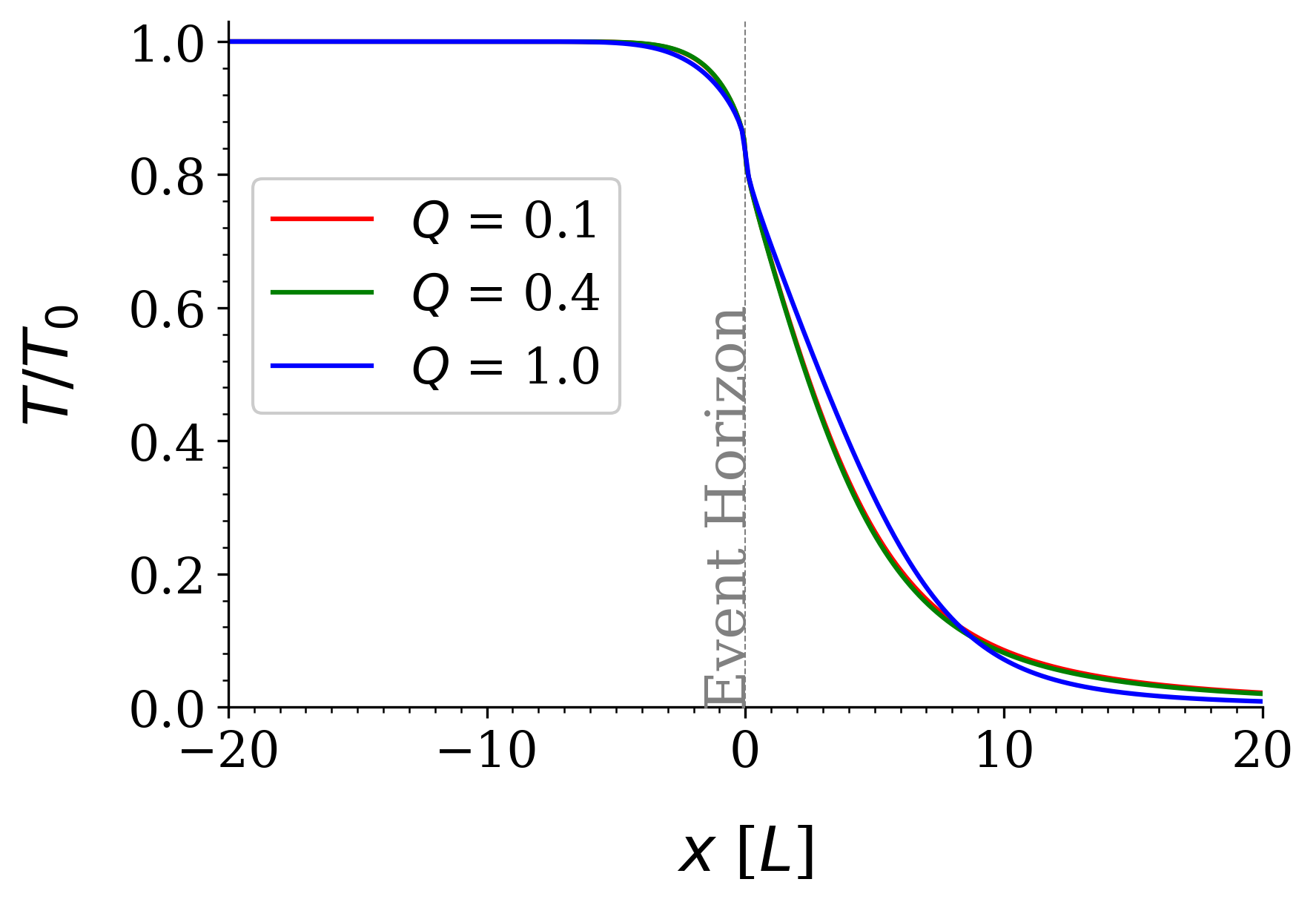}
        \caption{$\alpha = 0.1$ and $l=1$.}
        \label{fig:Ta1}
    \end{subfigure}\qquad\qquad
       \begin{subfigure}{0.42\textwidth}
        \centering
        \includegraphics[width=\linewidth]{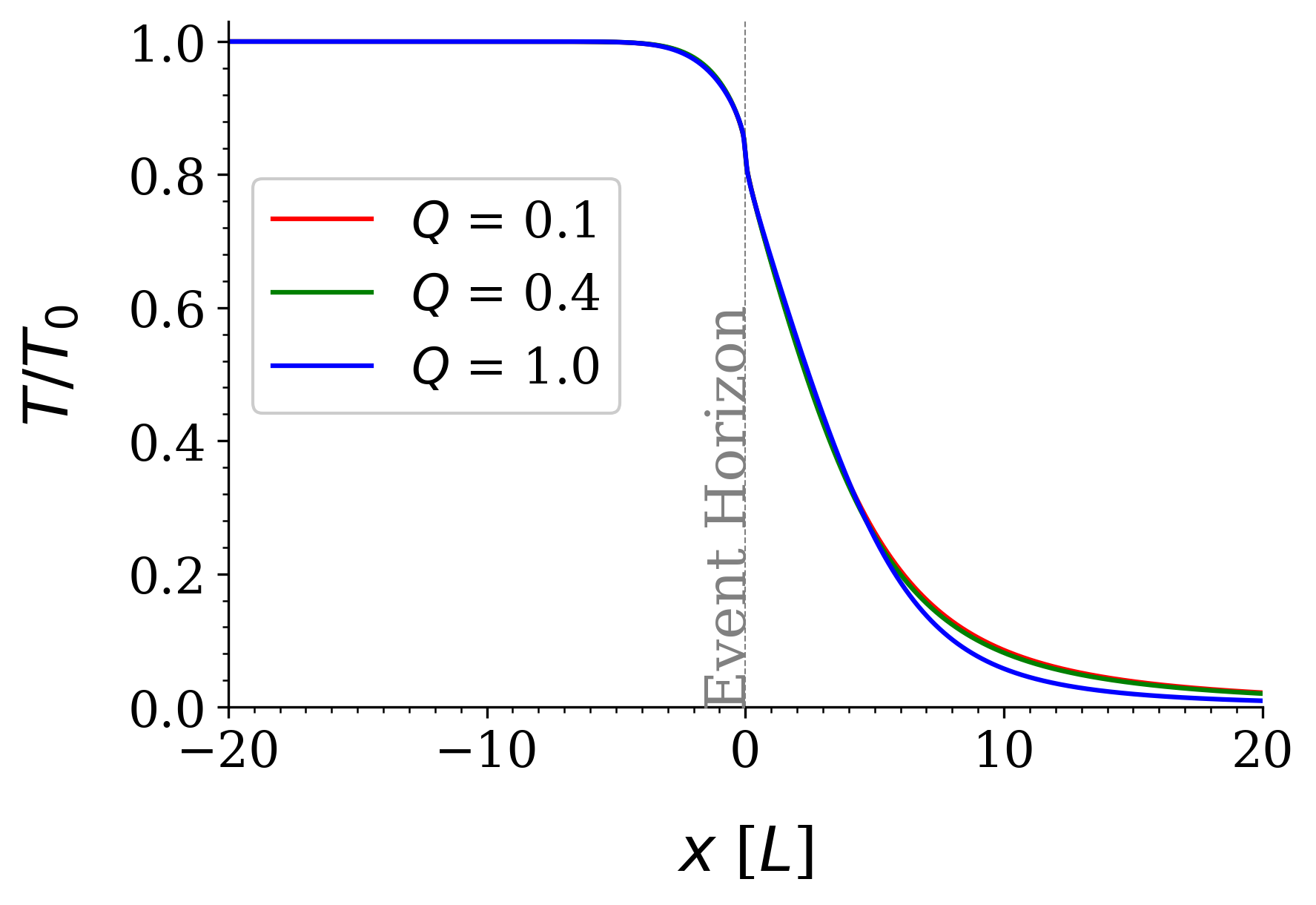}
        \caption{$\alpha = 0.5$ and $l=1$.}
        \label{fig:Tb1}
    \end{subfigure}\medbreak\medbreak
         \begin{subfigure}{0.42\textwidth}
        \centering
        \includegraphics[width=\linewidth]{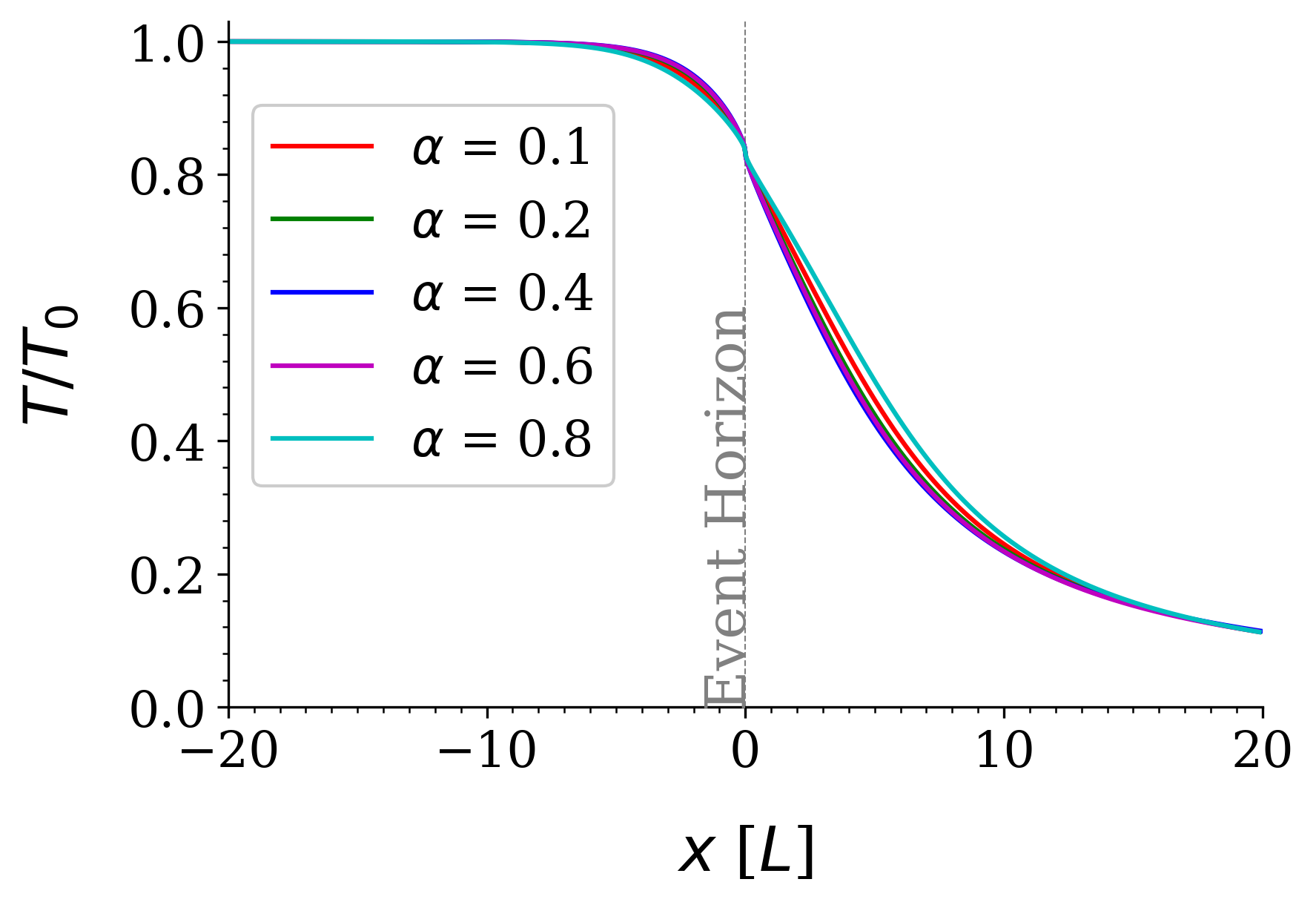}
        \caption{$Q = 1$ and $l=0$.}
        \label{fig:Tc0}
    \end{subfigure}\qquad\qquad
    \begin{subfigure}{0.42\textwidth}
        \centering
        \includegraphics[width=\linewidth]{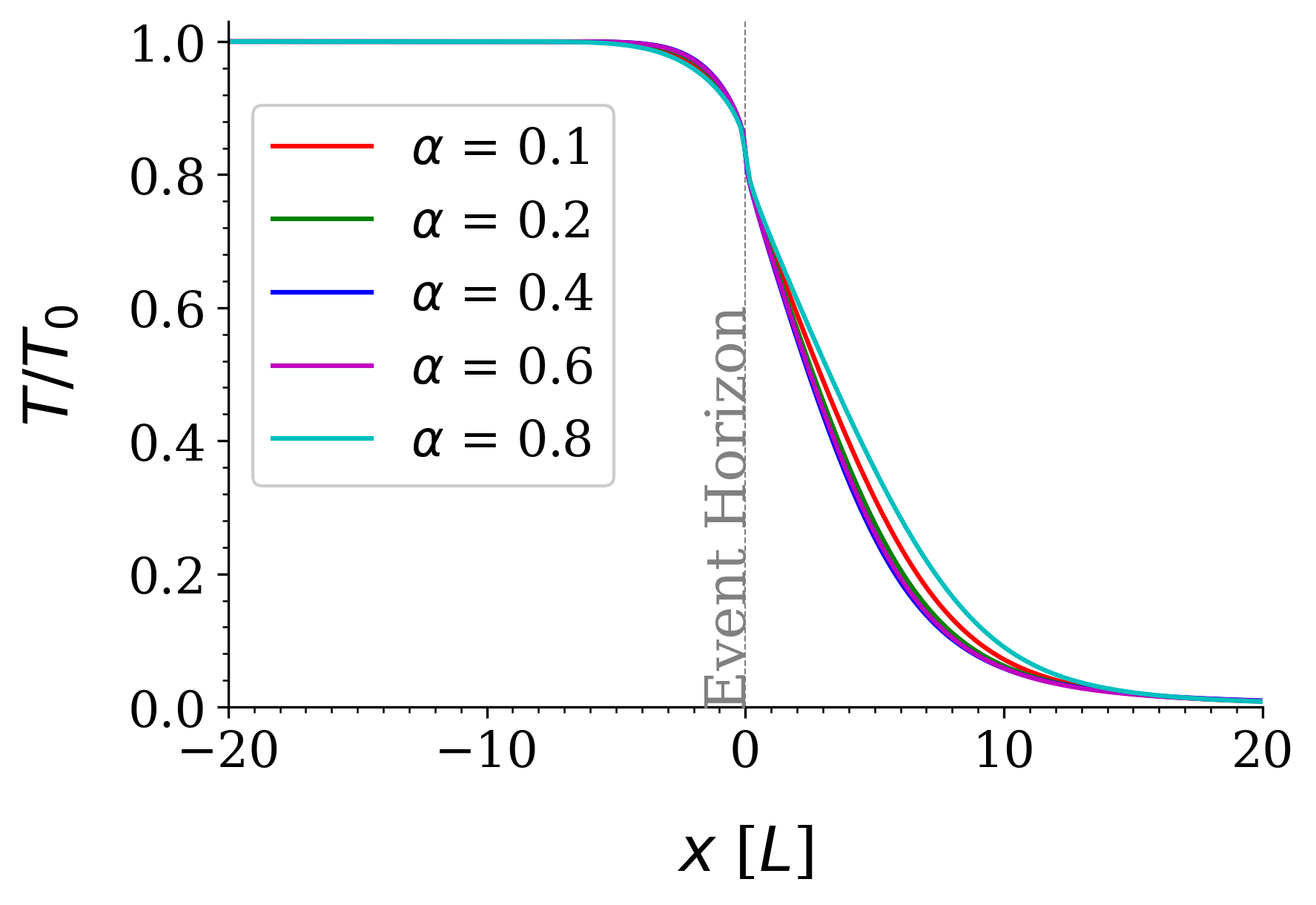}
        \caption{$Q = 1$ and $l=1$.}
        \label{fig:Tc1}
    \end{subfigure}
    \caption{Relative temperature as a function of the longitudinal direction of the de Laval nozzle, for the GD-extended hairy Reissner--Nordstr\"om metric  \eqref{eq23}.}
    \label{fig:temperature}
\end{figure}

\begin{figure}[H]
    \centering
    \begin{subfigure}{0.45\textwidth}
        \centering
        \includegraphics[width=\linewidth]{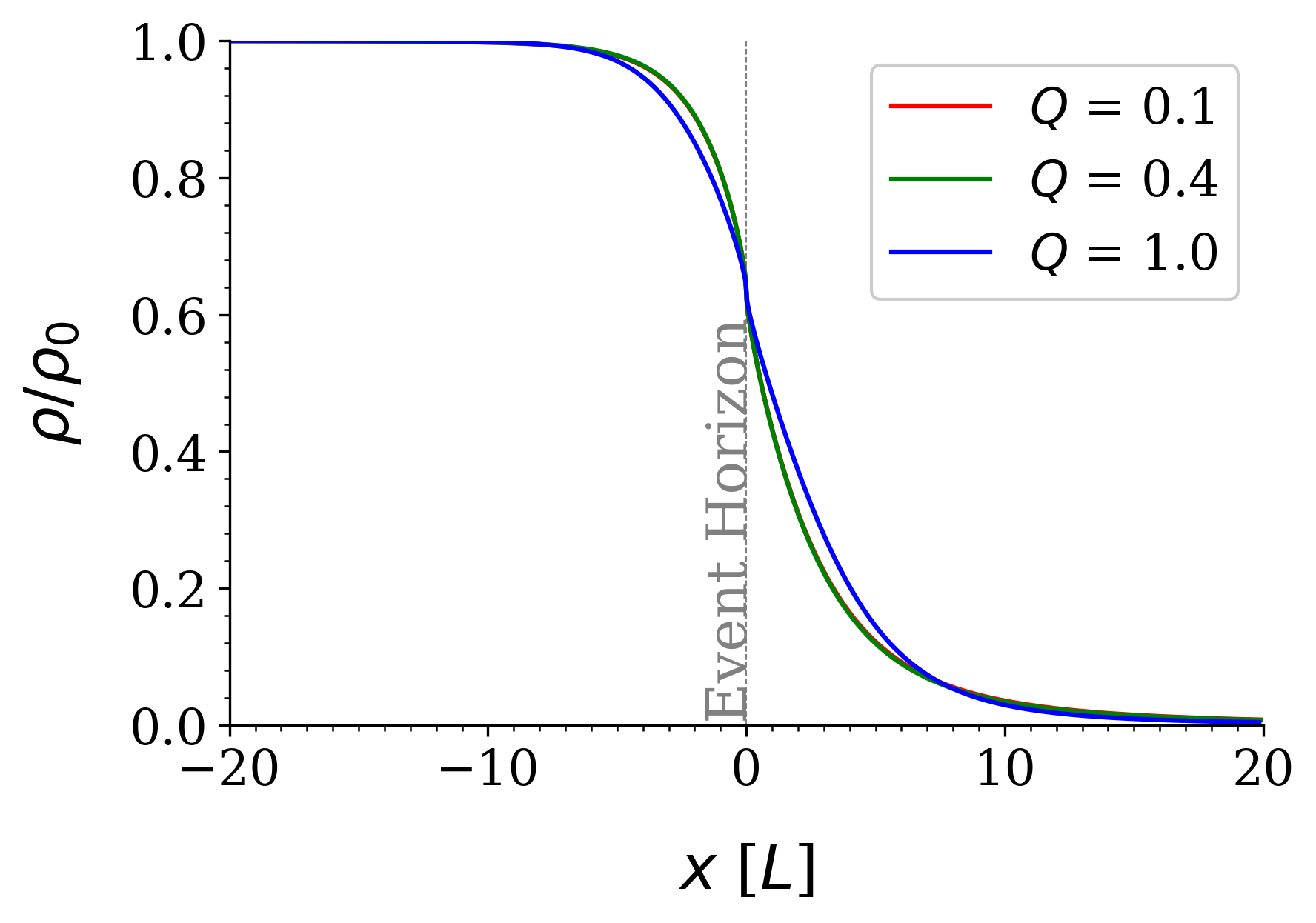}
        \caption{$\alpha = 0.1$ and $l=0$.}
        \label{fig:Da0}
    \end{subfigure}\qquad\qquad
    \begin{subfigure}{0.45\textwidth}
        \centering
        \includegraphics[width=\linewidth]{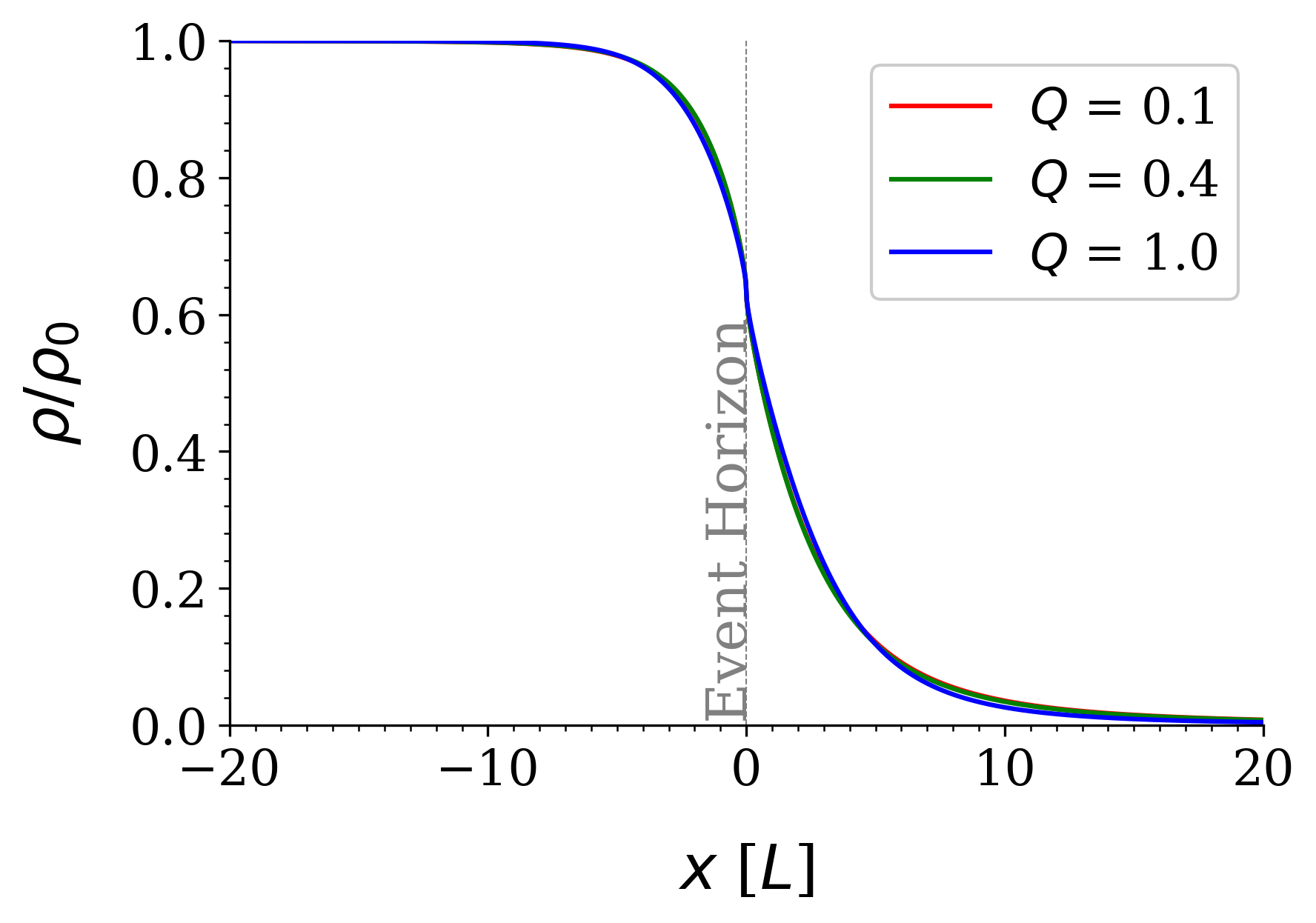}
        \caption{$\alpha = 0.5$ and $l=0$.}
        \label{fig:Db0}
    \end{subfigure}\medbreak\medbreak\medbreak
    \begin{subfigure}{0.45\textwidth}
        \centering
        \includegraphics[width=\linewidth]{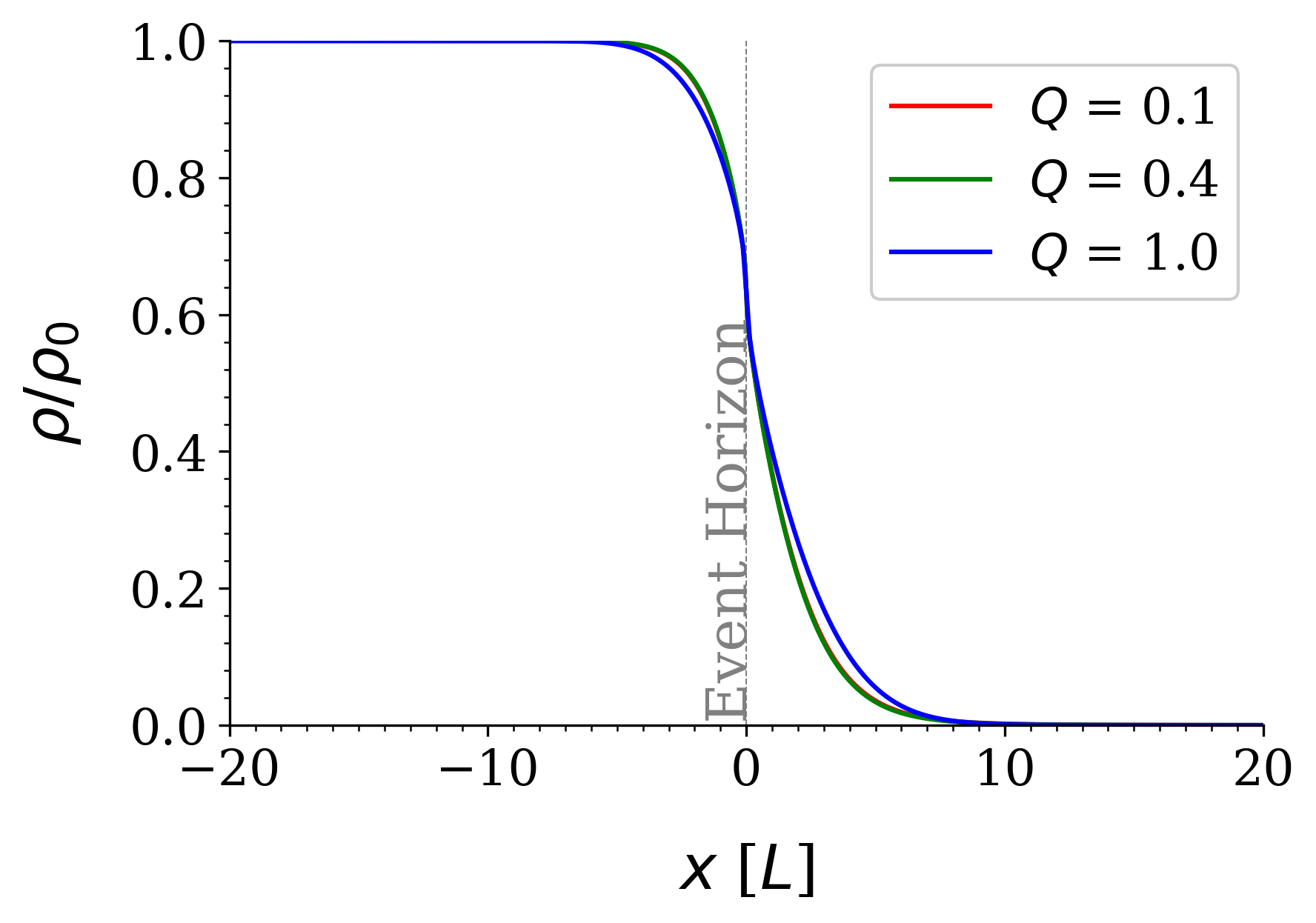}
        \caption{$\alpha = 0.1$ and $l=1$.}
        \label{fig:Da1}
    \end{subfigure}\qquad\qquad
    \begin{subfigure}{0.45\textwidth}
        \centering
        \includegraphics[width=\linewidth]{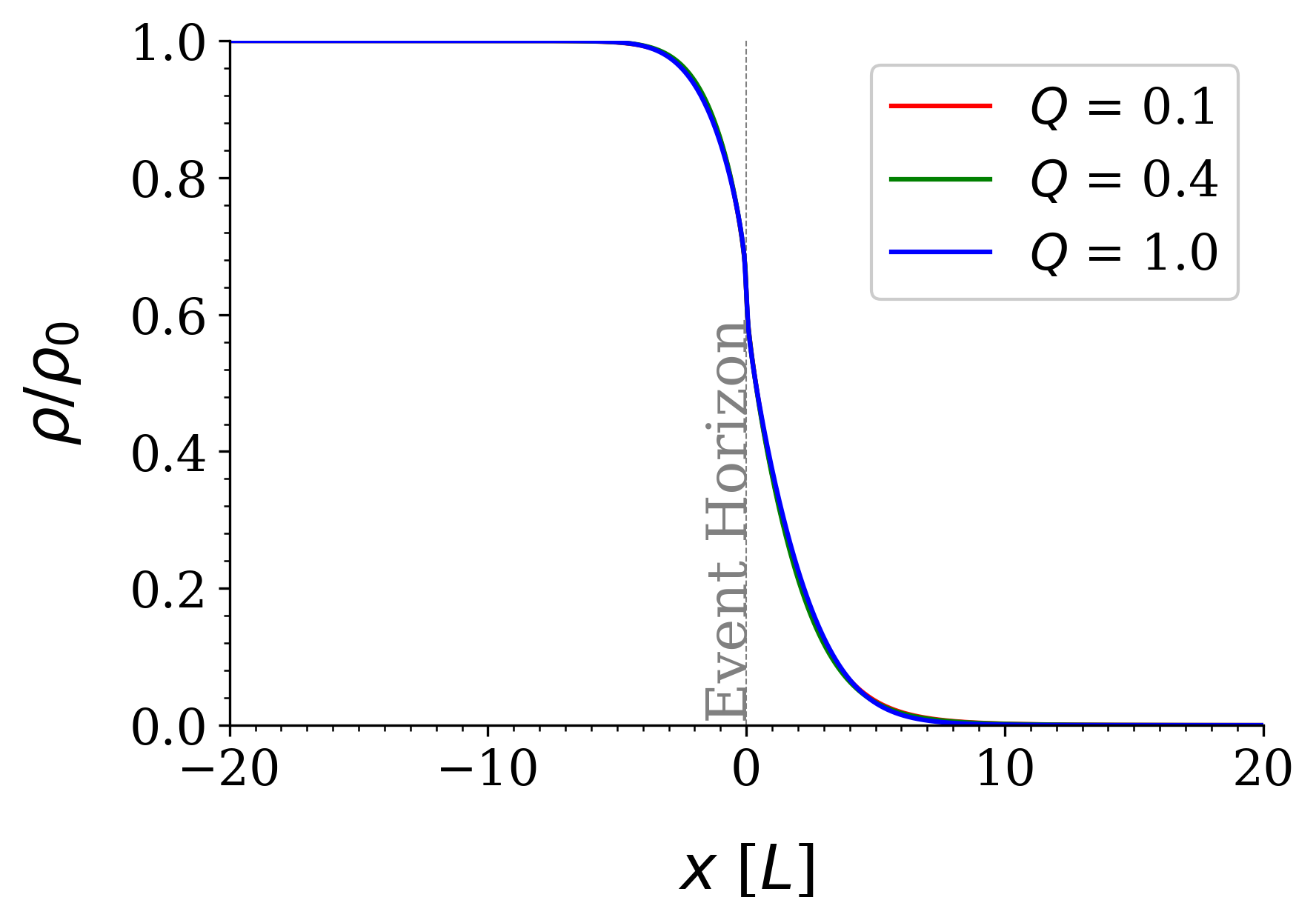}
        \caption{$\alpha = 0.5$ and $l=1$.}
        \label{fig:Db1}
    \end{subfigure}\medbreak\medbreak\medbreak
        \begin{subfigure}{0.45\textwidth}
        \centering
        \includegraphics[width=\linewidth]{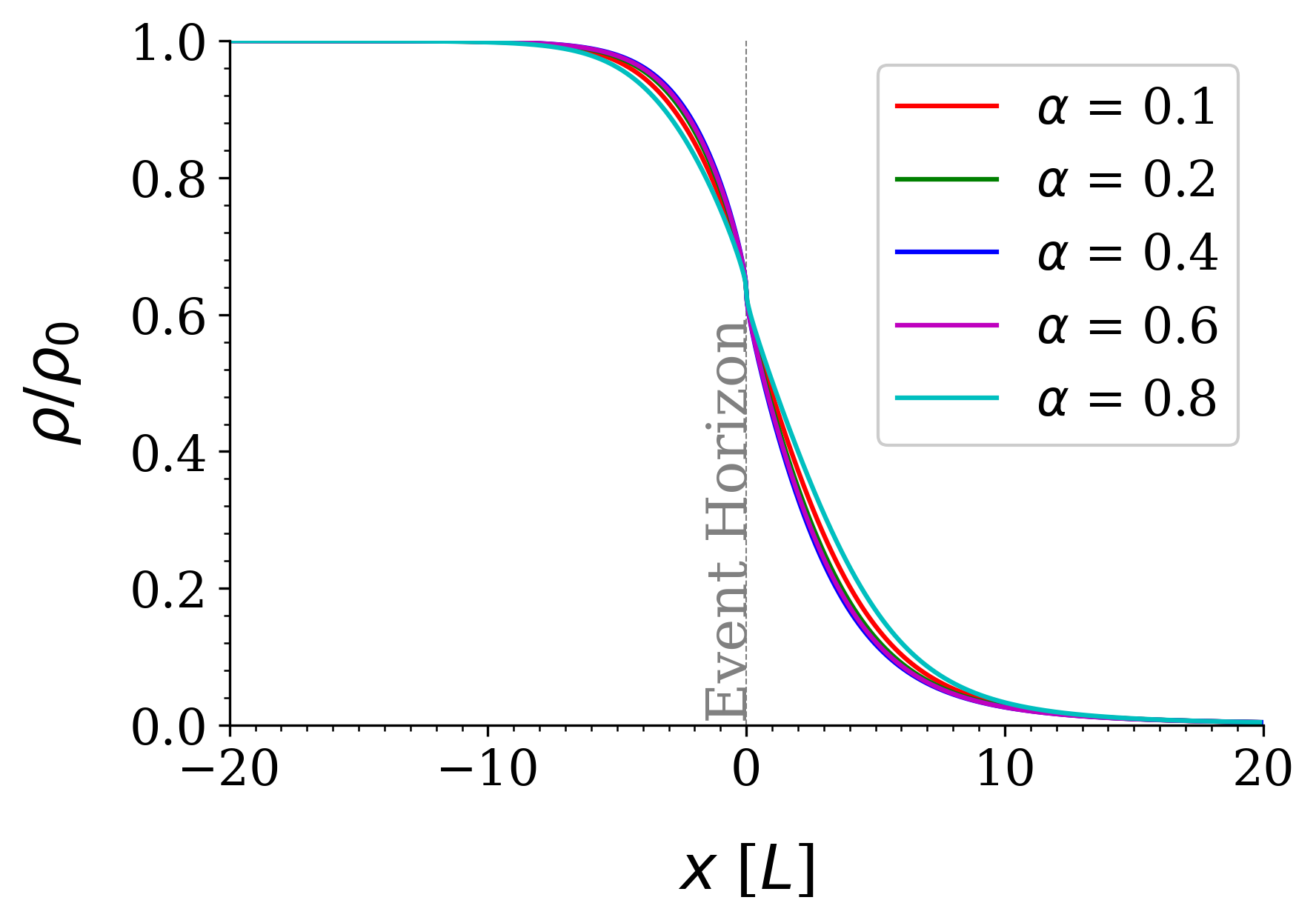}
        \caption{$Q = 1$ and $l=0$.}
        \label{fig:Dc0}
    \end{subfigure}\qquad\qquad
    \begin{subfigure}{0.45\textwidth}
        \centering
        \includegraphics[width=\linewidth]{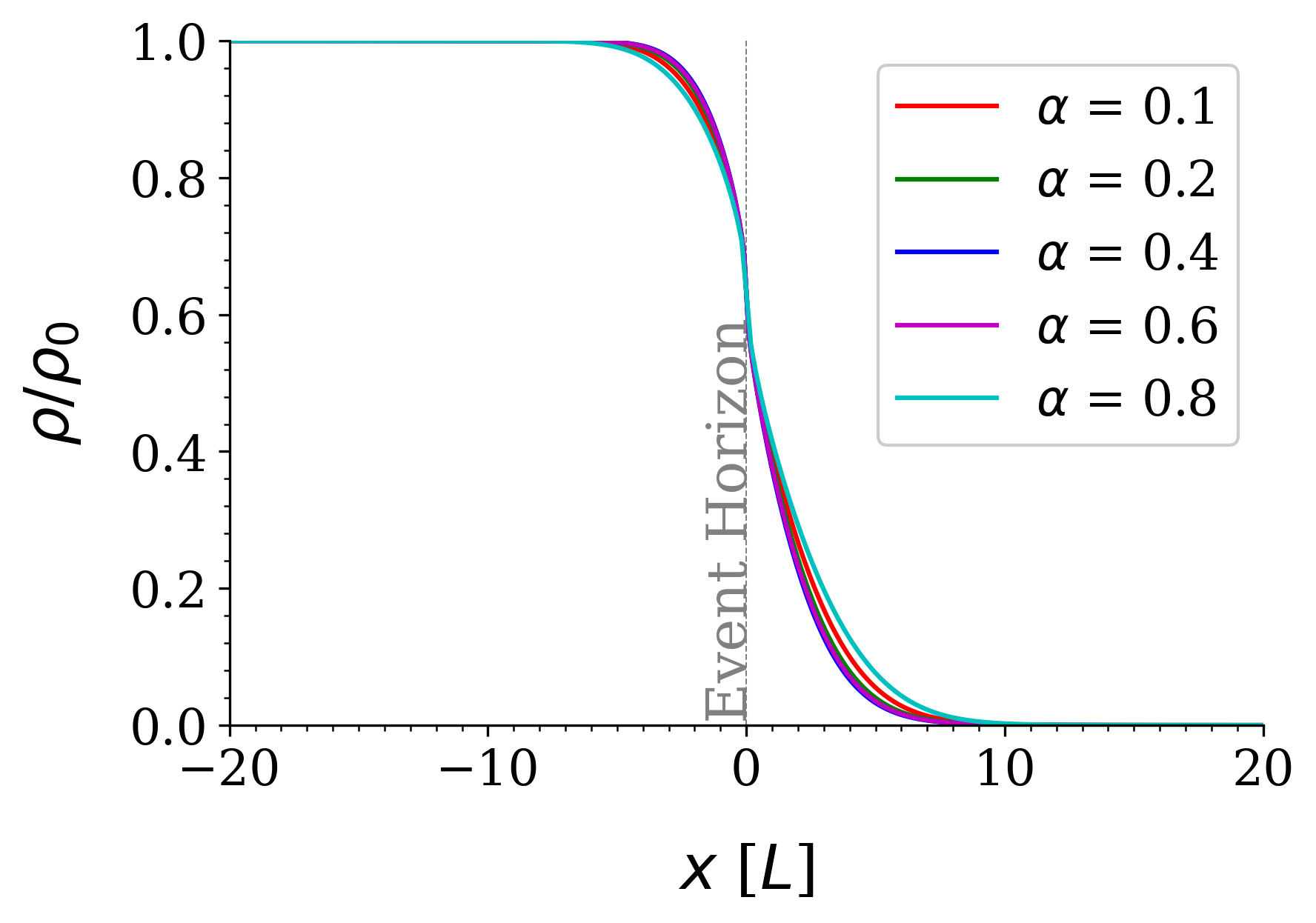}
        \caption{$Q = 1$ and $l=1$.}
        \label{fig:Dc1}
    \end{subfigure}
    \caption{Relative density as a function of the longitudinal direction of the de Laval nozzle, for the GD-extended hairy Reissner--Nordstr\"om metric  \eqref{eq23}.}
    \label{fig:Density}
\end{figure}

\begin{figure}[H]
     \centering
     \begin{subfigure}{0.42\textwidth}
         \centering
         \includegraphics[width=\linewidth]{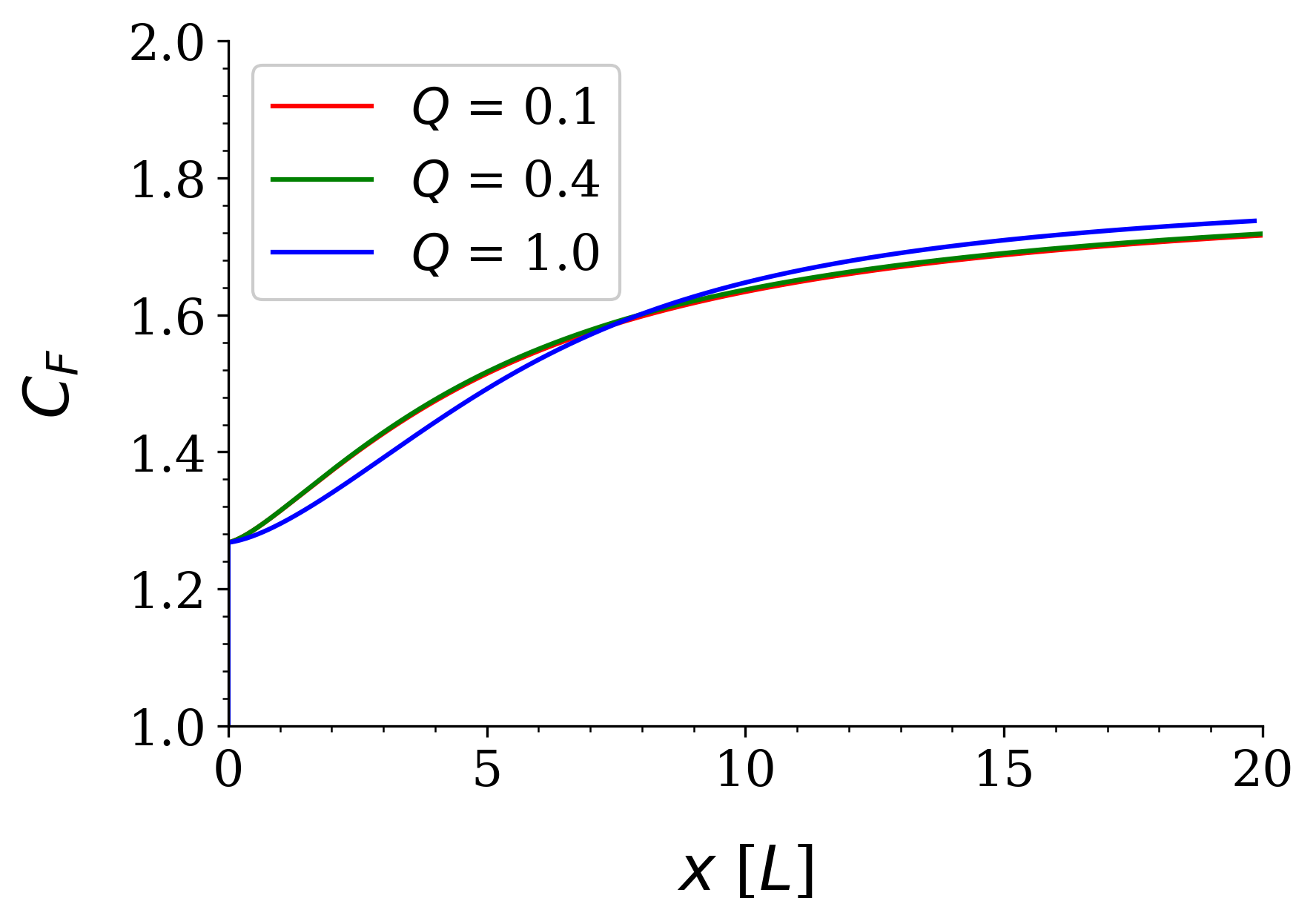}
         \caption{$\alpha = 0.1$ and $l=0$.}
         \label{fig:Cfa0}
     \end{subfigure}\qquad\qquad
     \begin{subfigure}{0.42\textwidth}
         \centering
         \includegraphics[width=\linewidth]{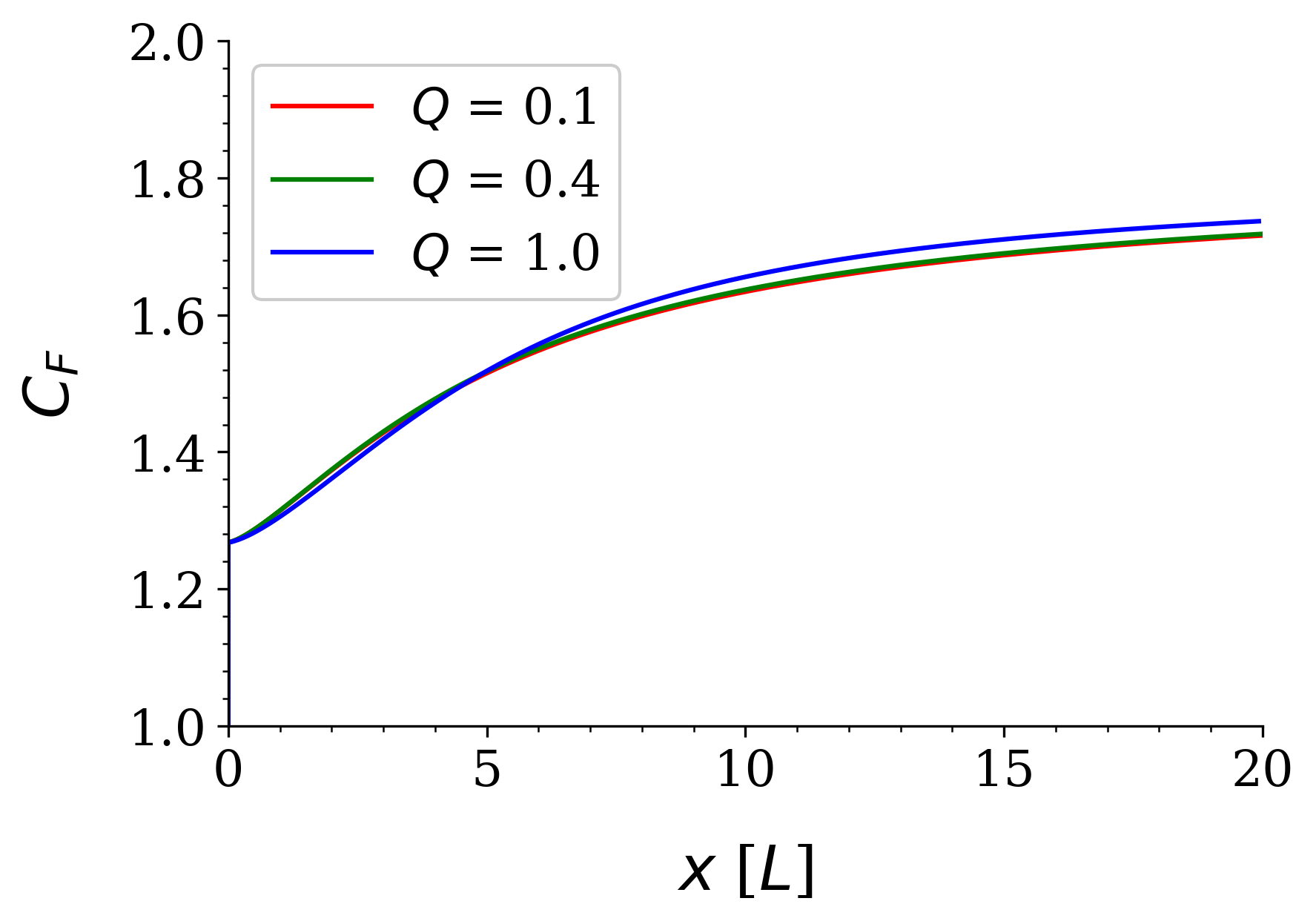}
         \caption{$\alpha = 0.5$ and $l=0$.}
         \label{fig:Cfb0}
     \end{subfigure}\medbreak\medbreak
     \begin{subfigure}{0.42\textwidth}
         \centering
         \includegraphics[width=\linewidth]{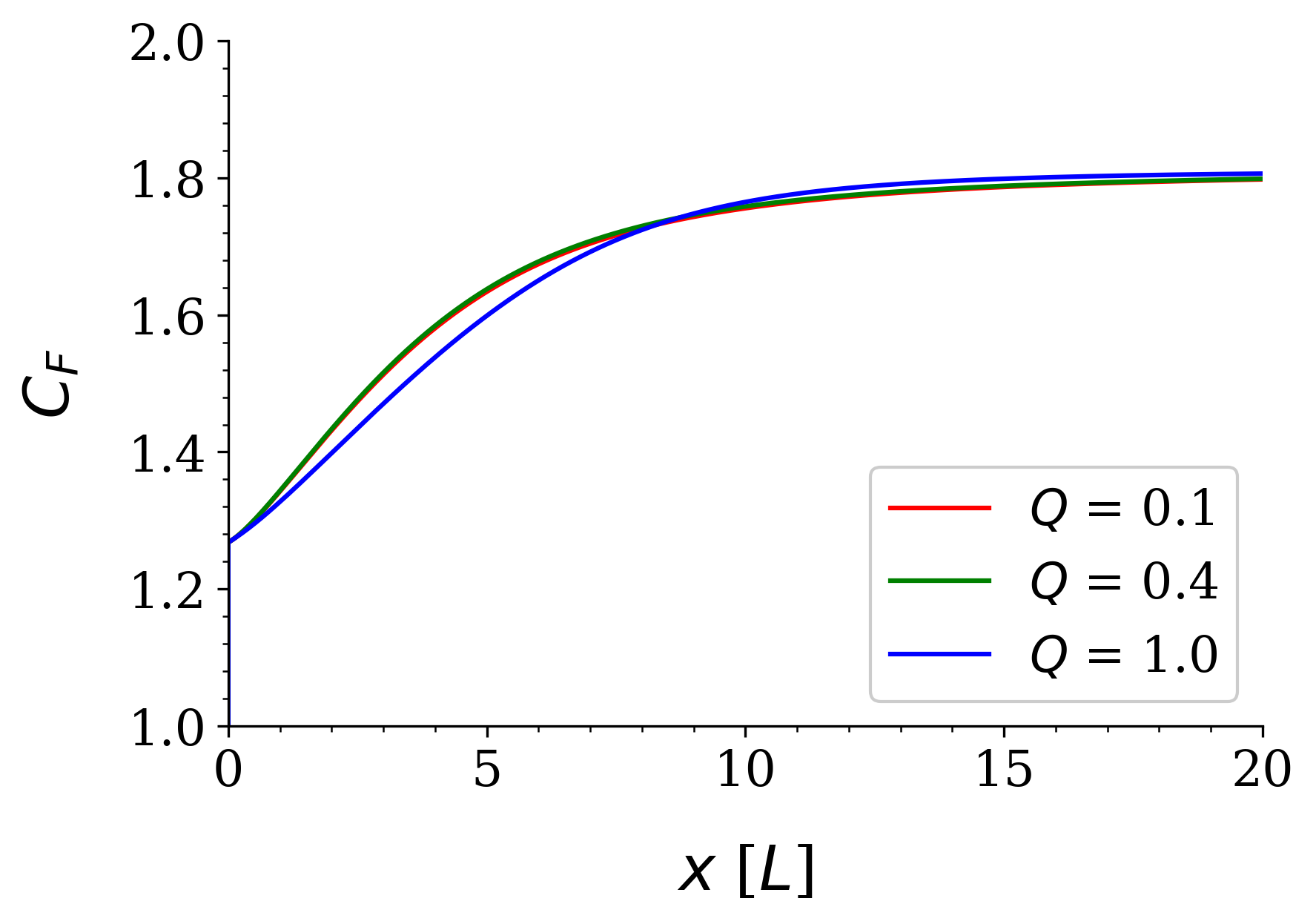}
         \caption{$\alpha = 0.1$ and $l=1$.}
         \label{fig:Cfa1}
     \end{subfigure}\qquad\qquad
     \begin{subfigure}{0.42\textwidth}
         \centering
         \includegraphics[width=\linewidth]{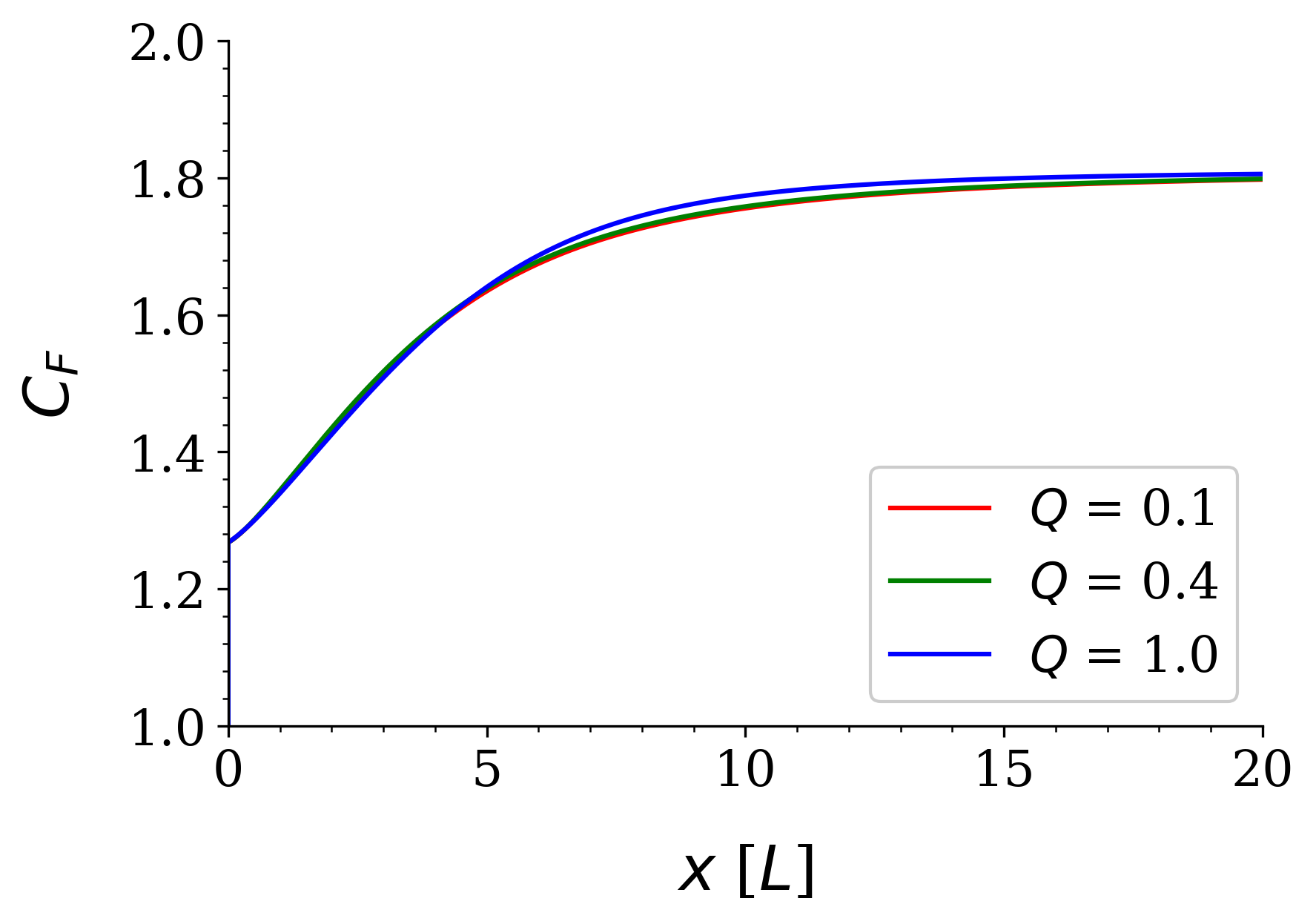}
         \caption{$\alpha = 0.5$ and $l=1$.}
         \label{fig:Cfb1}
     \end{subfigure}\medbreak\medbreak   \begin{subfigure}{0.42\textwidth}
         \centering
         \includegraphics[width=\linewidth]{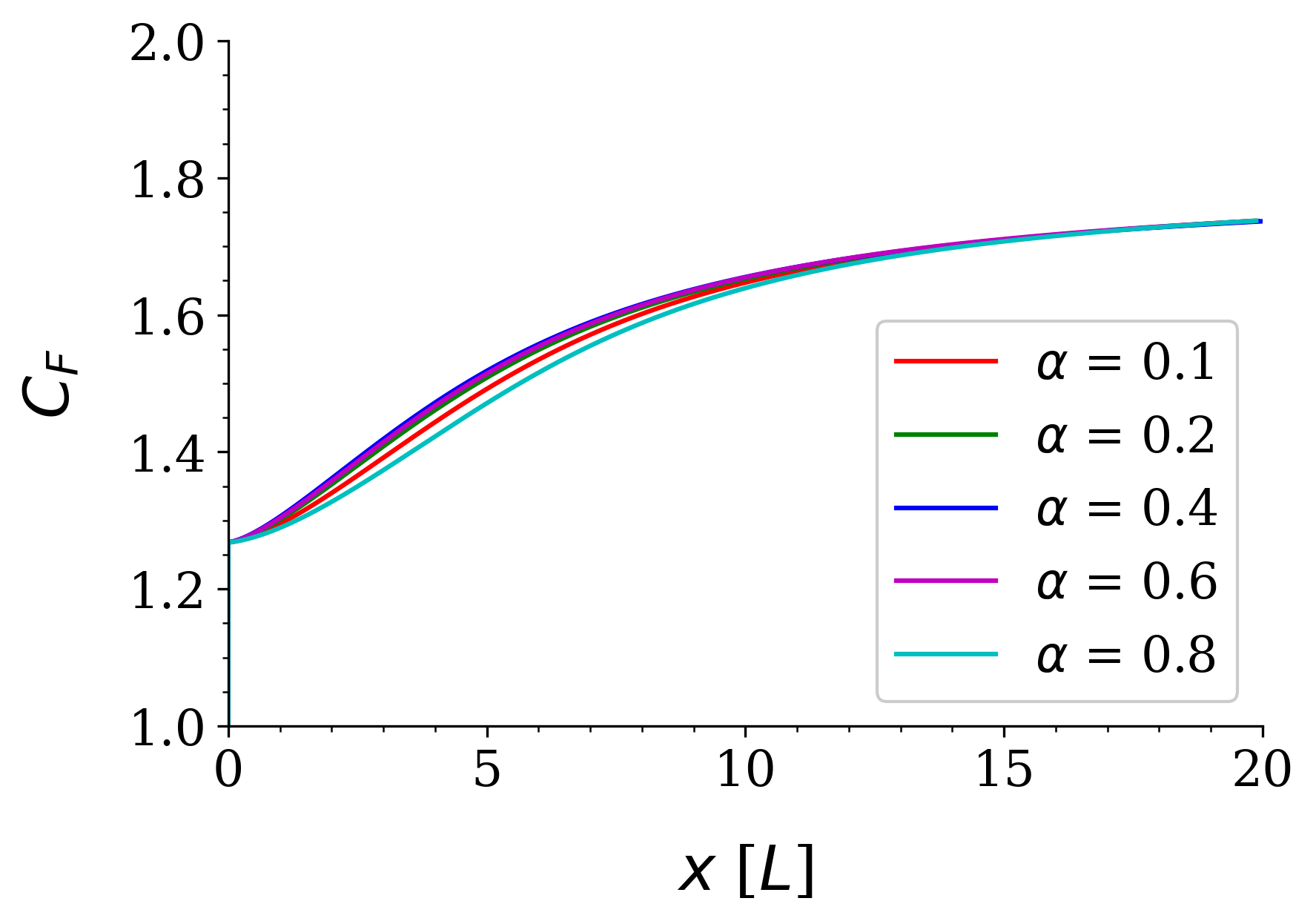}
         \caption{$Q = 1$ and $l=0$.}
         \label{fig:Cfc0}
     \end{subfigure}\qquad\qquad  
     \begin{subfigure}{0.42\textwidth}
         \centering
         \includegraphics[width=\linewidth]{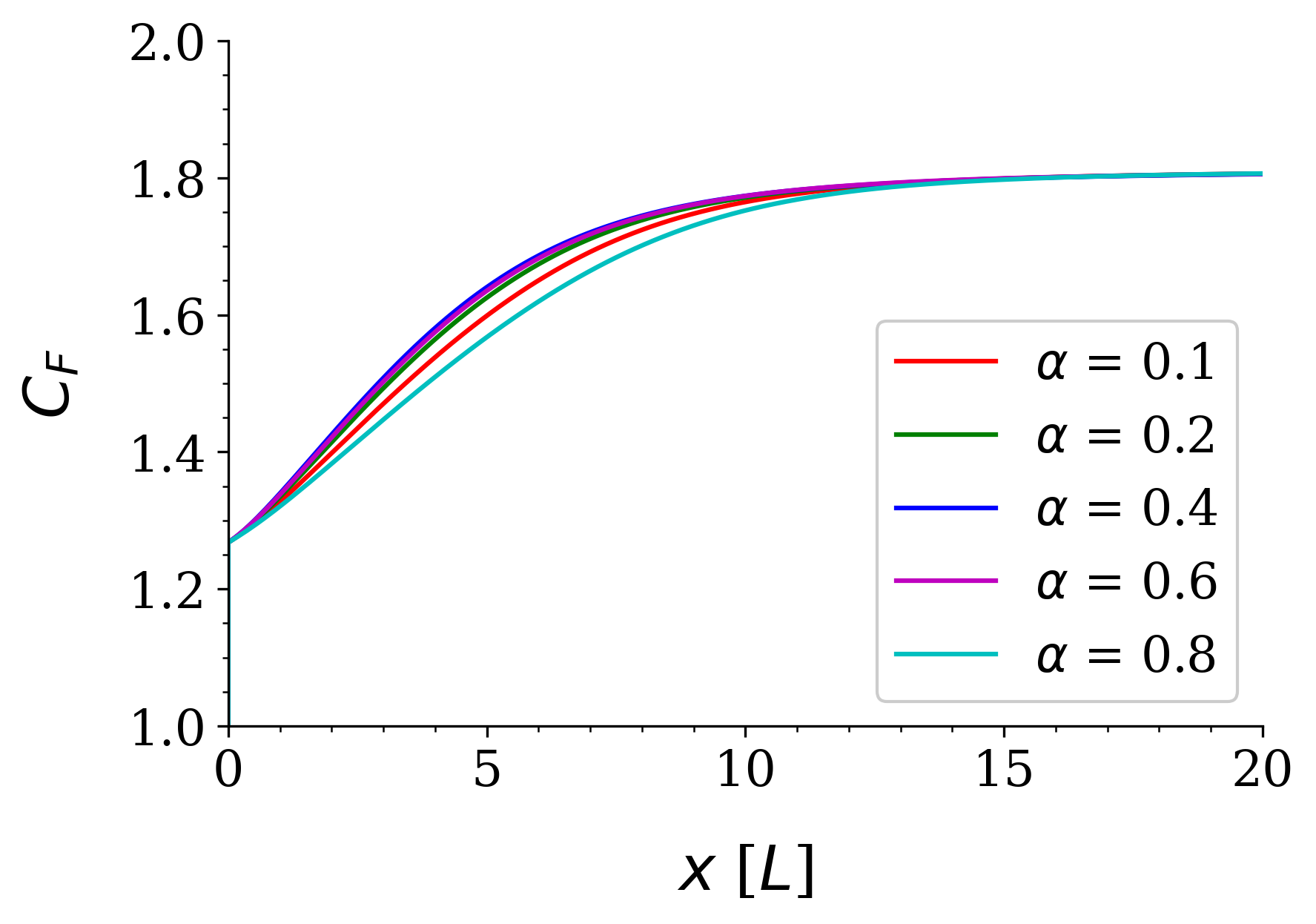}
         \caption{$Q = 1$ and $l=1$.}
         \label{fig:Cfc1}
     \end{subfigure}
     \caption{Thrust coefficient as a function of the longitudinal direction of the de Laval nozzle, for the GD-extended hairy Reissner--Nordstr\"om metric  \eqref{eq23}.}
     \label{fig:Cf}
 \end{figure}

{\color{black}
\subsection{Features of the analog de Laval nozzle}

To solve Eq. \eqref{eq: lambda} numerically, combinations of the GD parameters
$\alpha$, $\ell$, and $Q$ in the extended Reissner--Nordstr\"om metric~\eqref{eq23},
and the multipole orbital quantum number $l$ in Eq. \eqref{final effective potential}, must be chosen.}
Ref.~\cite{Ovalle:2020kpd} showed that the GD hairy parameter $\ell$ is proportional to $\alpha$. Therefore one can assume, without loss of generality, that $\ell = 2 \alpha$ for instance. Hence we will analyse the properties of the de Laval nozzle as a function of the GD hairy parameters $\alpha$, $Q$, for the wave modes of degree $l=0$ and $l=1$. After, when the QN modes are calculated and related to the nozzle quality factor, multipole values $l=2,\ldots,6$ will be also taken into account, since they are relevant for the analysis of the overtones. Nevertheless, to investigate the influence of the GD hairy charges on the profiles of the pressure, Mach number, nozzle shape, temperature, density, and thrust coefficient of the nozzle, the $s$- and $p$-wave modes are already sufficient.  When Eqs. \eqref{eq23} and \eqref{final effective potential} are analysed, the range $\alpha > 1$ does not provide a reasonable set of solutions, because either  it has no real event horizon radius $r_h$ or the effective potential $V_{\textsc{eff}}$ is not well-behaved. Therefore in all plots that follow, we vary $Q$ with small values of $\alpha<1$ fixed. After, we vary $\alpha$, for the fixed value $Q=1$. Each set of plots is implemented for the $s$-wave modes ($l=0$) and the $p$-wave modes ($l=1$). In all plots that follow in Figs.  \ref{fig:Veff} -- \ref{fig:Cf}, the value $x=0$ corresponds to the nozzle throat centre, corresponding to the analog event horizon. 

Fig. \ref{fig:Veff} displays the effective potential as a function of the longitudinal direction of the de Laval nozzle, for several choices of GD parameters $\alpha$ and $Q$. Fig. \ref{fig:Va0} shows the effective potential for $\alpha = 0.1$ and the $s$-wave modes $l=0$, for several values of $Q$, whereas  Fig. \ref{fig:Vb0} exhibits the case for $\alpha=0.5$. 
In these figures, for all values of $Q\lesssim 0.55$, the potential is essentially the same for both the values $\alpha=0.1$ and $\alpha=0.5$, 
showing an insignificant influence of the GD hairy parameter $\alpha$ on the effective potential for the range $Q\lesssim 0.55$, at least for the $s$-wave modes.  
For values $Q=1$, the effective potential reaches a slightly higher peak $V_{\textsc{eff}}(x)=0.0316$ at $x=4.03$ in Fig. \ref{fig:Va0}, whereas
$V_{\textsc{eff}}(x)=0.0331$ at $x=2.51$, in Fig. \ref{fig:Vb0}. For all values $Q\gtrsim 0.55$, in Figs. \ref{fig:Va0} -- \ref{fig:Vb1} we verify numerically that the higher the values of the GD parameter $\alpha$, the higher the effective potential peak is, and it occurs in smaller distances along the nozzle longitudinal axis. 
An analogous behaviour can be realized in Figs. \ref{fig:Va1} and \ref{fig:Vb1}, now regarding the $p$-wave mode $l=1$, with the only substantial difference residing in the fact that the difference between the effective potential for $Q=0.1$ and $Q=0.4$ is slightly higher for the $p$-wave mode than for the $s$-wave mode. Besides, for the $p$-wave mode and $Q=1$, 
the effective potential has two inflection points for $\alpha = 0.1$, whereas it has one inflection point for $\alpha = 0.5$. However, for the $p$-wave mode and $Q=1$, the peak of the effective potential is the same for both $\alpha = 0.1$ and $\alpha = 0.5$, with the only difference that occurs at different values of $x$. It is straightforward to see that the GD parameter $\alpha$ does not have the same influence on the results as $Q$. Although the effective potential for $l=1$ has significant differences for the cases $\alpha=0.1$ and $\alpha=0.5$ for the specific case $Q=1$, respectively in Figs. \ref{fig:Va1} and \ref{fig:Vb1}, the behavior of the  effective potential for $Q=0.1$ and $Q=0.4$ is practically equal irrespectively the value of $\alpha$ is.
However, for $Q=1$  for asymptotically high values of $x$, the effective potential profile is slightly higher for $\alpha = 0.1$, but still quite similar, following a similar pattern. 
The parameter $\alpha$ only changes the curve substantially when $Q=1$ and it is not linearly correlated to the point of inflection, according to Figs. \ref{fig:Vc0} and \ref{fig:Vc1}. It is also worth mentioning that all the peaks of the effective potential in Fig. \ref{fig:Vc0} are higher than the respective peaks for fixed values of $\alpha$ in Fig. \ref{fig:Vc1}, indicating that the quantum azimuthal number $l$ alters the peak configuration.  

As the corresponding effective potential governing of perturbations in a de Laval nozzle emulates the potential for perturbations for the GD-extended hairy Reissner--Nordstr\"om black hole  \eqref{eq23}, the specific shape of the de Laval nozzle is a useful quantity that can be obtained. The results are displayed in Fig. \ref{fig:shape}. 
\noindent Fig. \ref{fig:shape} shows the nozzle shape for a reasonable range of GD-hairy parameters.  The nozzle shape determines how much thermal energy may be converted into kinetic energy.  The wider the slope of the nozzle shape is, the faster the fluid flows throughout the nozzle, generating more thrust. Focusing on Fig. \ref{fig:Ac1}, even though the effective potential is roughly similar for different values of $\alpha$, the eccentricity seems to have an important role, as the farther from $r_h$, the less energetic flow $V_{\textsc{eff}}$ provides. 
Figs. \ref{fig:Aa0} and \ref{fig:Ab0} show that for the $l=0$ $s$-wave mode, the influence of the hairy parameter $\alpha$ is almost null, for each fixed value of $Q$. Instead,  the $l=1$ $p$-wave mode evinces 
an important influence of the hairy parameter $\alpha$: for $Q=1$, the nozzle is wider for $\alpha = 0.5$ than for $\alpha=0.1$, along the longitudinal direction, corresponding to a lower focal length and lower eccentricity.   On the other hand, the opposite behaviour is verified to the values $Q=0.1$ and $Q=0.4$, with the nozzle presenting a higher eccentricity for higher values of $\alpha$, as illustrated in Figs. \ref{fig:Aa1} and \ref{fig:Ab1}. Now, fixing the value $Q=1$, the  $l=0$ $s$-wave mode presents a very slight variation when $\alpha$ varies in the range $\alpha\in[0.1, 0.8]$, as shown in Fig. \ref{fig:Ac0}, although the nozzle shapes of higher eccentricity are attained for higher values of $\alpha$.  The case of the  $l=1$ $p$-wave \ref{fig:Ac1} presents a noticeable variation of the nozzle eccentricity with $\alpha$ varying.

Now the Mach number $\mathtt{M}$ can be addressed and analysed as a function of the longitudinal direction of the de Laval nozzle. Subsonic propagating fluid flows can reach sonic velocities at the nozzle throat, corresponding to $\mathtt{M}=1$, constituting the so-called choked flow. Going through the longitudinal 
direction along the nozzle, its cross-sectional area gets larger and the gas expands, making the flow velocity rise to supersonic patterns, with $\mathtt{M}>1$. Sound waves are not allowed to propagate in the reverse direction. The transition between subsonic and supersonic flows can be seen in all the plots in Fig. \ref{fig:mach}, around $\mathtt{M}\approxeq1$, characterizing the nozzle sonic point for sound waves throughout the fluid flow. 
The multipole $l$ drastically changes the amount of effective potential, and so the de Laval nozzle aerodynamical properties are dictated by the Mach number,  as seen in Fig. \ref{fig:mach}.  
Figs. \ref{fig:Ma0} and \ref{fig:Mb0} depict the Mach number for the $s$-wave mode. For $\alpha = 0.1$ fixed, Fig. \ref{fig:Ma0} shows that the lower the value of the charge $Q$, the faster the value of the Mach number increases, up to the sonic point at the nozzle throat. After the throat, the Mach number increases  almost indistinctly for $Q=0.1$ and $Q=0.4$ up to $x \approx 7.5$. At this point, the Mach number labelled by the green curve corresponding to $Q=0.4$ increases slightly faster than the one related to $Q=0.1$. In the point $x \approx 7.5$ along the longitudinal direction of the de Laval nozzle, the Mach number for $Q=1$ increases slower than the Mach number for $Q=0.1$ and $Q=0.4$ up to  $x \approx 7.5$. After that, the rate of increment of the Mach number for $Q=1$ escalates. This phenomenon also happens for $\alpha = 0.5$ fixed, Fig. \ref{fig:Mb1}, however at lower paces. It reflects the fact that increasing the hairy parameter $\alpha$ attenuates the rate of increment 
of the Mach number, when different values of the black hole charge are taken into account. 
Figs. \ref{fig:Ma1} and \ref{fig:Mb1} display the Mach number for the $p$-wave mode. For $\alpha = 0.1$ fixed, Fig. \ref{fig:Ma1} illustrates a faster increment of the Mach number for lower values of the charge $Q$, up to the sonic point. After it along the longitudinal direction, the Mach number increases  almost indistinctly for $Q=0.1$ and $Q=0.4$ up to $x \approx 4.4$, wherefrom the Mach number labelled by the green curve corresponding to $Q=0.4$ increases slightly faster than the one related to $Q=0.1$. Along the longitudinal direction, the Mach number for $Q=1$ increases at a slower rate when compared to the Mach number profile for $Q=0.1$ and $Q=0.4$ up to  $x \approx 8.2$. After this crossover point, the rate of increment of the Mach number for $Q=1$ is higher, when compared to the other values of $Q$ here reported. Differently of the $s$-wave mode, this phenomenon does not occur for $\alpha = 0.5$, as shown in Fig. \ref{fig:Mb1}, showing that the Mach number is practically invariant up to $x\approxeq 2.1$. However, at the crossover point $x\approxeq 4.2$, the influence of the charge $Q$ steps in and  one can realize that the rate of increment of the Mach number for $Q=1$ is higher than the one for $Q=0.4$, which by its turn is slightly higher than the one for $Q=0.1$. Therefore,  for the $p$-wave mode, different values of $\alpha$ produce a distinct profile. Now Figs. \ref{fig:Mc0} and \ref{fig:Mc1} portray the Mach number as a function of the longitudinal direction of the de Laval nozzle, for the GD-extended hairy Reissner--Nordstr\"om metric  \eqref{eq23}, for $Q=1$ fixed, respectively for $l=0$ and $l=1$. The $s$-wave mode  in Fig. \ref{fig:Mc0} show a homogeneous profile for values of $\alpha\in [0.1, 0.8]$. Before reaching the sonic point, the Mach number increases faster for lower values of $\alpha$ and it has a marginally bigger magnitude for higher values of $\alpha$. Beyond the nozzle throat, corresponding to the sonic point, the Mach number, for the value $\alpha = 0.8$, attains lower values when compared to other values of $\alpha$ up to the crossover point $x\approxeq 17.4$, at which it reaches the highest values, compared to other values of $\alpha$. On the other hand, despite the Mach number increasing faster for lower values of $\alpha$ up to the sonic point, for the $p$-wave mode in Fig. \ref{fig:Mc1} and it has a slightly bigger magnitude for higher values of $\alpha$, beyond the sonic point the Mach number starts to vary substantially for different values of $\alpha = 0.8$ and there is no crossover point.


In what follows the plots in Fig. \ref{fig:pressure} show the pressure relative to total quantities representing the thermodynamic state of the reservoir. As the gas exits the de Laval nozzle throat, the increment in the area permits it to undergo a throttling Joule--Thomson-type expansion, wherein the gas irreversibly expands at supersonic speeds from high to low pressure, thrusting the velocity of the mass flow beyond sonic speed without a considerable change in kinetic energy. Fig. \ref{fig:Pa0} shows the relative pressure as a function of the longitudinal direction of the nozzle, for the GD-extended hairy Reissner--Nordstr\"om metric  \eqref{eq23}, for the $s$-wave mode, for $\alpha=0.1$ and three different values of $Q$. The pressure profile is almost equal for the values $Q=0.1$ and $Q=0.4$, however, the relative pressure is lower for $Q=1$ up to the nozzle throat, starting to attain higher values after the sonic point when compared to $Q=0.1$ and $Q=0.4$. For all values of $Q$ addressed, the sonic point is an inflection point for the relative pressure. These properties also hold for $\alpha = 0.5$ in Fig. \ref{fig:Pb0} for the $s$-wave mode, with the only difference comprising the fact that the relative pressure profiles are quite similar for all values of $Q$ analysed. Fig. \ref{fig:Pa1} shows the relative pressure as a function of the longitudinal direction of the nozzle, for the GD-extended hairy Reissner--Nordstr\"om metric  \eqref{eq23}, for the $p$-wave mode, for $\alpha=0.1$ and three different values of $Q$. The pressure profile is again almost equal for the values $Q=0.1$ and $Q=0.4$, however, the relative pressure is lower for $Q=1$ up to the nozzle throat, starting to attain higher values after the sonic point when compared to $Q=0.1$ and $Q=0.4$. Similarly to the $s$-wave mode,  the sonic point is an inflection point for the relative pressure, irrespectively the value of $Q$. These properties also hold for $\alpha = 0.5$ in Fig. \ref{fig:Pb1} for the $p$-wave mode, being the relative pressure profiles quite similar for all values of $Q$ analysed. In all cases here discussed, as the gas enters the de Laval nozzle moving at subsonic velocities, as the cross-sectional area contracts along the nozzle, the gas is forced to accelerate until the velocity becomes sonic at the nozzle throat, where the cross-sectional area is the smallest. From the throat the cross-sectional area then increases along the longitudinal direction, letting the gas expand and the axial velocity become progressively more supersonic.
Figs. \ref{fig:Pc0} and \ref{fig:Pc1} show the relative pressure behavior for different values of $\alpha$, with $Q=1$ fixed, for the $l=0$ and $l=1$. For all the six plots in Fig. \ref{fig:pressure}, the relative pressure is asymptotically vanishing for $x\gg1$ and nearly unit for $x\to-\infty$. One can conclude that the de Laval nozzle will only choke at the throat if the pressure and mass flow through the nozzle are sufficient to reach sonic speeds. Otherwise, no supersonic flow can be achieved, and it will act as a Venturi tube. It demands that the entry pressure to the de Laval nozzle be significantly higher than the ambient one or, equivalently, that the stagnation pressure of the jet must be higher than the ambient back pressure.
The relative pressure profiles in Fig. \ref{fig:pressure} are also relevant to determine the exhaust velocity at the nozzle exit, $v_{\scalebox{.63}{\textsc{e}}}$, out of the de Laval nozzle, generating thrust as 
\beq\label{ve}
v_{\scalebox{.63}{\textsc{e}}} = \left[\frac{RT}{\mu}\frac{2\gamma}{\gamma - 1} \left[1 - \left(\frac{p_0}{p}\right)^{\frac{\gamma - 1}{\gamma}}\right]\right]^{1/2},
\eeq
where $\mu$ is the molecular weight of the gas under scrutiny. 
Having the profiles of the relative pressure appearing in Eq. (\ref{ve}) in Figs. (\ref{fig:Pa0}) -- (\ref{fig:Pc1}) for the GD-extended hairy Reissner--Nordstr\"om metric, the influence of the GD hairy parameters $\alpha$ and $Q$ in the exhaust velocity can be therefore read off Eq. (\ref{ve}). 

Now Fig. \ref{fig:temperature} shows the relative temperature as a function of the longitudinal direction of the de Laval nozzle, for the GD-extended hairy Reissner--Nordstr\"om metric  \eqref{eq23}.
Fig. \ref{fig:Ta0} shows the relative temperature as a function of the longitudinal direction of the nozzle, for the GD-extended hairy Reissner--Nordstr\"om metric  \eqref{eq23}, for the $s$-wave mode, for $\alpha=0.1$ and distinct values of $Q$. The relative temperature profile is almost equal for all the values of $Q$ analysed. Nevertheless, the relative temperature is marginally lower for $Q=1$ up to the nozzle throat, starting to reach higher values after the sonic point, when compared to $Q=0.1$ and $Q=0.4$, up to the crossover point $x\approxeq 6.3$. From that point on, the relative temperature is lower for $Q=1$ and also vanishes asymptotically at a lower rate. The analysis of the $s$-wave mode, for $\alpha=0.5$ in Fig. \ref{fig:Ta1} is quite similar, with the only substantial difference that the crossover point is $x\approxeq 4.0$. The $p$-wave mode analysis emulates the $s$-wave one qualitatively, with different crossover points, respectively for $\alpha = 0.1$ and $\alpha = 0.5$. On the other hand, Fig. \ref{fig:Tc0} illustrates the relative temperature for the $s$-wave mode for $Q=1$  fixed. Although different values of the GD-hairy parameter $\alpha$ present slightly different profiles near the sonic point and their asymptotic values are essentially the same, higher values of $\alpha$ have dominant temperature after the nozzle throat, $x=0$, up to $x\approxeq 17.2$. For $x\gtrsim 17.2$, then the relative temperature becomes indistinct for different values of $\alpha$. Fig. \ref{fig:Tc1}, showing the $l=1$ mode, is qualitatively similar to  Fig. \ref{fig:Tc0}, for the $l=0$ mode, with the main difference comprising the fact that the relative temperatures, for different values of the hairy parameter $\alpha$, are drastically lower along the nozzle longitudinal direction, in general, for the $p$-wave mode.

The plots in Fig. \ref{fig:Density} depict the relative density  as a function of the longitudinal direction of the de Laval nozzle, for the GD-extended hairy Reissner--Nordstr\"om metric  \eqref{eq23}.  All the profiles in Fig. \ref{fig:Density} mimic the relative pressure profiles, respectively in the plots in Fig.  \ref{fig:pressure}. 
These are the quantities aimed to be measured in a laboratory. Reinforcing the discussed behaviour for the multipole parameter $l$, Fig. \ref{fig:temperature} shows that the fluid flow for the $s$-wave mode ($l=0$), by the end of the de Laval nozzle still has  $15.0\%$ the reservoir temperature, whereas the fluid flow for the $p$-wave mode ($l=1$), in the same point, has about $0.9\%$ of the reservoir temperature. This temperature variation is converted into thrust and can be well used in comparing the QN modes. 
The analysis of the Figs. \ref{fig:pressure} and \ref{fig:Density} show that even though the shapes of the nozzles are different, the different gravitational perturbations from metric Eq. \eqref{eq23}, with the same multipole $l$, produce almost the same sound wave in an analog de Laval nozzle.

The thrust coefficients calculated for all the nozzles are shown in Fig. \ref{fig:Cf}. The efficiency of the nozzle in converting thermal energy into kinetic energy is related to the thrust coefficient, \(C_F\), defined as
\begin{equation}
C_F = \frac{{F_{\textsc{thrust}}}}{p_0 {\scalebox{.85}{\textsc{A}}}_\star}, 
\end{equation}
where the thrust can be described as \cite{sutton2016rocket}:
\begin{equation}
    F_{\textsc{thrust}} = p_0{\scalebox{.85}{\textsc{A}}}_\star\gamma\, \sqrt{\frac{2}{\gamma-1}}\qty(\frac{2}{\gamma+1})^{\frac{\gamma+1}{2\gamma-2}}\qty[1-\qty(\frac{p}{p_0})^\frac{\gamma-1}{\gamma}]^{1/2} + {\scalebox{.85}{\textsc{A}}}_{\scalebox{.63}{\textsc{e}}} \qty(p_{\scalebox{.63}{\textsc{e}}} - p_{\textsc{amb}}),
\end{equation}
where \(p_{\scalebox{.63}{\textsc{e}}}\) denotes the pressure at which the gas exits the de Laval nozzle, \(p_{\textsc{amb}}\) stands for the ambient pressure, and ${\scalebox{.85}{\textsc{A}}}_{\scalebox{.63}{\textsc{e}}}$ denotes the area of the nozzle endpoint. 
The thrust coefficient estimates the force that is amplified by the expansion of fluid as it flows through the nozzle, compared to the force triggered if the compression chamber were connected only to the convergent side and throat, but not to the divergent one. 
Dividing the equation by \(p_0 {\scalebox{.85}{\textsc{A}}}_\star\) and considering the nozzle in a vacuum, one obtains
\begin{equation}
    C_F\qty(x) = \gamma\, \sqrt{\frac{2}{\gamma-1}}\qty(\frac{2}{\gamma+1})^{\frac{\gamma+1}{2\gamma-2}}\qty[1-\qty(\frac{p(x)}{p_0})^\frac{\gamma-1}{\gamma}]^{1/2} + \frac{{\scalebox{.85}{\textsc{A}}}(x)}{{\scalebox{.85}{\textsc{A}}}_\star} \frac{p(x)}{p_0}
\end{equation}
 Again we use the area ${\scalebox{.85}{\textsc{A}}}$ and the pressure measured in units of the throat cross-sectional area, ${\scalebox{.85}{\textsc{A}}}_\star$, and total pressure, $p_0$, respectively, in such a way that ${\scalebox{.85}{\textsc{A}}}/{\scalebox{.85}{\textsc{A}}}_\star \mapsto {\scalebox{.85}{\textsc{A}}}$ and $p/p_0 \mapsto p$. The thrust coefficient in de Laval nozzles represents the efficiency of throwing gases out of the nozzle, measuring the de Laval nozzle capacity to turn internal pressure into velocity at the nozzle exit. A higher value of the thrust coefficient complies with a more effective performance of the de Laval nozzle. 
For the $s$-wave case in Figs. \ref{fig:Cfa0} and \ref{fig:Cfa1}, the asymptotic values of the thrust coefficient are essentially the same for $Q=0.1$ and $Q=0.4$, for $l=0$. The case $Q=1$ has a significant change for values in the range $x\gtrsim 5.0$: for $\alpha = 0.1$, Fig. \ref{fig:Cfa0},  the thrust coefficient has a steeper ascent with an inflection point occurring 
 at $x= 3.2$, whereas its inflection point occurs at $x = 1.2$ for Fig. \ref{fig:Cfa1}, for $\alpha = 0.5$. 
 Nevertheless, their asymptotic values are quite similar, comparing fixed values of $Q$ respectively for $\alpha = 0.1$ and $\alpha=0.5$.  
 For $Q=1$, for asymptotically large values of $x$ the thrust coefficient reaches a unique plateau, irrespectively the value of the GD hair parameter  $\alpha$, as shown in Figs. \ref{fig:Cfc0} and \ref{fig:Cfc1}, respectively for the $s$- and $p$-waves. We can numerically check that a similar plateau occurs for asymptotically large values of $x$, for fixed but arbitrary values of $\alpha\lesssim 1$, for $l=1$, as displayed in Figs. \ref{fig:Cfa1} and \ref{fig:Cfb1}, although the behaviour of the thrust coefficient along the longitudinal coordinate along the nozzle is less uniform for $\alpha  = 0.1$ than for $\alpha = 0.5$, respectively. For $\alpha = 0.1$, the $p$-wave mode reaches higher values faster for $Q=0.1$ and $Q=0.4$, when compared to $Q=1$.  Figs. \ref{fig:Cfc0} and \ref{fig:Cfc1} show the profile of the thrust coefficient for several values of $\alpha$, respectively for the $s$- and $p$-wave modes.

{\color{black}
\subsection{QN modes and frequencies}
}

\begin{figure}[H]
 \centering
 \begin{center}
\includegraphics[width=3.95in,height=2.95in]{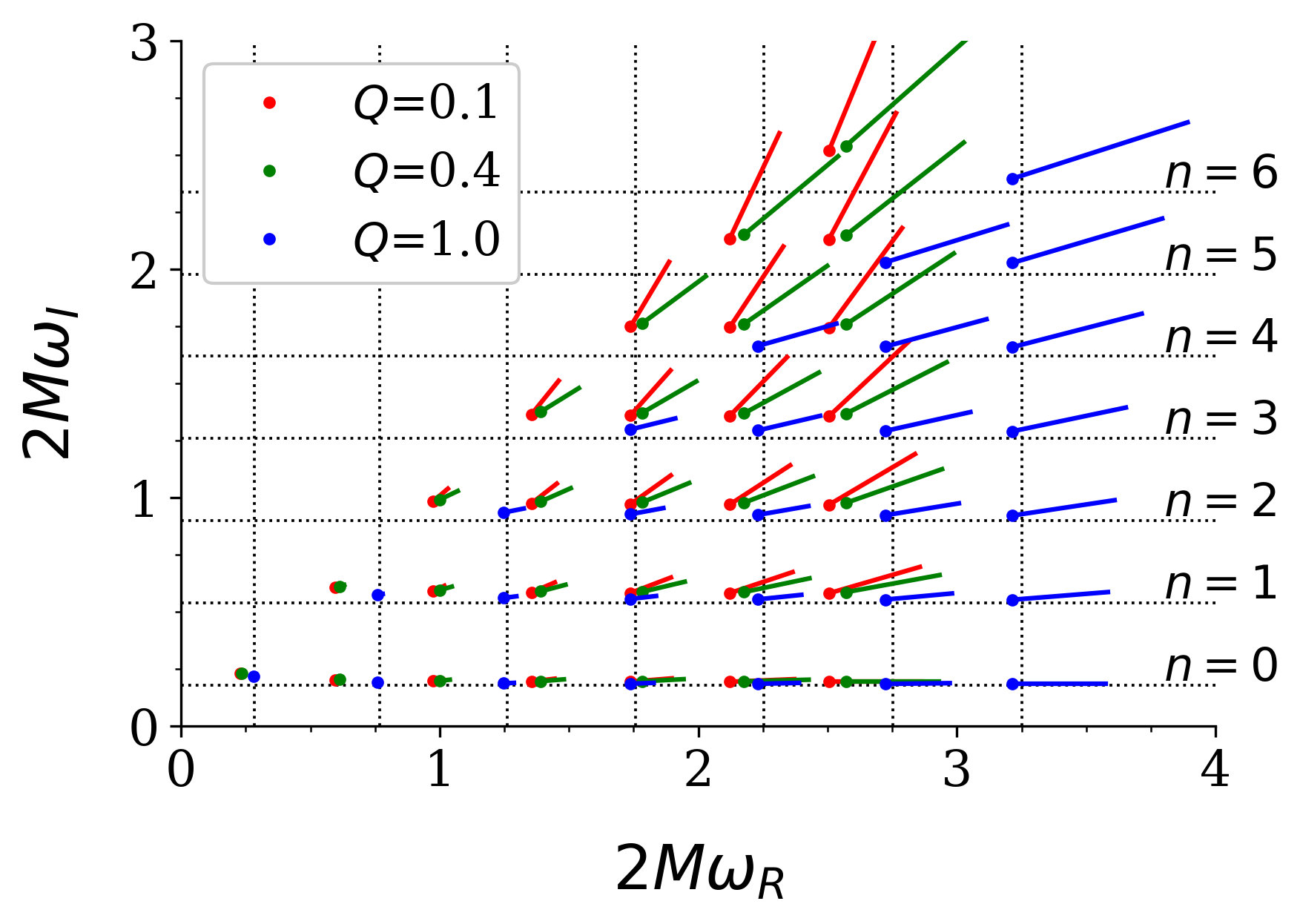}
\end{center}
 \caption{QN modes frequencies, for several overtones, for $Q=0.1, 0.4, 1.0$ and $\alpha = 0.1$. The dotted vertical lines correspond, from the left to the right, to
$l=0,1,\ldots,6$ and the first point on the left corresponds to the $s$-wave $l=0$ for $n=0$. }
 \label{qmnGDRN}
\end{figure}

\begin{figure}[H]
 \centering
 \begin{center}
\includegraphics[width=3.95in,height=2.95in]{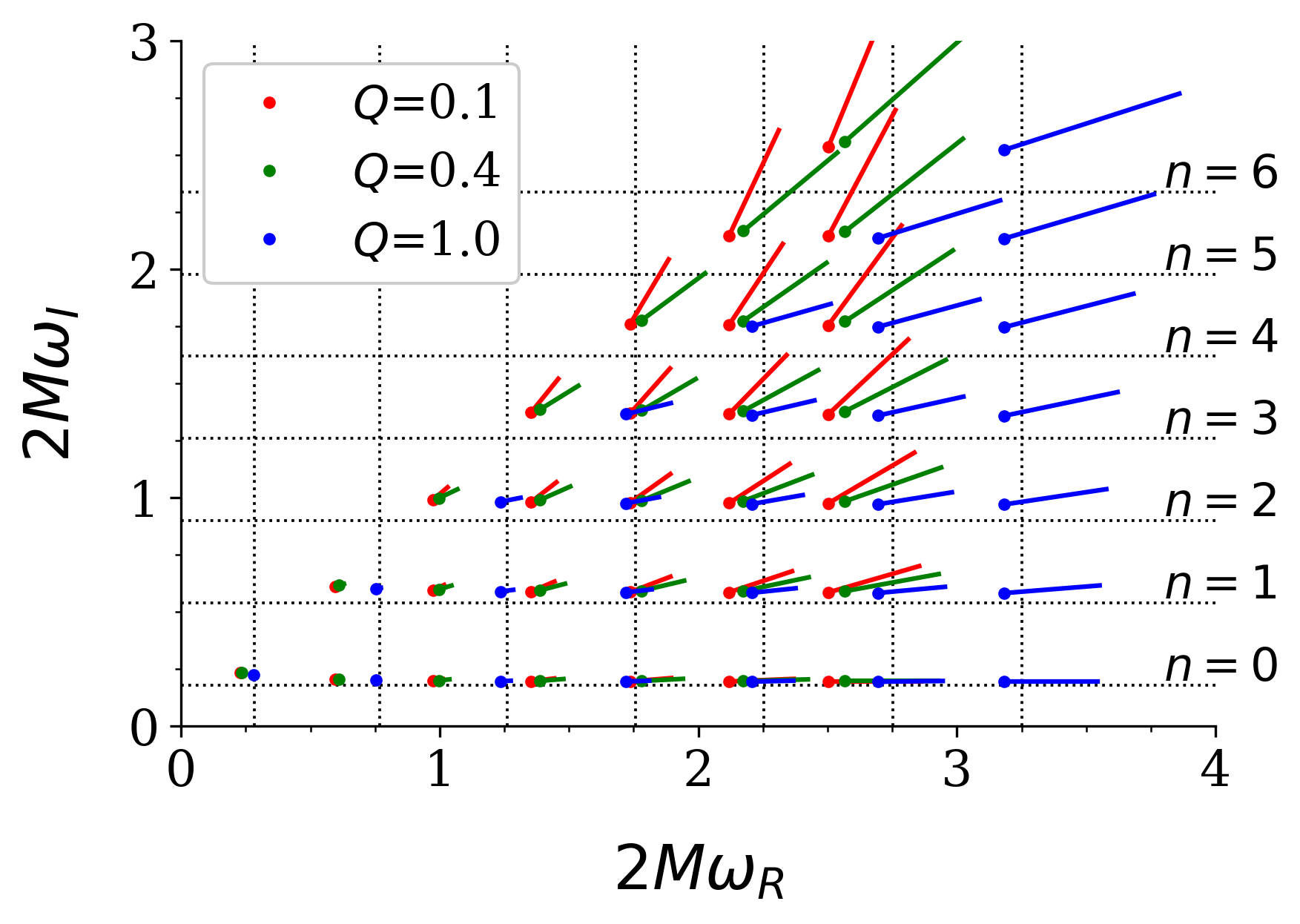}
\end{center}
 \caption{QN modes frequencies, for several overtones, for $Q=0.1, 0.4, 1.0$ and $\alpha = 0.5$. The dotted vertical lines correspond, from the left to the right, to
$l=0,1,\ldots,6$, and the first point on the left corresponds to the $s$-wave $l=0$ for $n=0$. }
 \label{qmnGDRN1}
\end{figure}

 Having addressed the main features of the analog de Laval nozzle, the QN mode frequencies can be calculated. 
A complex QN frequency $\omega_n$ characterises a QN mode, appearing in the Schr\"odinger-like equation \eqref{quasinormal modes equation}, in the gravity side, and in Eq. \eqref{eq:schrodinger acoustic BN}, regarding the aerodynamical analog. 
Although deriving the QN modes analytically can be an extremely intricate task, some approximation techniques can be employed. Among them, the  Mashhoon procedure will be used hereon, due to clarity and reasonability \cite{ferrari1984new}. The relevant QN boundary conditions for asymptotically flat black holes, as the one here investigated, are consistent from the astrophysical point of view \cite{KonoplyaZhidenko:2011}. Most of the complicacy involved in calculating QN modes of several black holes arises from the fact that the effective potential slowly decays when it approaches the radial infinity asymptotically. Due to a branch cut, GWs backscatter off the effective potential, producing backward tails. The Mashhoon method avoids ill features and QN mode frequencies can be accurately computed by a P\"oschl--Teller effective potential, which decays exponentially for $r_\star\to\infty$, 
\begin{equation}
V_{{\scalebox{.55}{\textsc{PT}}}}(r_\star) = \frac{V_0}{\cosh^2
[\upgamma(r_\star - r_{\star0})]},
\end{equation}
where $V_0=V(r_0)$ represents the height of the potential and 
\begin{align}
\upgamma = \sqrt{\frac1{2V_0}\lim_{r\to r_0}\dfrac{\dd^2V}{\dd r_\star^2}}.
\end{align}
The boundary conditions for the Schr\"odinger-like equation are compatible with a vanishing wave function at the boundary. Hence the QN modes are reduced to the bound states for an inverse $(V\mapsto -V)$ effective potential. 
The QN mode frequencies read \cite{ferrari1984new,KonoplyaZhidenko:2011,BertiCardoso:2009} 
\begin{align}
 \omega_n = \sqrt{V_0 - \frac{\upgamma^2}{4}} + i\gamma\left(n + \frac{1}{2}\right).\label{ot}
\end{align}
The natural number $n$ in Eq. (\ref{ot}) is the overtone \cite{KonoplyaZhidenko:2011,BertiCardoso:2009}. 
Table \ref{tabela} displays the QN mode frequencies for the GD-extended hairy Reissner--Nordstr\"om metric, for several values of $l$ and $n$, for different values of the GD hairy parameter $\alpha$ regulating the tensor-vacuum. For fixed values of $\alpha$, the higher the value of $l$, the greater Re$(\omega_n)$ is. For fixed values of the multipole $l$, corresponding to fixed values of Re$(\omega_n)$, the higher the overtone and the greater Im$(\omega_n)$ are. For both $l$ and $n$ simultaneously fixed, the higher the value of $\alpha>0$,  the lower both Re$(\omega_n)$ and Im$(\omega_n)$ are.  
\begin{table}[H]
{\small{\begin{tabular}{||c|c||c|c|c|c|c||}
\hline\hline
${l}$ & $\bf \emph{n}$ & $ {{\scalebox{.9}{${\alpha}$}}}=0.1$ & $ {{\scalebox{.9}{${\alpha}$}}}=0.2$ & ${{\scalebox{.9}{${\alpha}$}}}=0.4$ & $ {{\scalebox{.9}{${\alpha}$}}}=0.6$ & ${{\scalebox{.9}{${\alpha}$}}}=0.8$ \\ \hline\hline

\,0 \,&\, 0 \,&\, 0.2802 + 0.2192$i$ \,&\, 0.2802 + 0.2221$i$ \,&\, 0.2807 + 0.2255$i$ \,&\, 0.2826 + 0.2221$i$ \,&\, 0.2814 + 0.2156$i$ \,\\ \hline

\,1 \,&\, 0 \,&\, 0.7599 + 0.1918$i$ \,&\, 0.7536 + 0.1964$i$ \,&\, 0.7501 + 0.2011$i$ \,&\, 0.7581 + 0.1981$i$ \,&\, 0.7678 + 0.1872$i$ \,\\ \hline

\,1 \,&\, 1 \,&\, 0.7599 + 0.5753$i$ \,&\, 0.7536 + 0.5893$i$ \,&\, 0.7501 + 0.6033$i$ \,&\, 0.7581 + 0.5942$i$ \,&\, 0.7678 + 0.5617$i$ \,\\ \hline

\,2 \,&\, 0 \,&\, 1.2470 + 0.1870$i$ \,&\, 1.2361 + 0.1919$i$ \,&\, 1.2299 + 0.1967$i$ \,&\, 1.2437 + 0.1938$i$ \,&\, 1.2606 + 0.1825$i$ \,\\ \hline

\,2 \,&\, 1 \,&\, 1.2470 + 0.5609$i$ \,&\, 1.2361 + 0.5756$i$ \,&\, 1.2299 + 0.5900$i$ \,&\, 1.2437 + 0.5814$i$ \,&\, 1.2606 + 0.5474$i$ \,\\ \hline

\,2 \,&\, 2 \,&\, 1.2470 + 0.9349$i$ \,&\, 1.2361 + 0.9593$i$ \,&\, 1.2299 + 0.9833$i$ \,&\, 1.2437 + 0.9690$i$ \,&\, 1.2606 + 0.9123$i$ \,\\ \hline

\,3 \,&\, 0 \,&\, 1.7375 + 0.1855$i$ \,&\, 1.7222 + 0.1904$i$ \,&\, 1.7135 + 0.1953$i$ \,&\, 1.7330 + 0.1925$i$ \,&\, 1.7568 + 0.1810$i$ \,\\ \hline

\,3 \,&\, 1 \,&\, 1.7375 + 0.5565$i$ \,&\, 1.7222 + 0.5713$i$ \,&\, 1.7135 + 0.5859$i$ \,&\, 1.7330 + 0.5775$i$ \,&\, 1.7568 + 0.5430$i$ \,\\ \hline

\,3 \,&\, 2 \,&\, 1.7375 + 0.9275$i$ \,&\, 1.7222 + 0.9522$i$ \,&\, 1.7135 + 0.9765$i$ \,&\, 1.7330 + 0.9625$i$ \,&\, 1.7568 + 0.9050$i$ \,\\ \hline

\,3 \,&\, 3 \,&\, 1.7375 + 1.2985$i$ \,&\, 1.7222 + 1.3331$i$ \,&\, 1.7135 + 1.3671$i$ \,&\, 1.7330 + 1.3475$i$ \,&\, 1.7568 + 1.2669$i$ \,\\ \hline

\,4 \,&\, 0 \,&\, 2.2295 + 0.1849$i$ \,&\, 2.2097 + 0.1898$i$ \,&\, 2.1985 + 0.1947$i$ \,&\, 2.2236 + 0.1919$i$ \,&\, 2.2544 + 0.1804$i$ \,\\ \hline

\,4 \,&\, 1 \,&\, 2.2295 + 0.5546$i$ \,&\, 2.2097 + 0.5695$i$ \,&\, 2.1985 + 0.5842$i$ \,&\, 2.2236 + 0.5758$i$ \,&\, 2.2544 + 0.5411$i$ \,\\ \hline

\,4 \,&\, 2 \,&\, 2.2295 + 0.9244$i$ \,&\, 2.2097 + 0.9492$i$ \,&\, 2.1985 + 0.9736$i$ \,&\, 2.2236 + 0.9597$i$ \,&\, 2.2544 + 0.9018$i$ \,\\ \hline

\,4 \,&\, 3 \,&\, 2.2295 + 1.2942$i$ \,&\, 2.2097 + 1.3289$i$ \,&\, 2.1985 + 1.3630$i$ \,&\, 2.2236 + 1.3436$i$ \,&\, 2.2544 + 1.2626$i$ \,\\ \hline

\,4 \,&\, 4 \,&\, 2.2295 + 1.6639$i$ \,&\, 2.2097 + 1.7086$i$ \,&\, 2.1985 + 1.7525$i$ \,&\, 2.2236 + 1.7275$i$ \,&\, 2.2544 + 1.6233$i$ \,\\ \hline

\,5 \,&\, 0 \,&\, 2.7221 + 0.1846$i$ \,&\, 2.6978 + 0.1895$i$ \,&\, 2.6841 + 0.1944$i$ \,&\, 2.7150 + 0.1917$i$ \,&\, 2.7526 + 0.1800$i$ \,\\ \hline

\,5 \,&\, 1 \,&\, 2.7221 + 0.5537$i$ \,&\, 2.6978 + 0.5686$i$ \,&\, 2.6841 + 0.5833$i$ \,&\, 2.7150 + 0.5750$i$ \,&\, 2.7526 + 0.5401$i$ \,\\ \hline

\,5 \,&\, 2 \,&\, 2.7221 + 0.9228$i$ \,&\, 2.6978 + 0.9477$i$ \,&\, 2.6841 + 0.9721$i$ \,&\, 2.7150 + 0.9583$i$ \,&\, 2.7526 + 0.9002$i$ \,\\ \hline

\,5 \,&\, 3 \,&\, 2.7221 + 1.2919$i$ \,&\, 2.6978 + 1.3267$i$ \,&\, 2.6841 + 1.3609$i$ \,&\, 2.7150 + 1.3416$i$ \,&\, 2.7526 + 1.2603$i$ \,\\ \hline

\,5 \,&\, 4 \,&\, 2.7221 + 1.6610$i$ \,&\, 2.6978 + 1.7058$i$ \,&\, 2.6841 + 1.7498$i$ \,&\, 2.7150 + 1.7249$i$ \,&\, 2.7526 + 1.6204$i$ \,\\ \hline

\,5 \,&\, 5 \,&\, 2.7221 + 2.0302$i$ \,&\, 2.6978 + 2.0849$i$ \,&\, 2.6841 + 2.1386$i$ \,&\, 2.7150 + 2.1083$i$ \,&\, 2.7526 + 1.9805$i$ \,\\ \hline

\,6 \,&\, 0 \,&\, 3.2151 + 0.1844$i$ \,&\, 3.1864 + 0.1894$i$ \,&\, 3.1702 + 0.1942$i$ \,&\, 3.2067 + 0.1915$i$ \,&\, 3.2512 + 0.1799$i$ \,\\ \hline

\,6 \,&\, 1 \,&\, 3.2151 + 0.5531$i$ \,&\, 3.1864 + 0.5681$i$ \,&\, 3.1702 + 0.5827$i$ \,&\, 3.2067 + 0.5745$i$ \,&\, 3.2512 + 0.5396$i$ \,\\ \hline

\,6 \,&\, 2 \,&\, 3.2151 + 0.9219$i$ \,&\, 3.1864 + 0.9468$i$ \,&\, 3.1702 + 0.9712$i$ \,&\, 3.2067 + 0.9575$i$ \,&\, 3.2512 + 0.8993$i$ \,\\ \hline

\,6 \,&\, 3 \,&\, 3.2151 + 1.2906$i$ \,&\, 3.1864 + 1.3255$i$ \,&\, 3.1702 + 1.3597$i$ \,&\, 3.2067 + 1.3405$i$ \,&\, 3.2512 + 1.2590$i$ \,\\ \hline

\,6 \,&\, 4 \,&\, 3.2151 + 1.6594$i$ \,&\, 3.1864 + 1.7042$i$ \,&\, 3.1702 + 1.7482$i$ \,&\, 3.2067 + 1.7234$i$ \,&\, 3.2512 + 1.6188$i$ \,\\ \hline

\,6 \,&\, 5 \,&\, 3.2151 + 2.0281$i$ \,&\, 3.1864 + 2.0829$i$ \,&\, 3.1702 + 2.1367$i$ \,&\, 3.2067 + 2.1064$i$ \,&\, 3.2512 + 1.9785$i$ \,\\ \hline

\,6 \,&\, 6 \,&\, 3.2151 + 2.3969$i$ \,&\, 3.1864 + 2.4616$i$ \,&\, 3.1702 + 2.5252$i$ \,&\, 3.2067 + 2.4894$i$ \,&\, 3.2512 + 2.3382$i$ \,\\ \hline

\hline\hline
\end{tabular}}}

\caption{QN modes frequencies $(2M\omega)$ for the GD-extended hairy Reissner--Nordstr\"om metric, for varying value of $\alpha$ and $Q=1$, for several overtones and quantum azimuthal numbers $l=0,1,\ldots,6$. }

\label{tabela}

\end{table}
Since the value $\alpha = 0$ corresponds to the well-known Schwarzschild case, to avoid redundancy with the literature it is not listed in Table \ref{tabela}. Compared to the Schwarzschild case and even to other GD-black hole solutions, as the one in Ref. \cite{Cavalcanti:2022cga}, the quality factor for the GD-extended hairy Reissner--Nordstr\"om metric (\ref{eq23}) is the largest for higher overtones, indicating the influence of the GD hairy parameters here are consonant with experimental aspects of the analog de Laval nozzle. 
The QN mode frequencies of  GD-extended hairy Reissner--Nordstr\"om metric have a noticeable deviation when compared to the Schwarzschild QN mode frequencies, increasing for greater values of $n$. 
Figs. \ref{qmnGDRN} and \ref{qmnGDRN1} depict the real and imaginary values of the QN mode frequencies for several values of $l$, for different overtones, respectively for $\alpha = 0.1$ and $\alpha = 0.5$. 

\begin{figure}[H]
 \centering
 \begin{center}
\includegraphics[width=3.95in,height=2.95in]{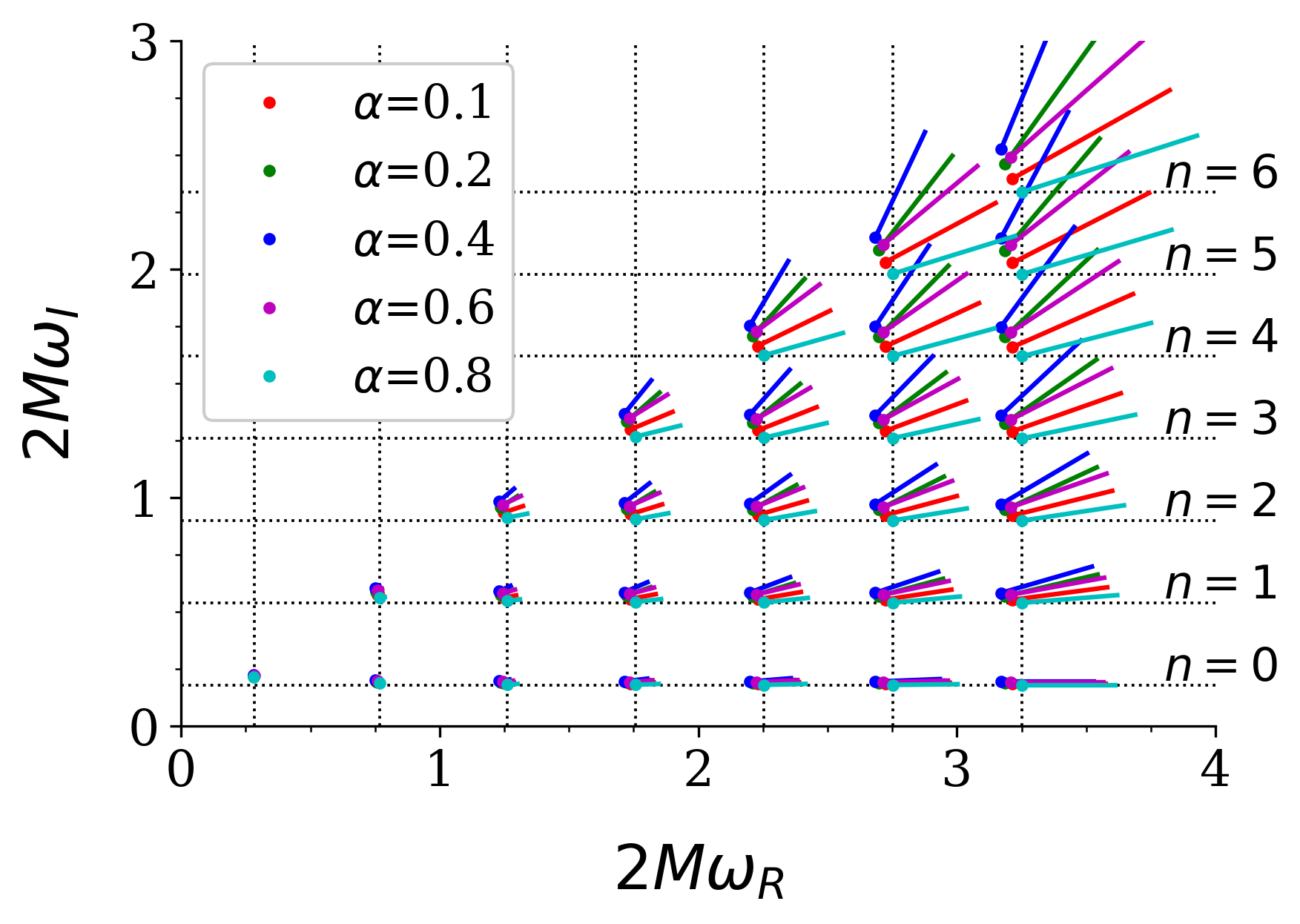}
\end{center}
 \caption{QN modes frequencies, for several overtones. The coloured lines for a given pair $(l,n)$ correspond to a fixed $\alpha$, for $Q=1$. The dotted vertical lines correspond, from the left to the right, to
$l=0,1,\ldots,6$, and the first point on the left corresponds to the $s$-wave $l=0$ for $n=0$.}
 \label{qmnGDRN2}
\end{figure}

As posed in Ref. \cite{Okuzumi:2007hf}, the nozzle quality factor $q_n \sim {\rm Re}(\omega_n)/{\rm Im}(\omega_n)$ is a quantity that is proportional to the number of oscillation cycles in the damping process. Since the QN ringing is everywhere concealed by noise after a few damping periods,  it is crucial to engineer a de Laval nozzle that produces QN modes compatible with higher values of the quality factor, to effectively detect the QN ringing. 
In what concerns  Figs. \ref{qmnGDRN} -- \ref{qmnGDRN2}, we can analyse the spectrum of QN modes that is more favourable to providing a quality factor that is appropriate for experimental apparatuses. 
Fig. \ref{qmnGDRN2} depicts the QN modes for the GD-extended hairy Reissner--Nordstr\"om metric, with a large superposition of QN modes for increasing values of the overtones $n$, contrasted to the Schwarzschild QN mode frequencies, for several values of the GD hairy parameter $\alpha$ regulating the tensor-vacuum. For each fixed value of $n$, the quality factor increases as a function of the multipole $l$ to a great extent. For fixed values of $l$, the higher the overtone, the lower the quality factor is. Interestingly, for fixed values of $l$ and $n$, the higher the GD hairy charge $\alpha$, the higher the quality factor of the nozzle is. In this way, the QN ringing can be better detected for higher values of the GD parameter $\alpha$, for each fixed value of $l$ in all overtones.

{\color{black}
\section{Conclusions}
\label{sec5}

Acoustic black holes can be realized by stable transonic fluid flows in a de~Laval nozzle. 
GD-extended Reissner--Nordstr\"om black hole solutions were used here to define the properties 
of an analog de Laval nozzle in the laboratory. 
The QN modes and frequencies of sound waves in the analog de Laval nozzle
were then computed and that could provide ways of testing experimentally potential features 
of the corresponding GD-extended black holes.  
In fact, the wave equation for gravitational perturbations can be mapped into the
wave equation of the aerodynamical systems with similar effective potentials.
It was then shown how the values of the black hole hairy parameters $\alpha$ and $Q$ 
determines the nozzle shape and consequently affects the pressure, Mach number,
temperature, density, and the thrust coefficient, for both $s$- and $p$-wave modes. 
The QN modes were finally computed for different values of the parameters $\alpha$
and $Q$, for several overtones, and compared to the Schwarzschild case. 
The quality factor of the nozzle was also analysed in terms of the QN mode frequencies. 
The QN modes correspond to spin-0 perturbations. Any realistic GW, corresponding to spin-2 perturbations, 
have no analog gravity in the nozzle. 
  \par
A possible extension of the present work would be to consider analog models of
GD-extended black holes with rotation.
This will require finding a way to map the gravitational perturbations into quasi-one-dimensional
transonic fluid flows that can be realised in a laboratory.

\subsection*{Acknowledgements}
R.C. is partially supported by the INFN grant FLAG and his work has also been carried out
in the framework of activities of the National Group of Mathematical Physics (GNFM, INdAM).
C.N.S.~thanks CAPES (Grant No.~001).
RdR~thanks the S\~ao Paulo Research Foundation -- FAPESP (Grants No.~2021/01089-1
and No.~2022/01734-7),  
The National Council for Scientific and Technological Development (CNPq) (Grants No. 303390/2019-0 and No. 401567/2023-0), and the Coordination for the Improvement of
Higher Education Personnel (CAPES-PrInt  88887.897177/2023-00), for partial financial support.
R.dR~thanks R.C.~and DIFA -- Universit\`a di Bologna, for the hospitality,
and R.T.~Cavalcanti, for fruitful discussions.

}

\appendix
\section{Analytical Solution for $\lambda\qty(r)$}
From Eq. \eqref{eq: recurrence eq}, written again below, our goal is to expand $\lambda(r)$ by the Frobenius method.
\begin{equation}
    \qty(n+2)\qty(n+1) a_{n+2} = \frac{V_{\text{eff}}(r)}{\mathcal{F}(r)^2} a_n - \frac{\mathcal{F}'(r)}{\mathcal{F}(r)} \qty(n+1)a_{n+1}
\end{equation}
For a convergent series, one can separate it into even and odd powers, as:
\begin{align}
    \lambda\qty(r) &= \sum_{k=0}^{\infty} a_k \qty(r-r_h)^k\nonumber\\ 
    &= a_0 + a_1\qty(r-r_h) + \sum_{k=1}^{\infty} a_{2k} \qty(r-r_h)^{2k}+\sum_{k=1}^{\infty} a_{2k+1} \qty(r-r_h)^{2k+1},
\end{align}
With initial conditions (\ref{eq:condinilamb}), 
the series can be expanded as:
\begin{equation*}
a_k =
   \frac{\qty(-1)^k V a_0}{k! \ \mathcal{F}^k} \ 
   \textcolor{lightgray}{\begin{matrix}
2 \\ 
3 \\ 
4 \\ 
5 \\ 
6 \\ 
7 \\ 
8 \\ 
9 \\ 
10 \\ 
11 \\ 
\vdots
\end{matrix}}
\begin{pmatrix}
{1} \\
{\mathcal{F}'} \\
{V+\mathcal{F}'^2} \\
{\mathcal{F}'^3+2V\mathcal{F}'} \\
{V^2+\mathcal{F}'^4+3V\mathcal{F}'^2} \\
{\mathcal{F}'^5+4V\mathcal{F}'^3+3V^2\mathcal{F}'} \\
{V^3+\mathcal{F}'^6+5V\mathcal{F}'^4+6V^2\mathcal{F}'^2} \\
{\mathcal{F}'^7+6V\mathcal{F}'^5+10V^2\mathcal{F}'^3+4V^3\mathcal{F}'} \\
{V^4+\mathcal{F}'^8+7V\mathcal{F}'^6+15V^2\mathcal{F}'^4+10V^3\mathcal{F}'^2} \\
{\mathcal{F}'^9+8V\mathcal{F}'^7+21V^2\mathcal{F}'^5+20V^3\mathcal{F}'^3+5V^4\mathcal{F}'} \\
\vdots
\end{pmatrix}_{\text{position }k-2}
\text{, for }k\geq 2.
\end{equation*}
For $k$ even, it wits
\begin{equation*}
a_{2k} =
   \frac{\qty(-1)^{2k} V a_0}{(2k)! \ \mathcal{F}^{2k}} \ 
\begin{pmatrix}
{1} \\
{V+\mathcal{F}'^2} \\
{V^2+\mathcal{F}'^4+3V\mathcal{F}'^2} \\
{V^3+\mathcal{F}'^6+5V\mathcal{F}'^4+6V^2\mathcal{F}'^2} \\
{V^4+\mathcal{F}'^8+7V\mathcal{F}'^6+15V^2\mathcal{F}'^4+10V^3\mathcal{F}'^2} \\
\vdots
\end{pmatrix}_{\text{position }(k-1)}
\text{, for }k\geq 2 \text{ even},
\end{equation*}
whereas for $k$ odd it yields:
\begin{equation*}
a_{2k+1} =
   \frac{\qty(-1)^{2k+1} a_0 V \mathcal{F}'}{(2k+1)! \ \mathcal{F}^{2k+1}} \ 
\begin{pmatrix}
{1} \\
{\mathcal{F}'^2+2V} \\
{\mathcal{F}'^4+4V\mathcal{F}'^2+3V^2} \\
{\mathcal{F}'^6+6V\mathcal{F}'^4+10V^2\mathcal{F}'^2+4V^3} \\
{\mathcal{F}'^8+8V\mathcal{F}'^6+21V^2\mathcal{F}'^4+20V^3\mathcal{F}'^2+5V^4} \\
\vdots
\end{pmatrix}_{\text{position }(k-1)}
\text{, for }k\geq 2 \text{ odd}.
\end{equation*}
Hence, the equations for each parity of $k$ can be obtained. 

\noindent For even coefficients, one can write
\begin{equation}
    a_{2k} = \frac{\qty(-1)^{2k} V a_0}{(2k)! \ \mathcal{F}^{2k}} \sum_{n=0}^{k-1} V^n \mathcal{F}'^{2k-2n-2} \mathcal{C}^\text{ even}_{n}\qty({k-1})
\end{equation}
where $\mathcal{C}^\text{ even}_{n}\qty(x)$ reads
\begin{align}
\mathcal{C}^\text{ even}_n\qty(x) = \lim_{m \rightarrow x}\frac{\qty(-1)^n\Gamma\qty(2n-2m)}{\Gamma\qty(n+1)\Gamma\qty(n-2m)} \text{ , for }x\geq n.
\end{align}
It is worth illustrating the lowest degree terms
\begin{align*}
    \mathcal{C}^\text{ even}_{0}\qty(x) &= 1 \\
    \mathcal{C}^\text{ even}_{1}\qty(x) &= \qty(2x-1) \\
    \mathcal{C}^\text{ even}_{2}\qty(x) &= \qty(x-1)\qty(2x-3) \\
    \mathcal{C}^\text{ even}_{3}\qty(x) &= \qty(x-2)\qty(2x-5)\qty(2x-3)/3 \\
    \mathcal{C}^\text{ even}_{4}\qty(x) &= \qty(x-3)\qty(x-2)\qty(2x-7)\qty(2x-5)/6 \\
    \mathcal{C}^\text{ even}_{5}\qty(x) &= \qty(x-4)\qty(x-3)\qty(2x-9)\qty(2x-7)\qty(2x-5)/30 \\
    \vdots &
\end{align*}
Therefore, the coefficient $a_{2k}$ reads
\begin{equation}
    a_{2k} = \frac{a_0 }{\qty(2k)!} \frac{V_{\text{eff}}}{\mathcal{F}'^2} \qty( \frac{\mathcal{F}'}{\mathcal{F}})^{2k} \  \sum_{n=0}^{k-1} \qty(\frac{V_{\text{eff}}}{\mathcal{F}'^{2}})^n  \mathcal{C}^\text{ even}_{n}\qty(k-1).
\end{equation}


\noindent Now, for $k$ odd, following a similar methodology,  
the coefficient $a_{2k+1}$ wits
\begin{equation}
    a_{2k+1} =  \frac{- a_0  }{\qty(2k+1)!} \frac{V_{\text{eff}}}{\mathcal{F}'^2} \qty(\frac{\mathcal{F}'}{\mathcal{F}})^{2k+1} \ \sum_{n=0}^{k-1} \qty(\frac{V_{\text{eff}}}{\mathcal{F}'^{2}})^n \mathcal{C}^\text{ odd}_{n}\qty(k-1),
\end{equation}
where 
\begin{align}
\mathcal{C}^\text{ odd}_n\qty(x) = \lim_{m \rightarrow x}\frac{\qty(-1)^n \Gamma\qty(2n-2m-1)}{\Gamma\qty(n+1)\Gamma\qty(n-2m-1)} \text{ , for }x\geq n,
\end{align}
whose lowest degree terms read
\begin{align*}
    \mathcal{C}^\text{ odd}_{0}\qty(x) &= 1 \\
    \mathcal{C}^\text{ odd}_{1}\qty(x) &= 2x \\
    \mathcal{C}^\text{ odd}_{2}\qty(x) &= \qty(x-1)\qty(2x-1) \\
    \mathcal{C}^\text{ odd}_{3}\qty(x) &= \qty(x-2)\qty(x-1)\qty(2x-3) \ 2/3 \\
    \mathcal{C}^\text{ odd}_{4}\qty(x) &= \qty(x-3)\qty(x-2)\qty(2x-5)\qty(2x-3)/6 \\
    \mathcal{C}^\text{ odd}_{5}\qty(x) &= \qty(x-4)\qty(x-3)\qty(x-2)\qty(2x-7)\qty(2x-5) \ 2/30 \\
    \vdots &
\end{align*}
Putting these results together yields
\begin{align}
    \frac{\lambda\qty(r)}{a_0} = 1 
    + \sum_{k=1}^{\infty} &\left\{ \qty[\frac{1}{\qty(2k)!} \frac{V_{\text{eff}}}{\mathcal{F}'^2} \qty(\frac{\mathcal{F}'}{\mathcal{F}})^{2k} \  \sum_{n=0}^{k-1} \qty(\frac{V_{\text{eff}}}{\mathcal{F}'^{2}})^n \mathcal{C}^\text{ even}_{n}\qty(k-1)] \qty(r-r_h)^{2k}- \right.\nonumber\\
    &\left.- \qty[\frac{1}{\qty(2k+1)!} \frac{V_{\text{eff}}}{\mathcal{F}'^2} \qty(\frac{\mathcal{F}'}{\mathcal{F}})^{2k+1} \ \sum_{n=0}^{k-1} \qty(\frac{V_{\text{eff}}}{\mathcal{F}'^{2}})^n \mathcal{C}^\text{ odd}_{n}\qty(k-1)] \qty(r-r_h)^{2k+1}\right\}.\label{eqrt}
\end{align}
To make the construction of Eq. (\ref{eqrt}), Tables \ref{ii} and \ref{iii}   contain values of $\mathcal{C}^\text{ even}_{n}\qty(x)$ and $\mathcal{C}^\text{ odd}_{n}\qty(x)$:
\begin{table}[H]
\centering
\begin{tabular}{|c|cccccccccccc|}
\hline
\textbf{$x$↓  $n$→} & 0 & 1  & 2   & 3   & 4     & 5     & 6     & 7     & 8   & 9  & 10 & $\hdots$ \\
\hline
0               & 1 &    &     &     &       &       &       &       &     &    &   & \\
1               & 1 & 1  &     &     &       &       &       &       &     &    &   & \\
2               & 1 & 3  & 1   &     &       &       &       &       &     &    &   & \\
3               & 1 & 5  & 6   & 1   &       &       &       &       &     &    &   & \\
4               & 1 & 7  & 15  & 10  & 1     &       &       &       &     &    &   & \\
5               & 1 & 9  & 28  & 35  & 15    & 1     &       &       &     &    &   & \\
6               & 1 & 11 & 45  & 84  & 70    & 21    & 1     &       &     &    &   & \\
7               & 1 & 13 & 66  & 165 & 210   & 126   & 28    & 1     &     &    &   & \\
8               & 1 & 15 & 91  & 286 & 495   & 462   & 210   & 36    & 1   &    &   & \\
9               & 1 & 17 & 120 & 455 & 1,001 & 1,287 & 924   & 330   & 45  & 1  &   & \\
10              & 1 & 19 & 153 & 680 & 1,820 & 3,003 & 3,003 & 1,716 & 495 & 55 & 1 & \\ 
$\vdots$ &&&&&&&&&&&&\\
\hline
\end{tabular}
\caption{Values of $\mathcal{C}^\text{ even}_{n}\qty(x)$.}
\label{ii}
\end{table}
\begin{table}[H]
\centering
\begin{tabular}{|c|cccccccccccc|}
\hline
\textbf{$x$↓  $n$→} & 0 & 1  & 2   & 3   & 4     & 5     & 6     & 7     & 8     & 9   & 10 & $\hdots$  \\
\hline
0               & 1 &    &     &     &       &       &       &       &       &     &  &  \\
1               & 1 & 2  &     &     &       &       &       &       &       &     & &   \\
2               & 1 & 4  & 3   &     &       &       &       &       &       &     &  &  \\
3               & 1 & 6  & 10  & 4   &       &       &       &       &       &     &   & \\
4               & 1 & 8  & 21  & 20  & 5     &       &       &       &       &     & &   \\
5               & 1 & 10 & 36  & 56  & 35    & 6     &       &       &       &     & &   \\
6               & 1 & 12 & 55  & 120 & 126   & 56    & 7     &       &       &     &  &  \\
7               & 1 & 14 & 78  & 220 & 330   & 252   & 84    & 8     &       &     &  &  \\
8               & 1 & 16 & 105 & 364 & 715   & 792   & 462   & 120   & 9     &     &  &  \\
9               & 1 & 18 & 136 & 560 & 1,365 & 2,002 & 1,716 & 792   & 165   & 10  &  &  \\
10              & 1 & 20 & 171 & 816 & 2,380 & 4,368 & 5,005 & 3,432 & 1,287 & 220 & 11 &\\
$\vdots$ &&&&&&&&&&&&\\
\hline
\end{tabular}
\caption{Values of $\mathcal{C}^\text{ odd}_{n}\qty(x)$.}
\label{iii}
\end{table}
A test of convergence can proceed, assuming that $V_\text{eff} = \mathcal{F} = \mathcal{F}' = 1$. Therefore, the series converges if 
\beq
    \lim_{x\rightarrow \infty} \frac{1}{\qty(2x)!} \sum_{n=0}^{x} \mathcal{C}^\text{ even}_{n}\qty(x) = 0, \label{conv1}\\
    \lim_{x\rightarrow \infty} \frac{1}{\qty(2x+1)!} \sum_{n=0}^{x} \mathcal{C}^\text{ odd}_{n}\qty(x) = 0.\label{conv2}
\eeq
Eqs. (\ref{conv1}, \ref{conv2}) can be numerically verified, as 
\begin{table}[H]
\centering 
\begin{tabular}{|c|c|c|}
\hline
$x$ & $\frac{1}{\qty(2x)!} \sum_{n=0}^{x} \mathcal{C}^\text{ even}_{n}\qty(x)$  &$ \frac{1}{\qty(2x+1)!} \sum_{n=0}^{x} \mathcal{C}^\text{ odd}_{n}\qty(x)$  \\
\hline
0          & 1 & 1 \\
1          & 1 & $5.0\times 10^{-1}$ \\
2          & $2.1\times 10^{-1}$ & $6.7\times 10^{-2}$ \\
3          & $1.8\times 10^{-2}$ & $4.2\times 10^{-3}$ \\
4          & $8.4\times 10^{-4}$ & $1.5\times 10^{-4}$ \\
5          & $2.5\times 10^{-5}$ & $3.6\times 10^{-6}$ \\
6          & $4.9\times 10^{-7}$ & $6.1\times 10^{-8}$ \\
7          & $7.0\times 10^{-9}$ & $7.5\times 10^{-10}$ \\
8          & $7.6\times 10^{-11}$ & $7.3\times 10^{-12}$ \\
9          & $6.5\times 10^{-13}$ & $5.6\times 10^{-14}$ \\
10         & $4.5\times 10^{-15}$ & $3.5\times 10^{-16}$ \\
$\vdots$ &$\vdots$&$\vdots$ \\
\hline
\end{tabular}
\end{table}
As we can see, it numerically converges.

\bibliography{bib_analog_GD}
\end{document}